\documentstyle[psfig]{mn}

\newif\ifAMStwofonts

\newcommand{\ltsimeq}{\raisebox{-0.6ex}{$\,\stackrel
{\raisebox{-.2ex}{$\textstyle <$}}{\sim}\,$}}
\newcommand{\gtsimeq}{\raisebox{-0.6ex}{$\,\stackrel
{\raisebox{-.2ex}{$\textstyle >$}}{\sim}\,$}}

\ifoldfss
  \ifCUPmtlplainloaded \else
    \NewTextAlphabet{textbfit} {cmbxti10} {}
    \NewTextAlphabet{textbfss} {cmssbx10} {}
    \NewMathAlphabet{mathbfit} {cmbxti10} {} 
    \NewMathAlphabet{mathbfss} {cmssbx10} {} 
  \fi
  \ifAMStwofonts
    \ifCUPmtlplainloaded \else
      \NewSymbolFont{upmath} {eurm10}
      \NewSymbolFont{AMSa} {msam10}
      \NewMathSymbol{\upi}     {0}{upmath}{19}
      \NewMathSymbol{\umu}     {0}{upmath}{16}
      \NewMathSymbol{\upartial}{0}{upmath}{40}
      \NewMathSymbol{\leqslant}{3}{AMSa}{36}
      \NewMathSymbol{\geqslant}{3}{AMSa}{3E}

       \let\le=\leqslant
       \let\ge=\geqslant
    \fi
  \fi
\fi 

\ifnfssone
  \newmathalphabet{\mathit}
  \addtoversion{normal}{\mathit}{cmr}{m}{it}
  \addtoversion{bold}{\mathit}{cmr}{bx}{it}
  \newmathalphabet{\mathbfit} 
  \addtoversion{normal}{\mathbfit}{cmr}{bx}{it}
  \addtoversion{bold}{\mathbfit}{cmr}{bx}{it}
  \newmathalphabet{\mathbfss} 
  \addtoversion{normal}{\mathbfss}{cmss}{bx}{n}
  \addtoversion{bold}{\mathbfss}{cmss}{bx}{n}
  \ifAMStwofonts
    \ifCUPmtlplainloaded \else
      \UseAMStwoboldmath
      \makeatletter
      \new@mathgroup\upmath@group
      \define@mathgroup\mv@normal\upmath@group{eur}{m}{n}
      \define@mathgroup\mv@bold\upmath@group{eur}{b}{n}
      \edef\UPM{\hexnumber\upmath@group}
      \new@mathgroup\amsa@group
      \define@mathgroup\mv@normal\amsa@group{msa}{m}{n}
      \define@mathgroup\mv@bold\amsa@group{msa}{m}{n}
      \edef\AMSa{\hexnumber\amsa@group}
      \makeatother
      \mathchardef\upi="0\UPM19
      \mathchardef\umu="0\UPM16
     \mathchardef\upartial="0\UPM40
      \mathchardef\leqslant="3\AMSa36
      \mathchardef\geqslant="3\AMSa3E

       \let\le=\leqslant
       \let\ge=\geqslant
    \fi
  \fi
\fi 

\ifnfsstwo
  \DeclareMathAlphabet{\mathbfit}{OT1}{cmr}{bx}{it}
  \SetMathAlphabet\mathbfit{bold}{OT1}{cmr}{bx}{it}
  \DeclareMathAlphabet{\mathbfss}{OT1}{cmss}{bx}{n}
  \SetMathAlphabet\mathbfss{bold}{OT1}{cmss}{bx}{n}
  \ifAMStwofonts
    \ifCUPmtlplainloaded \else
      \DeclareSymbolFont{UPM}{U}{eur}{m}{n}
      \SetSymbolFont{UPM}{bold}{U}{eur}{b}{n}
      \DeclareSymbolFont{AMSa}{U}{msa}{m}{n}
      \DeclareMathSymbol{\upi}{0}{UPM}{"19}
      \DeclareMathSymbol{\umu}{0}{UPM}{"16}
      \DeclareMathSymbol{\upartial}{0}{UPM}{"40}
      \DeclareMathSymbol{\leqslant}{3}{AMSa}{"36}
      \DeclareMathSymbol{\geqslant}{3}{AMSa}{"3E}

       \let\le=\leqslant
       \let\ge=\geqslant
    \fi
  \fi
\fi 

\ifCUPmtlplainloaded \else
  \ifAMStwofonts \else 
    \def\upi{\pi}
    \def\umu{\mu}
    \def\upartial{\partial}
  \fi
\fi

\date{Accepted 2001 May, Received 2001 March
}
\pagerange{\pageref{firstpage}--\pageref{lastpage}}
\pubyear{2001}

\title{Infrared Spectroscopy of Substellar Objects in Orion}
\author[P.W. Lucas, P.F. Roche, France Allard, and Peter.H. Hauschildt]
{P.W.Lucas$^{1}$, P.F.Roche$^{2}$, France Allard$^{3}$ and P.H.Hauschildt$^{4}$\\
$^1$Dept. of Physical Sciences, University of Hertfordshire, College Lane,
Hatfield AL10 9AB.\\ email: pwl@star.herts.ac.uk\\
$^2$Astrophysics Dept., University of Oxford, 1 Keble Road, Oxford OX1 3RH.\\
$^3$CRAL, Ecole Normale Superieure, 46 Allee d'Italie, Lyon, 69364 France, Cedex 07\\
$^4$Dept. of Physics and Astronomy \& Centre for Simulational Physics, University of
Georgia, Athens, GA 30602-2451}

\begin{document}
\maketitle

\label{firstpage}

\begin{abstract}

	We present broad band spectra of a sample of 21 low luminosity sources 
in the Trapezium Cluster, with masses in the range 0.008 - 0.10 M$_{\odot}$
(assuming an age of 1~Myr). These were selected for low extinction in most cases
and are located west of the brighter nebulosity. The spectra 
are in the $H$ bandpass (1.4-1.95~$\mu$m) and $K$ bandpass (1.9-2.5~$\mu$m) also for 
most of the brighter sources, with a resolution of 50~nm. They were taken with the United 
Kingdom Infrared Telescope (UKIRT) using the CGS4 spectrometer. Absorption by water 
vapour bands is detected in all the substellar candidates except one, which is a 
highly reddened object with strong H$_{2}$ emission and an anomalously blue \it{(I-J)} \rm 
colour, implying that it is a very young cluster member with circumstellar
matter. The observation of prominent water vapour bands confirms the low Effective 
Temperatures implied by our \it{(I-J)} \rm colour measurements in an earlier paper and 
would imply late M or L spectral types if these were older field dwarfs. However, the 
profiles of the $H$ bandpass spectra are very different from those of field dwarfs
with similar water absorption strength, demonstrating that they are not foreground
or background objects. In addition, the CO absorption bands at 2.3~$\mu$m and the
NaI absorption feature at 2.21~$\mu$m are very weak for such cool sources. 
All these features are quite well reproduced by the AMES-Dusty-1999 model atmospheres of 
Allard et al.(2000,2001), and arise from the much lower gravities predicted for the 
Trapezium sources ($3.5<log(g)<4.0$) compared to evolved objects (log g $\sim 5.5$), 
This represents a new proof of the substellar status of our sources, independent
of the statistical arguments for low contamination, which are reexamined here.
The very late spectral types of the planetary mass objects and very low mass brown dwarfs 
demonstrate that they are cluster members, since they are too luminous to be 
field dwarfs in the background. We also present additional UKIRT photometry of a small 
region in the south of the Trapezium cluster where the extinction and nebular 
brightness are low, which permitted the detection of objects with 1~Myr masses 
slightly lower than our previous least massive source at 8 M$_{Jup}$. Following a 
minor update to our previous $J$ band photometry, due to a new UKIRT filter calibration, 
there are $\sim15$ planetary mass candidates in the full dataset.

\end{abstract}


\section{Introduction}

	Brown dwarf candidates have been discovered in large numbers
in many star formation regions and very young stellar clusters, through
deep infrared photometric searches (eg. Comeron et al. 1993,1996; 
McCaughrean et al. 1995; Bejar, Zapatero-Osorio \& Rebolo 1999). Some 
of these surveys have yielded objects of planetary mass, in the Chamaeleon I 
molecular cloud (Tamura et al.1998; Oasa et al.1999) in the $\sigma$-Orionis cluster
(Zapatero-Osorio et al.2000) and in the Trapezium Cluster in Orion
(Lucas \& Roche 2000, hereafter Paper I; Hillenbrand \& Carpenter 2000). 
In most cases there is a very high probability that photometric candidates are indeed 
substellar cluster members. Some candidates with near infrared $JHK$ photometry 
exhibit $K$ bandpass excesses due to hot circumstellar dust, which demonstrate their 
youth (eg. Oasa et al.1999), although this criterion will miss the majority of 
young stellar objects (YSOs). In the Trapezium cluster, even substellar
candidates without $K$ bandpass excesses can generally be assumed to be cluster members
because any measured extinction precludes confusion with foreground field dwarfs 
while the dense backdrop of the OMC-1 cloud and the high galactic latitude 
(b=18$^{\circ}$) both mitigate against background contamination. 

	It is still important to observe brown dwarf candidates spectroscopically,
both to confirm their status and to explore the rich physics of their
atmospheres, which have much lower gravities than their evolved counterparts.
There are effects specific to very young sources which complicate the interpretation 
of their spectra. These are: (1) light scattered and emitted by circumstellar matter, 
which modifies the signal from the photosphere; (2) reddening by the cluster 
medium, which can only be imperfectly removed; (3) the lower gravity of the
atmosphere. The low gravities mean that the spectra cannot be simply compared to those of
Class V dwarfs and such spectra have not been successfully modelled in the near infrared
until now. However, the new AMES-Dusty-1999 models of Allard et al.(2000, 2001)
which employ the NASA Ames water line list and a more detailed 
treatment of dust, appear to be considerably more successful than previous models 
(see Section 4). Evidence for low gravity spectral features has been seen in the
Pleiades (Martin, Rebolo \& Zapatero-Osorio 1996) which have an age of $\sim
100$~Myr. Luhman \& Rieke (1998) and Luhman et al.(1998a) published optical 
spectroscopy of large samples of YSOs in IC348 and L1495E, including some substellar 
objects. They allowed for the effects of low gravity by fitting spectra 
to templates formed from the average of Class V dwarfs and Class III giants,
which produced a good match to the data. However this approach is not appropriate 
for the very low mass sources in the Trapezium and $\sigma$-Orionis, since the few
good quality infrared spectra of late-M type Class III and Class I giants in the 
literature (eg. Lancon \& Rocca-Volmerange 1992) do not have similar profiles to 
those presented here, and correspond to even lower gravity objects 
($-2 < log(g) < 1$. Moreover there are no L-type giants to compare with the
cooler sources found in Orion.

Wilking, Greene \& Meyer (1999) analysed $K$ bandpass spectra of brown dwarfs in 
$\rho$ Ophiuchus by comparison with Class V dwarfs, which may lead to errors in the 
derived effective temperatures. We experienced the same problem in Paper I when deriving 
temperatures and luminosities by comparison with Class V $IJH$ colours. Fortuitously 
this does not appear to have caused significant errors because: (a) many sources with
low extinction were 
observed with \it{(J-H)}=0.5 \rm to 0.6, indicating that this colour does not increase to
Red Giant Branch values (\it{(J-H)} \rm $\sim 0.9$) in very young brown dwarfs and (b) the 
AMES-Dusty-1999 models indicate that low gravity causes only small changes in 
the broad band colours, comparable to our typical photometric uncertainty of 0.1 magnitudes.

	In this paper we present low resolution $H$ and $K$ bandpass spectra of a sample 
of Trapezium cluster members including low mass stars, brown dwarfs and planetary mass 
objects. The spectra are analysed by comparison with both the AMES-Dusty-1999 models and
published spectra of Class V dwarfs. Optical spectra will be presented in a future 
paper. New photometric data are also presented covering the southern region of the 
Trapezium cluster. We also readdress the arguments concerning the nature of these 
sources, in light of the spectroscopic data.

\subsection{Note on Nomencalature}

	In this paper we use the less controversial but not very descriptive
term 'planetary mass objects' to describe sources below the deuterium burning
threshold. Some in the extrasolar planets community object to the term 'free
floating planets' to describe objects which are not orbiting a star 
and probably form in a molecular cloud core in a manner similar to stars 
(although the word planet originally meant 'wanderer'). As a possible alternative, we 
suggest the more compact term 'planetar' as a contraction of 'planetary mass object 
believed to have formed like a star'. A large amount of theoretical work is now 
in progress, examining a variety of different formation scenarios. There may prove
to be more than one mode of formation for these low mass bodies, as is believed
to be the case for stars. 

\section{Observations}

\subsection{Spectroscopy}

	Long slit spectroscopy of sources in the Trapezium cluster was carried out 
at the United Kingdom Infrared Telescope (UKIRT) on 27-30 November 1999, using CGS4,
a near infrared Cooled Grating Spectrograph. The observer was PWL. Sources were 
observed in pairs aligned along the slit of the spectrograph and the sample was 
selected from the dataset of Paper I and the new imaging data presented here, with the 
intention of sampling the mass range from 0.008 to 0.10 M$_{\odot}$, i.e. from
planetary mass to very low mass stars. 21 such sources were observed and 2 sources
of higher mass. There were the following additional criteria. 
(1) Location in a region of low nebular surface brightness in 
order to improve sensitivity. (2) Low extinction towards at least one member of a 
pair, defined by \it{(J-H)} \rm$< 1.5$ or A(V)$\ltsimeq7.5$, also to improve 
sensitivity. 
(3) Most sources were selected for 'normal' \it{(I-J)} \rm colours in an attempt to 
minimise contamination by light scattered from circumstellar matter (see Paper I) 
but 2 sources with the anomalously blue colours shared by 14\% of the cluster 
population were observed to see whether their spectra were different from the 
others. (4) Location within $\sim 40 \arcsec$ of a sufficiently bright star 
($J$ mag $<15$) to permit precise location of optically invisible targets within the 
slit by offsetting to the target. In practice, all 21 of the low mass targets were 
located at least 105 arcsec west of $\theta_{1}$ Orionis C at the cluster centre.

	CGS4 has a 256$\times$256 InSb array sensitive from 1-5.5~$\mu$m and a 
spatial scale of 0.62 arcsec/pixel. The 40 l/mm grating was used in first order, 
which yields a spectral dispersion of 25 nm/pixel but a 1.2\arcsec wide slit 
was employed in most cases, yielding an instrumental profile with a width of 50 nm.
This corresponds to a spectral resolution of R=330 in the $H$ bandpass and R=440 at $K$.
The wide slit was used to ensure that most of the flux from both members of a 
widely separated pair would fall in the slit. Each spectrum covers a 0.62~$\mu$m 
wavelength range. The $H$ bandpass spectra were centred at 1.65~$\mu$m and the $K$ 
bandpass 
spectra at 2.2~$\mu$m, overlapping in the 1.9~$\mu$m region. One pair of sources was 
observed at 25 nm resolution with a 0.6\arcsec slit, though not in very good conditions; 
no additional narrow line features were revealed. 
This pair of sources was observed with $H$ bandpass spectra centred at 1.58~$\mu$m on the 
first night, in attempt to include the steep $J$ band water edge at 1.34~$\mu$m, but the 
instrument efficiency in this configuration was found to be essentially zero for 
wavelengths $\lambda < 1.4~\mu$m, so the longer wavelength setting was used thereafter. 
The relative throughput of the instrument and the atmosphere in the $H$ bandpass is shown 
in Figure 1. The observations and source notes are detailed in Table 1. For all other 
pairs, the array was stepped over 2 pixels in a whole pixel step to remove bad pixels and 
the telescope was nodded along the slit in order to subtract the 
background, with background-limited individual integrations of 30 to 60~s duration. 
Total exposures time ranged from 10 minutes for the brightest sources to 1.5 hours
for the faintest. 21 primary sources were observed in the $H$ bandpass and 11 of these in 
the $K$ bandpass also. We also present $H$ bandpass spectra of 2 fairly low mass sources 
which appeared in the long slit by chance. The smaller number of $K$ bandpass observations
was due to constraints on observing time (weather and computer failures).
Priority was given to $H$ bandpass observations because of anticipated complications owing
to emission by hot dust in the $K$ bandpass and because the water absorption troughs are 
slightly more prominent in the $H$ bandpass. A small number of red main sequence stars 
were also observed in order to check the reliability of the data. Their spectra 
agree very well with those published by Leggett et al.(2000) for the same objects, 
also obtained with CGS4 at UKIRT.

\begin{figure}
    \leavevmode
\psfig{file=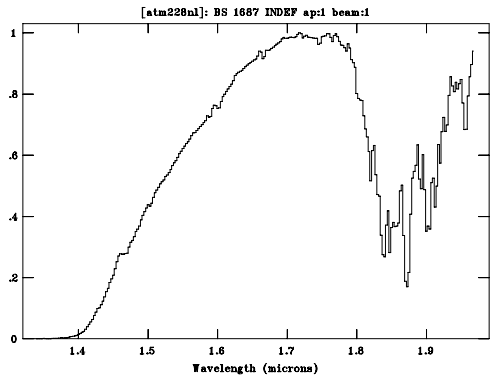,clip=,width=8cm}
\caption{Relative transmission of Atmosphere and Spectrograph per unit wavelength.}

\end{figure}

	The data were coadded into background subtracted 'reduced groups' using the 
{\small CGS4dr} software package and spectra were optimally extracted using the 
{\small TWODSPEC}
package in {\small IRAF}, the Image Reduction and Analysis Facility. Argon arc spectra, 
tungsten lamp flatfields and standard stars were observed frequently to compensate
for changes in the wavelength calibration when the slit orientation was changed
and to accurately correct for the changing telluric water vapour column density.
F-type stellar standards were generally used in preference to A-types since they
have weaker HI absorption lines. The fairly weak Brackett series lines were removed 
from the stellar standards by fitting the continuum with a high order cubic
spline, dividing the fit by the data to produce a line spectrum, and multiplying 
the data by the line spectrum. The strong Paschen $\alpha$ line at 1.875~$\mu$m 
coincides with a deep trough due to telluric water absorption and could not be
adequately removed, so we have removed a small section of the spectra near this 
wavelength. The Brackett series limit occurs at 1.46~$\mu$m and coincides with
a narrow telluric feature at the short wavelength edge of the $H$ bandpass, making it
impossible to accurately remove. Inspection of the NextGen models of Hauschildt et 
al.(1997) indicates that the Brackett series limit is weaker than the telluric 
feature in F-type standards and it does not appear to be noticeable in the spectra 
presented here.

	In addition to background shot-noise, a significant source of noise arises
from the spatial gradients in the nebular surface brightness, which make the 
background subtraction difficult. Owing to these gradients the reduced group images
are not flat, but show the effects of subtracting two 
frames separated by distance of the nod along the sky: either 10 or 20 arcsec 
was used depending on the separation between the members of a pair. The images also 
show the spectral residuals due to fluctuations in the strength of telluric OH 
emission lines. A polynomial was interactively fitted to the residual background at 
each wavelength and subtracted, using the {\small APALL} package in {\small TWODSPEC}, 
and this procedure was generally very successful in removing both nebulosity gradients and 
residual OH emission. The problem was greatest for the fainter sources, whose peak 
brightness is comparable to fluctuations in nebular suface brightness on a scale of 
a few arcsec, even for sources selected in regions of low nebular background. Where 
necessary, 2 iterations were used to produce a flat background. For the 
faintest sources, the background may be slightly over or under subtracted, but 
agreement between spectra taken at the two nod positions was used as a check to 
prevent this occuring. Only one spectrum was lost due to this problem: the $K$ bandpass 
spectrum of the faintest source, 051-147, was rendered useless by a small step in 
the surface brightness on the same scale as the nod length, but an $H$ bandpass spectrum
was successfully extracted. We recommend that a nod length of 5 arcsec or less 
be used in star formation regions when taking long slit spectra of faint sources.

\subsection{Photometry}	

	Additional photometry in the \it{IJH} \rm bands was obtained at UKIRT,
using the infrared camera UFTI, in the manner described in Paper I, using 
900~s exposures. The new data were taken on the nights of 11-12 November 1999 by 
observatory staff. They consisted of three contiguous 1.5 arcmin UFTI fields 
located 3 arcmin south of $\theta_{1}$ Ori C, oriented east-west as shown in 
Figure 2. This region was selected because our existing data indicated that the 
nebular surface brightness and average extinction would be low in this region, and 
published $K$ bandpass maps showed that the stellar density is reasonably high 
(Hillenbrand 1997). These features assisted in the detection of some new very faint 
sources, as described in Section 3.1. The seeing conditions were fairly good, with 
a FWHM of 0.5-0.6\arcsec in all three filters, aided by the tip/tilt image 
stabilisation of the UKIRT secondary mirror. 

\begin{figure*}
    \leavevmode
\psfig{file=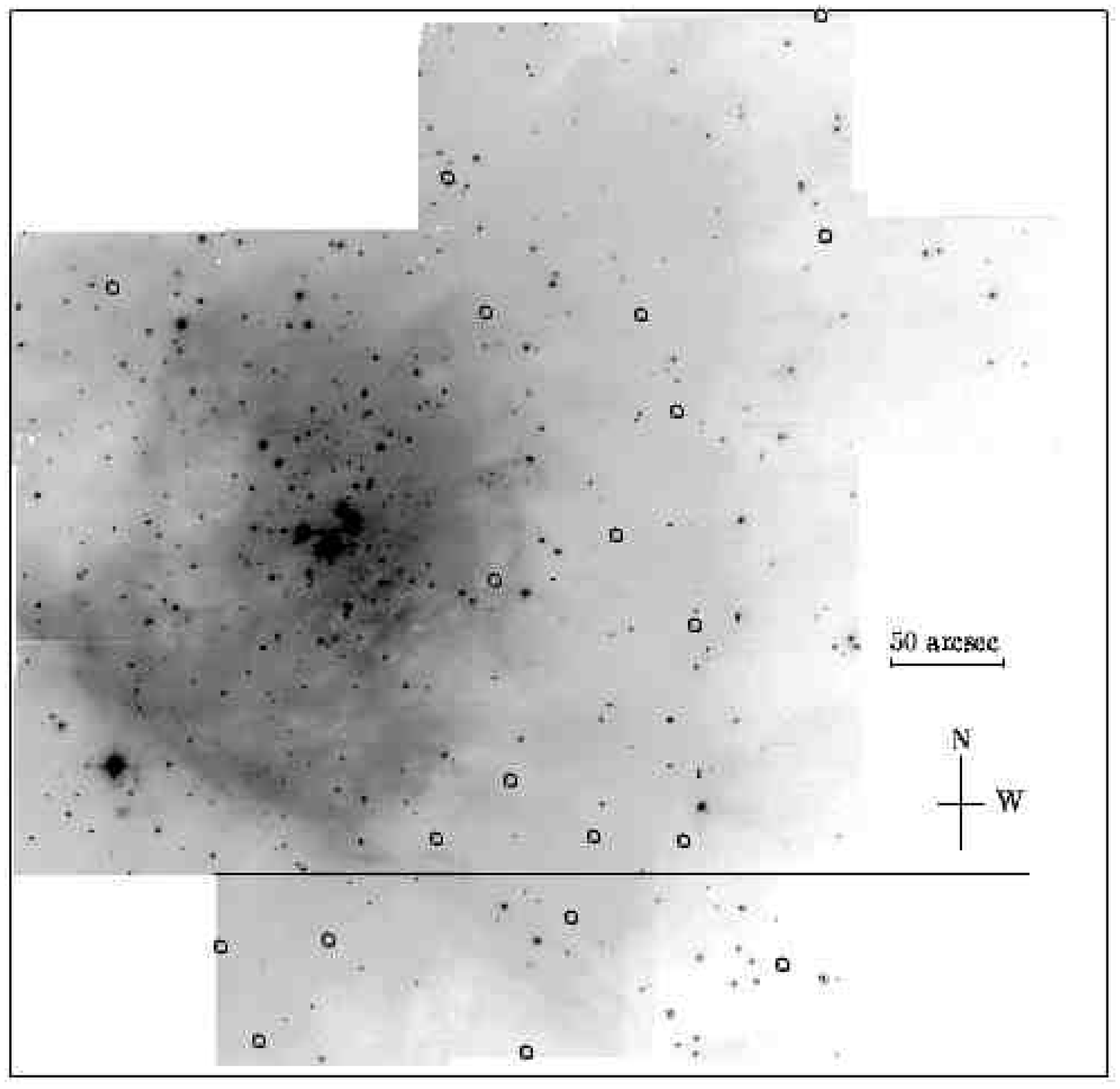,clip=,height=11.8cm,bbllx=0cm,bburx=21cm,bblly=0.2cm,bbury=20.6cm}

Figure 2: $J$ Bandpass image of the full photometric dataset. The region of new data 
is below the horizontal line. The spectroscopic sample was taken from sources to the 
west of the brightest nebulosity. The locations of very low mass candidates
(dereddened mag J$_{dr} > 17.5$, or M$\ltsimeq$0.016~M$_{\odot}$) are marked with circles.
\end{figure*}

	Photometry was performed using the DAOPHOT crowded field photometry package
in IRAF as in Paper I, using both manual photometry and automated crowded field 
photometry. The former was generally preferred in order to accurately subtract the 
highly structured nebulous background, except when the image profiles of sources 
overlapped significantly. Several sources were contained in the overlap region
between the new data and the dataset of Paper I. The measured magnitudes typically
differed by 0.1 mag or less, which is approximately the photometric error described
in Paper I. The main source of error in the brighter sources is the image 
profiles, which are spatially variable in the overlap region since this is 
t the very edge of both the old and new datasets. Each UFTI field is made up of a
9-point image mosaic with small jitters.

\subsubsection{Calibration Update}

	The $J$ band magnitude system for UFTI at UKIRT has recently been updated	
by observatory staff, who discovered that the transformations from the old
IRCAM system to the new UFTI system are different for cool red stars than other
main sequence stars. This is because these stars have water absorption bands in 
similar locations to the water bands in the earth's atmosphere and UFTI employs 
the new $J$ band filter developed for the Mauna Kea Consortium, optimised for the low 
water column above this observing site. The new transformation has been thoroughly
calibrated via measurements of late M type and L type stars.
Since our Paper I dataset was calibrated using red stars (in order to calculate 
transformations from Cousins $I$ to the UFTI $I$ band) this requires us to update 
our $J$ band photometry, and consequently the I band transformations also (see UKIRT 
web pages for the details). The new $J$ band magnitudes are typically 
0.1 magnitudes brighter than before, which leads to a slightly lower luminosity for 
sources with significant extinction, since the calculated extinction is reduced. 
Sources with near zero extinction are slightly more luminous than had been thought. 
However changes of this order are of very little consequence, since the 
mass-luminosity relation is quite steep throughout the brown dwarf regime and 
the change is similar to both the typical photometric uncertainty and the theoretical
uncertainty in the mass-luminosity conversion.

	The bolometric correction,$BC_{J}$, formerly used to convert dereddened $J$ 
magnitudes, $J_{dr}$, to luminosities is also updated in consequence:

\begin{eqnarray*}
	BC_{J} = 2.01;    J_{dr} \ge 13.71 \hspace{1cm} (1a)\\.
	BC_{J} = 0.209 J_{dr} -0.855;  J_{dr} < 13.71 \hspace{1cm} (1b)
\end{eqnarray*}

	The new models used here predict the fluxes in each filter directly without the 
need for a bolometric correction. 

\section{Results}

\subsection{Photometry}

	43 new sources were detected at both $J$ and $H$ bands in the new data,
of which 31 were detected at $I$ band. A colour-luminosity plot, with the new
data added to that of Paper I, is shown in Figure 3 for all 557 unsaturated sources. 
We plot 2 different 1~Myr isochrones as almost vertical solid lines at the left of the 
picture, to which sources may be dereddened. Both of these are self consistent 
predictions using the same model atmospheres for colour predictions and to provide 
the boundary condition for the evolutionary calculations. This removes the need
to use uncertain bolometric corrections. Furthest to the left is an isochrone 
supplied by Baraffe et al.(2001, in preparation), using the non-gray AMES-Dusty-1999 model 
atmospheres (Allard et al.2000, 2001). The isochrone slightly to the right is the 
prediction of Baraffe et al.(1998), recently extended to 1~Myr and to the deuterium 
burning threshold (Chabrier et al.2000). This second isochrone has evolution and colours 
based upon the NextGen model atmospheres. It is useful to see both because the Nextgen
atmospheres correctly predict the colours of main sequence stars with effective 
temperatures above 2500~K, whereas the AMES-Dusty models are successfully fit the colours
of field brown dwarfs, below 2200~K, where dust has an important influence on the
photosphere. The temperature at the deuterium burning threshold lies near the low 
temperature end of the NextGen regime but we cannot be sure which model will
predict the \it{(J-H)} \rm colours better in very young, low gravity photospheres. 
The true colour probably lies between the 2 predictions, as discussed below.

\begin{figure}
\begin{center}
\begin{picture}(200,290)

\put(0,0){\includegraphics{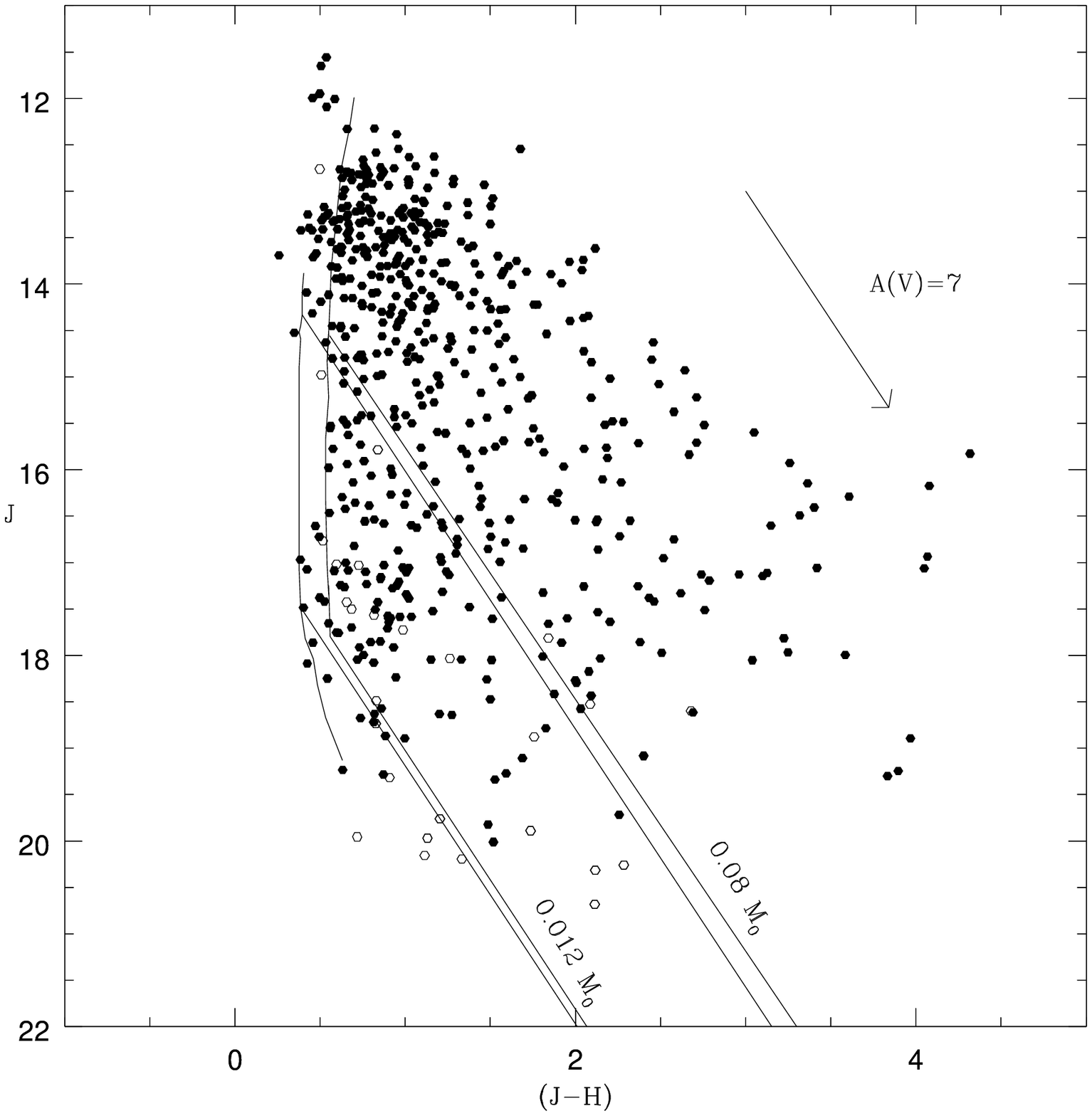}}

\end{picture} 
\end{center}
Figure 3: Colour-magnitude diagram for the 557 unsaturated $J$ bandpass sources.
Two theoretical 1~Myr isochrones are plotted at left, with reddening lines indicated
for 0.08~M$_{\odot}$ and the 0.012~M$_{\odot}$ deuterium threshold. Open circles represent
sources with large $J$ band photometric uncertainties ($\ge$0.2 mag).
\end{figure}

Recalibrated photometry of the full dataset is available electronically\footnote{
Photometry and spectra are available via anonymous FTP to star.herts.ac.uk, 
in pub/Lucas/Orion.}. The O'Dell \&
Wong (1996) coordinate based naming system is used as in Paper I. The advantageous 
features of the new dataset, as described above, permitted the detection of some 
very low luminosity sources: 4 new planetary mass candidates, 3 of which appear to 
be less massive than any of the sources described in Paper I, assuming an age of 
1~Myr and employing either the isochrones of Burrows et al.(1997, hereafter BM97)
or the new isochrone of Baraffe et al. However only 1 of these 3 faint
sources has good photometry: 189-659 has $m_{J}=19.23, (J-H)=0.63$, which indicates a 
notional mass of 6.2 M$_{Jup}$ at 1~Myr (on the BM97 isochrone) and apparently negligible 
extinction. The other 
2 sources have large photometric uncertainties in both the $J$ and $H$ bands, 
so their masses are highly uncertain. Intriguingly, these 
3 faint sources are all located close together on the sky, in a 1 arcmin square 
region 3 arcmin due south of $\theta_{1}$ Ori C, on the far side of the 
Orion Bar from the cluster centre. The Orion Bar is the ionisation front of the 
OMC-1 cloud, viewed edge-on as it curves into the line of sight. Sources which are 
located beyond it lie on lines of sight 
passing deep into the molecular cloud. Hence we might expect such sources to be younger
than the rest of the cluster and hence less massive for a given luminosity. However, 
the very low extinction toward 189-659 indicates that this source probably lies in 
front of the cloud and might possibly be a minimum mass star in the foreground. The 
locations of all the sources with dereddened $J$ magnitudes $J_{dr} \ge 17.5$ 
(i.e. M$ \le$ 0.012 to 0.016~M$_{\odot}$ at 1 Myr) are plotted in Figure 2. 

	There are now $approximately$ 15 planetary mass candidates in the dataset,
following the filter recalibration. 
The number of sources below the deuterium burning threshold cannot be specified 
very precisely, mainly because of the uncertainty in the \it{(J-H)} \rm colour of their 
photospheres mentioned earlier. Most evolutionary calculations are in very good 
agreement concerning the minimum mass for deuterium burning, at 0.012-0.013~M$_{\odot}$ 
(Saumon et al.1996; BM97; Chabrier et al.2000). In addition, the 1~Myr isochrones
of BM97, D'Antona \& Mazzitelli (1998) (herafter DM98, update to 1997 publication) and 
Baraffe et al.(1998) are in fair agreement
(within 0.2 to 0.3 mag) with regard to source luminosities for low mass 
brown dwarfs. However, the small uncertainty in the \it{(J-H)} \rm colours
can slightly change the number of planetary mass objects by moving the lines of 
constant mass in Figure 3 to the right or to the left by 0.1-0.15 mag, since a number
of sources appear to lie very close to the deuterium burning threshold. 
Applying the new Baraffe et al. isochrone, there are $\sim$13 planetary mass objects
(below the M$<0.012~$M$_{\odot}$ threshold calculated by Chabrier et al.), whereas the
NextGen-based Baraffe et al. ischrone indicates there are 16 such objects.
Using the main sequence colours and bolometric corrections as in Paper I, and the BM97 
isochrone for luminosity, indicates that there are 15 such objects. Given that the $J$
bandpass photometric errors are typically 0.1-0.2 mag for these faint candidates, the 
important issue is not the precise number but whether the IMF extends below the 
deuterium burning threshold, which appears to be the case.

	The NextGen colour predictions, using the Miller-Tennyson water line list, are 
only very slightly too red for main sequence stars but these atmospheres poorly 
reproduce the observed profiles of both field dwarfs with strong water absorption (Leggett 
et al. 2001) and the low gravity spectra presented here. The AMES-Dusty-1999 colour 
predictions, using the NASA AMES water line list (Partridge \& Schwenke 1997), are 0.1 to 
0.2 mag too blue in main 
sequence stars (see Allard, Hauschildt \& Schwenke 2000) but reproduce the profiles of 
the spectra presented here quite well. A small colour discrepancy in \it{(J-H)} \rm
would appear to arise from incompleteness in the water line list in the 1.7-2.2~$\mu$m 
water absorption band, which is shown directly for Trapezium sources in Section 4.1.

	Note that we have made a conservative choice of reddening law (see Paper I).
The choice has little effect in any case because the planetary mass candidates
necessarily have low extinction and because extinction laws have little variation in the 
infrared (see Cardelli, Clayton \& Mathis (1989)) except in extremely dense environments 
such as the accretion
disks of embedded protostars, which may contain dust grains larger than 1~$\mu$m in 
radius.

	There may be small but unquantified effects on the \it{(J-H)} \rm colours due to 
scattering by circumstellar matter and a small percentage of emission by hot 
circumstellar dust in some cases. Hot dust would 
make sources appear slightly redder (see Section 4) and overmassive; the direction 
of any scattering effect depends very much on assumptions about the circumstellar 
structure and viewing angle. The additional uncertainties of cluster membership and 
candidate ages are discussed in Section 6.

\it{Erratum} \rm

	We note that 1 planetary mass candidate from Paper I was found to be double 
counted owing to a typing error. An internal cross check of all the coordinates shows 
that this is the only such case in the dataset of 557 unsaturated sources detected in 
the $J$ and $H$ bandpasses. 

\subsection{Spectroscopy}
\subsubsection{$H$ band}
	21 $H$ band spectra were taken of sources with masses 
M $\ltsimeq 0.1$M$_{\odot}$. These are plotted in Figure 4(a-c), as F$_{\nu}$
spectra in order to suppress the apparent noise at short wavelengths. Dereddened 
spectra are plotted above the observed spectra for those sources with non-zero 
extinction. 
These spectra are not flux calibrated since for some observations observing 
conditions were not photometric. The fainter sources are Gaussian-smoothed
to enhance low resolution features. Nearly all of the spectra are dominated by
strong water absorption bands to either side of a sharp peak located between
1.68 and 1.70~$\mu$m. The exceptions are plotted in the lower part of Figure 4(b).
No real narrow spectral features are unambiguously detected
in the $H$ bandpass. Pixel to pixel variations therefore illustrate the noise in
each spectrum, which has wavelength dependent contributions from narrow telluric OH
emission lines and broad band telluric water absorption, both of which vary on a 
timescale of minutes during an exposure. The noise is most obvious in very faint 
sources such as 084-104 (Figure 4(c)) in which many spurious narrow features remain 
despite some smoothing. The data quality is poor in the 1.8-1.95$~\mu$m region owing 
to strong telluric water absorption and also becomes less reliable at $\lambda 
< 1.5~\mu$m, due to both telluric water absorption and the declining instument 
efficiency. Some of the spectra are strongly reddened, with the
result that the water absorption appears weaker on the long wavelength side
of the peak and stronger on the short wavelength side, eg. object 013-306.
In most cases the effect is minor owing to our preferential selection of sources with 
low extinction. The extinction toward each source is calculated using the $IJH$ 
photometry, by comparison with measured main sequence star colours as described in 
Paper I, and noted in Figure 4(a-c). The {\small DEREDDEN} task 
in IRAF was used, which is based on the Cardelli, Clayton \& Mathis extinction 
law, with reddening parameter R$_{V}$=5.5 as before and A($\lambda$) $ \propto 
\lambda^{-1.61}$ in the infrared. This is a somewhat imprecise procedure, because 
our typical photometric uncertainty of 0.1 mag in each filter leads to an uncertainty 
of 0.14 mag in the \it{(J-H)} \rm colour, which corresponds to an uncertainty of 
0.25 mag in $H$ band extinction and 1.2 mag in $V$ band extinction. 
However, the errors will be less for the brighter sources in the region of the 
spectroscopic sample, which was selected for faint nebulous background.
The use of main sequence colours gives similar results to the NextGen atmospheres: 
the atmospheric models predict changes of no more than $\sim$ 0.1 mag in \it{(J-H)} \rm 
due to lower gravity in young sources, so we do not think this uncertainty will be 
larger than measurement error.

\onecolumn
\begin{figure}
\leavevmode
\psfig{file=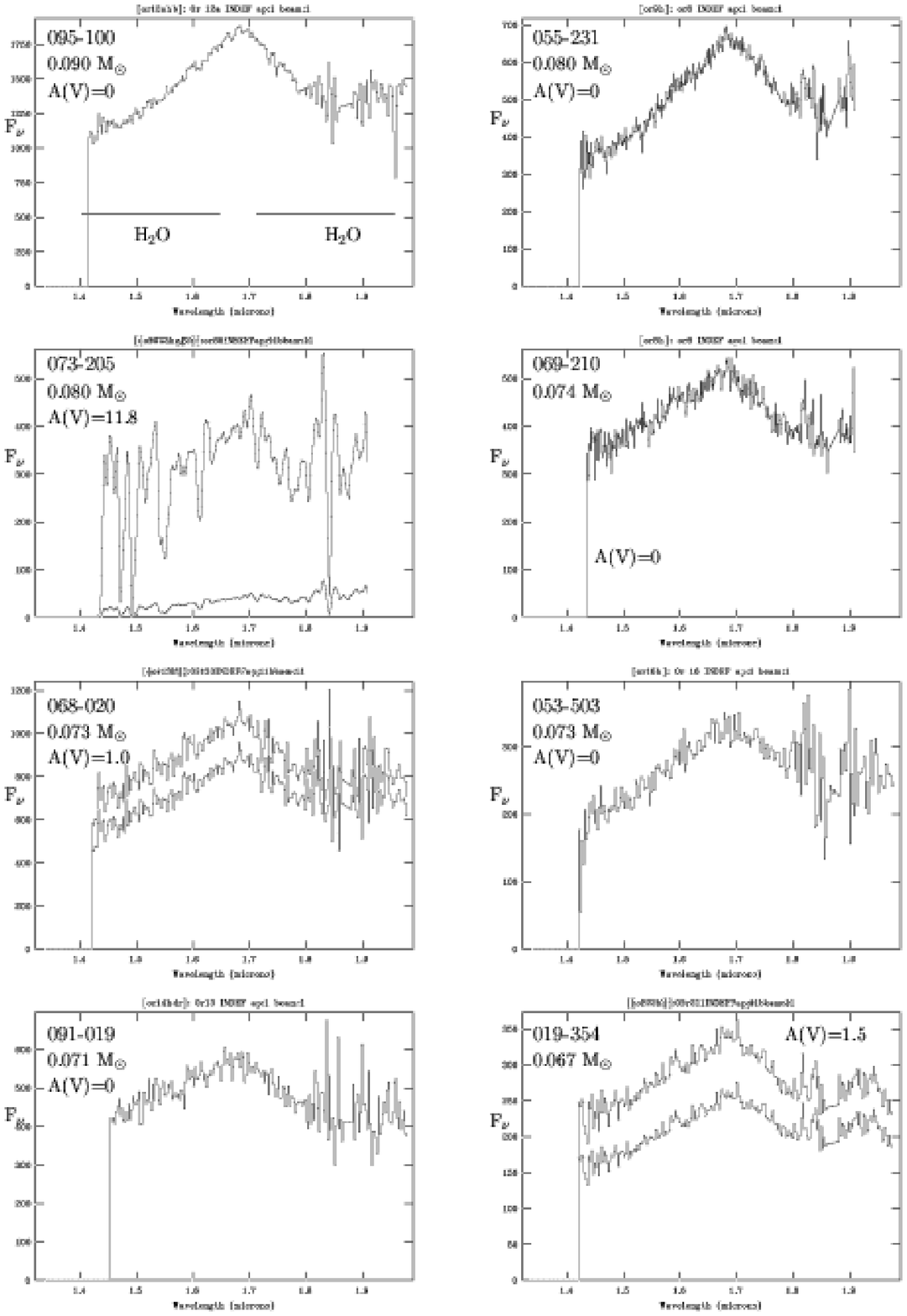,clip=,height=23cm,bbllx=1cm,bblly=1cm,bburx=21cm,bbury=28cm}

Figure 4(a): Observed $H$ bandpass spectra of Trapezium sources, with dereddened 
spectra overplotted as the upper line for each source when extinction is non-zero.
Masses derived from the BM97 1~Myr isochrone are indicated. The noise is indicated
by the pixel to pixel variations, since no narrow spectral features are detected
in the H bandpass.
\end{figure}

\pagebreak

\begin{figure}
\psfig{file=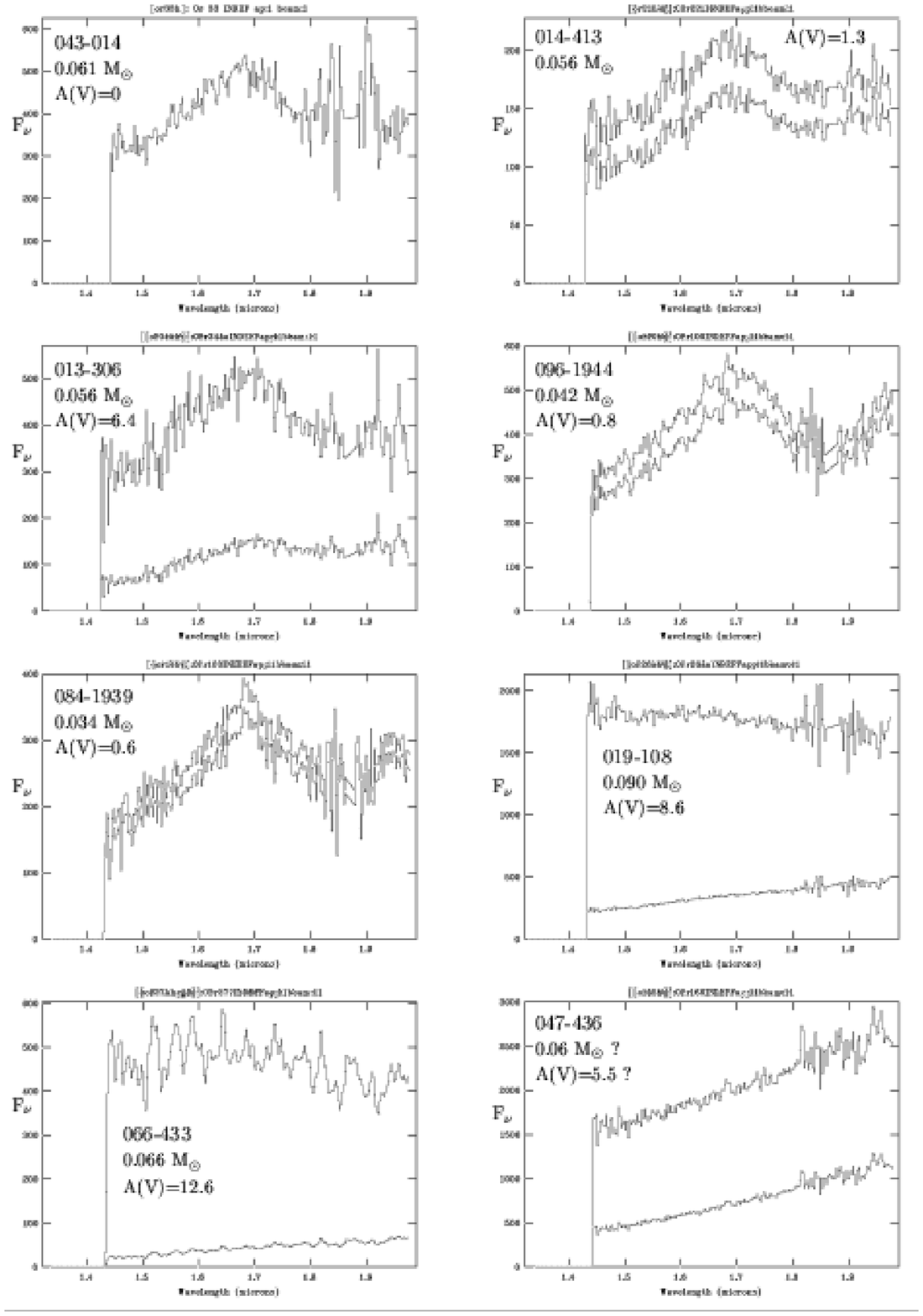,clip=,height=23cm,bbllx=1cm,bblly=1cm,bburx=21cm,bbury=28cm}

Figure 4(b): Observed and Dereddened $H$ bandpass spectra, as Figure 4(a).
047-436 shows no water absorption and 2 others have very weak water absorption.
\end{figure}

\pagebreak
\onecolumn

\begin{figure}
\psfig{file=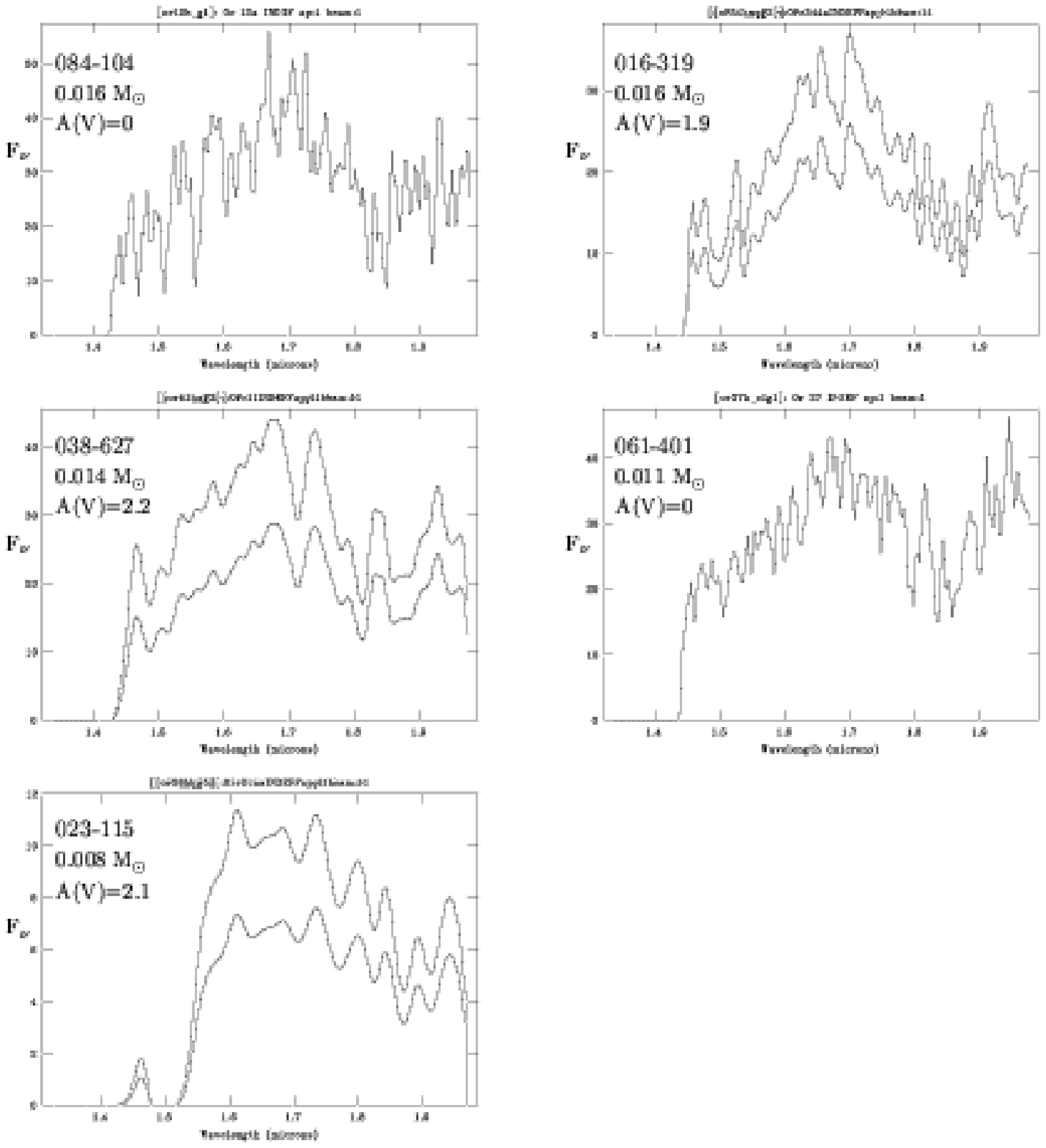,clip=,height=18cm,bbllx=2cm,bblly=5cm,bburx=22cm,bbury=23cm}

Figure 4(c): Observed and Dereddened $H$ bandpass spectra for the least massive
candidates, with some gaussian smoothing applied. The apparent absence of flux at 
$\lambda \ltsimeq 1.5~\mu$m in 023-115 is attributed to the decline in signal/noise 
at those wavelengths, due to atmosphere, instrument efficiency, extinction and 
(presumed) water absorption in the source itself. 
\end{figure}
\twocolumn
	The peaked $H$ bandpass profiles in Figure 4(a-c) are very different 
from those of late M and L-type dwarfs in the Local Neighborhood with strong water
absorption bands, as shown by our spectra of 2 red standards in Figure 5(a-b), 
compared with a composite spectrum of the Orion sources. The composite spectrum is an 
average of 12 reasonably high quality spectra with fairly strong water absorption, 
weighted according to the apparent noise in each spectrum. The composite has
slightly weaker water absorption than the average of the sample, since it is weighted
toward warmer, brighter sources near the stellar/substellar boundary.
Comparison with the data of Leggett 
et al.(2001) (overplotted in Figure 5(a)) for the same red standards demonstrates 
that the spectra are reliable and the differences from field stars are real. In 
spectra of field dwarfs there is strong 
water absorption at $\lambda < 1.55~\mu$m and $\lambda > 1.70~\mu$m, but a shallower
slope in between, which appears as a flat plateau in F$_{\lambda}$ plots of the data 
(Figure 5(b)). This is clearly seen in the F$_{\lambda}$ plots of a large sample 
of late M and L dwarfs presented by Leggett et al.2001.
marked by clear changes in slope at 1.70~$\mu$m and in the 1.55-1.60~$\mu$m interval.
The peaked profiles of the Trapezium objects, which show no plateau, appear to be a 
consistent signature of youth and cluster membership. They are modelled in Section 4.
There are no other clearly seen atomic or molecular features in the Trapezium
sources at $H$ band, which is unsurprising given the low spectral resolution
and modest signal to noise. 
Atomic absorption lines of Na and Ca have been observed in field dwarfs, eg. 
Jones et al.(1994) in high signal to noise spectra but many atomic lines appear
to be weaker in very young brown dwarfs (eg. Bejar, Zapatero-Osorio \& Rebolo 1999.) 
The water absorption depths are quantified in Table 2, and discussed in Section 4.

\begin{figure*}
\psfig{file=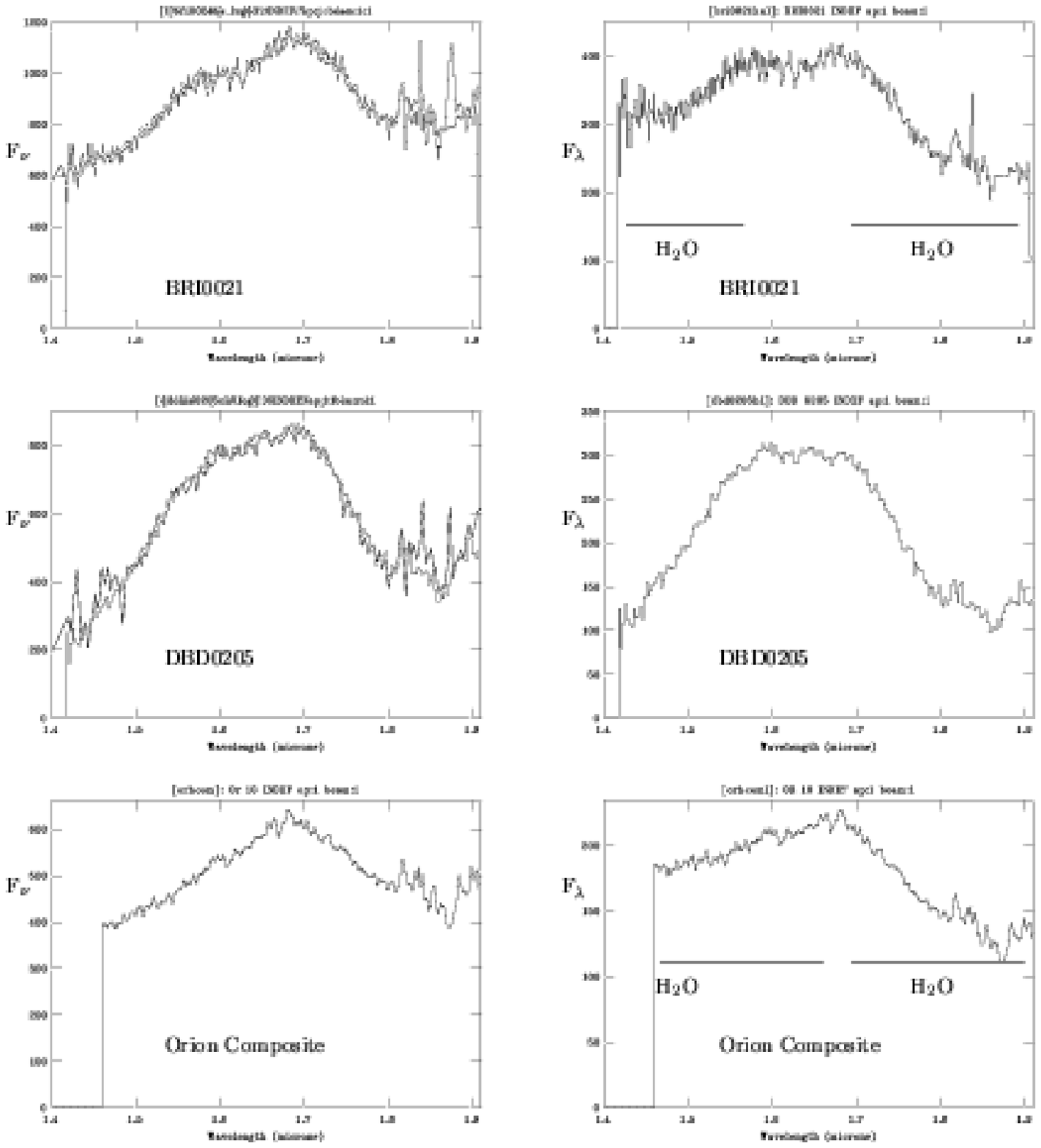,clip=,height=18cm,bbllx=2cm,bblly=5cm,bburx=22cm,bbury=23cm}

\vspace{-1cm}
Figure 5(a-b): Red Standards and Orion composite spectrum. (a)(left) F$_{\nu}$
spectra. Leggett et al.(2001) data are overplotted as a solid line for the field 
dwarfs to demonstrate the reliability of the spectra. (b)(right) spectra plotted as 
F$_{\lambda}$, illustrating the plateau seen in field dwarfs in the $H$ bandpass. 
The composite is a weighted average of the warmer, more massive sources, so it has 
slightly weaker water absorption than the average of the sample. Weak 
narrow absorption features in the composite are generally attributed to telluric OH
lines.
\end{figure*}

	Only 3/21 sources have weak or zero water absorption. 019-108 and
066-433 suffer very high extinction (A(V) = 8.6, 12.6 respectively) and have very weak 
water absorption, consistent with late K-type or very early M-type spectra. These 
earlier spectral types are not consistent with the temperatures expected at age 1~Myr
even for these relatively luminous sources. (At 1~Myr masses, M=0.090~M$_{\odot}$, 
0.066~M$_{\odot}$ and T$_{eff}$=3400~K, 2800~K respectively are derived photometrically,
using the BM97 isochrone). Therefore the $H$ band spectra indicate
that they are either even younger and hotter systems (age $\sim 0.1$~Myr)
or they are background field stars. The dereddened fluxes of these stars, 
are faint enough to be consistent with late K-type dwarfs at d$\ge440$~pc 
(m$_{J}=13.72, 14.45$ for 019-108 and 066-408 respectively) judging by comparison
with both the Luminosity Class V sequence of Tokunaga (2000) and the models of BCAH
for low mass stars at late K dwarf temperatures. 019-108 would have to lie just behind 
the OMC-1 cloud at 440~pc, while 066-433 would lie at only a slightly greater distance. 
It therefore seems slightly more probable that these 2 hotter, highly reddened sources 
are background stars but the alternative of extreme youth is not excluded,
especially since the stellar mass proplyd source 178-232 shows a similar spectrum
(see below).

	Only one source, 047-436, shows no water absorption in the $H$ bandpass, 
simply a very red continuum in Figure 4(b). This source is one of the 14\% from Paper I 
with an anomalously blue \it{(I-J)} \rm colour (though not an extreme case, I=18.277,
{\it(I-J)}=3.21) and so its extinction cannot be well measured. At present, we 
attribute such colours to light scattered by circumstellar matter. An attempt to 
apply the standard procedure leads to an insufficiently dereddened spectrum, shown 
in Figure 4(b). The youth and cluster membership of 047-436 is demonstrated by the $K$ 
bandpass spectrum, discussed below. 

	2 additional $H$ bandpass spectra of more massive sources are presented in 
Figure 6. 1 of these, 017-410 is a YSO with a mass of $\sim 0.3$M$_{\odot}$ which 
appeared in the 
slit by chance, owing to the crowded nature of the cluster. It shows fairly strong 
water absorption but more on the long wavelength side of the 1.7~$\mu$m than the short
wavelength side, which is different from the less massive sources, which may reflect
the higher gravity of this source. Luhman et al.(2000)
have obtained $K$ bandpass and optical spectra of stellar mass Trapezium sources
but there do not appear to be many $H$ bandpass spectra of Trapezium stars in
the literature to use for comparison. The second source 
is 178-232, a more massive star (M$\sim 0.4 M_{\odot}$) with an anomalously blue 
\it{(I-J)} \rm colour around which an emission line 'proplyd' structure was resolved 
by the Hubble Space Telescope (O'Dell \& Wong 1996). This has a very red spectrum, 
perhaps due to extinction by circumstellar matter rather than dust in the cluster 
medium and shows relatively weak water absorption, which would be expected for a 
hotter, younger object.

\begin{figure*}
\begin{center}
\begin{picture}(200,200) 
\put(0,0){\includegraphics{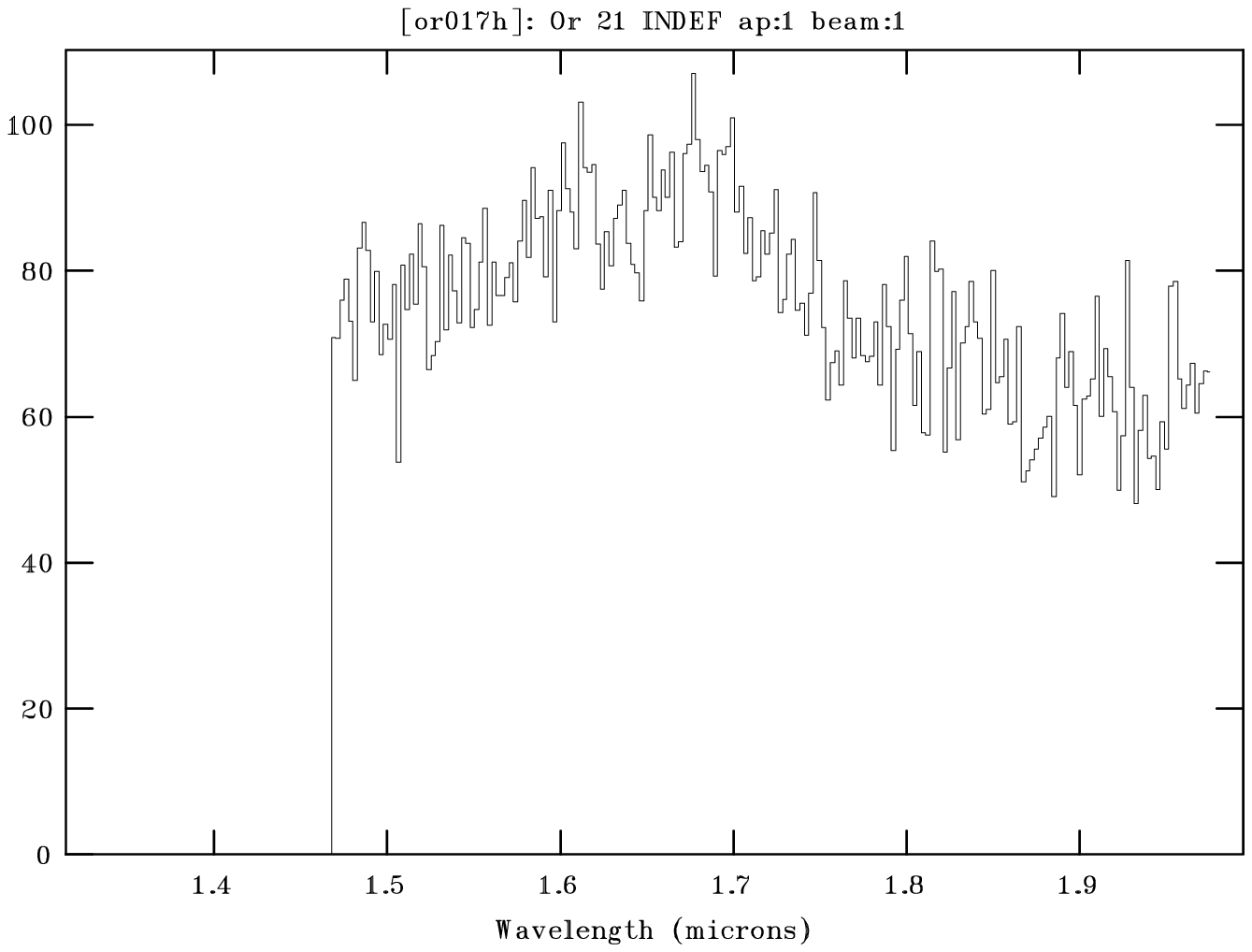}}
\put(-110,165){017-410}
\small
\put(-110,153){0.30 M$_{\odot}$}
\put(20,165){A(V)=1.2}
\put(-130,125){F$_{\nu}$}
\normalsize

\put(0,0){\includegraphics{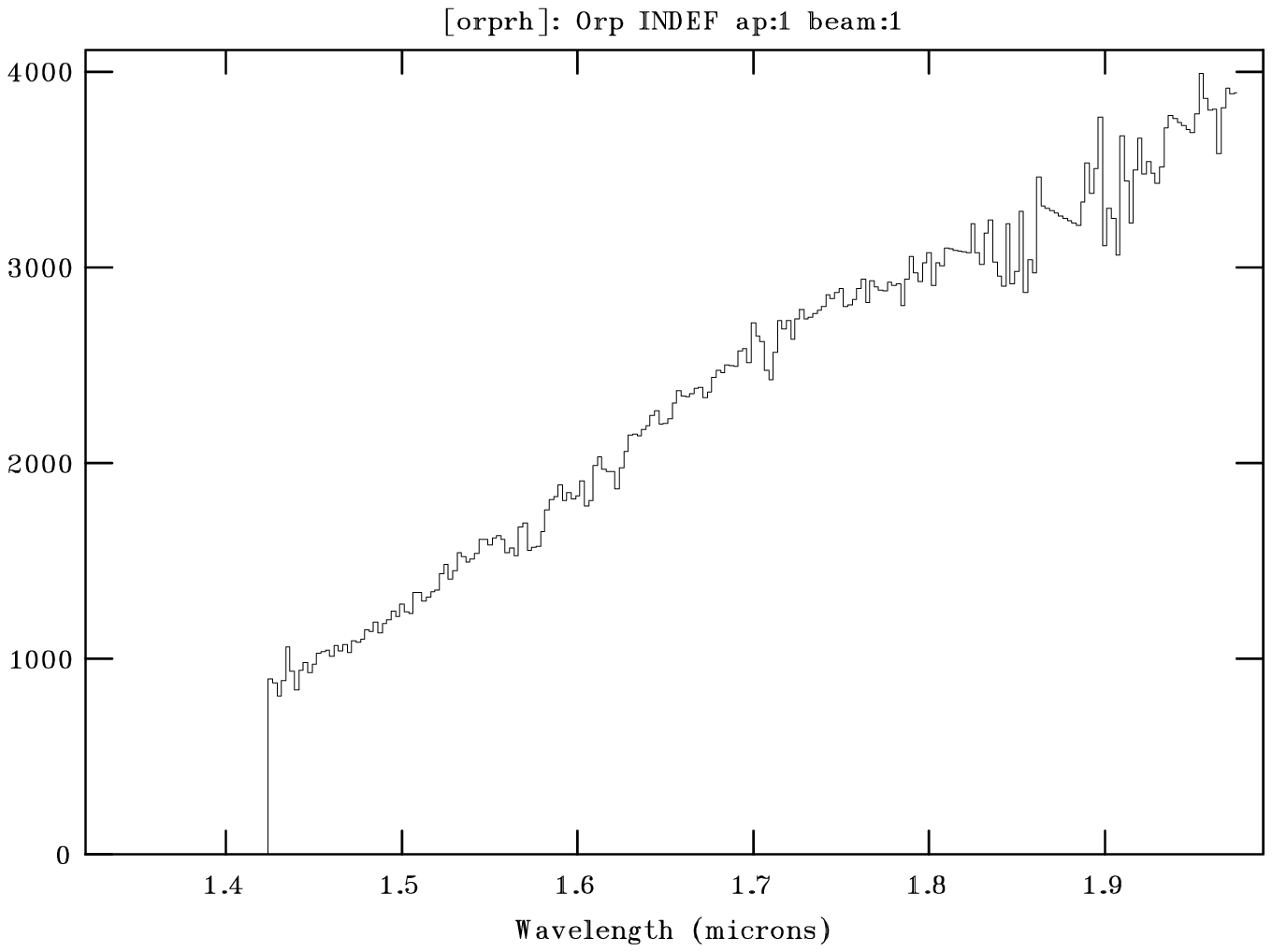}}
\put(150,165){178-232}
\small
\put(150,153){0.4 M$_{\odot}$ ?}
\put(150,141){A(V)=14.5 ?}
\put(125,125){F$_{\nu}$}
\normalsize

\end{picture} 
\end{center}
\vspace{-1cm}
Figure 6. Observed $H$ bandpass spectra of 2 more massive sources (0.3-0.4~M$_{\odot}$). 
178-232 is a proplyd; the weak water absorption and red spectrum is consistent with an
embedded, warmer source.
\end{figure*}

\subsubsection{$K$ band}	

	11 $K$ band spectra are shown in Figure 7(a-b), and dereddened $H$+$K$ spectra
are plotted in Figure 8(a-b). The $H$ and $K$ spectra are joined by averaging the overlap 
region from 1.89-1.97~$\mu$m, with some uncertainty ($\sim 10\%$) in the flux level
owing to the telluric water absorption at those wavelengths. 9/11 of the $K$ band 
spectra show strong water absorption bands on either side of a maximum near 
2.25~$\mu$m, as expected for young brown dwarfs. The other 2 spectra, of 019-108 
and 047-436, are consistent with very weak or zero water absorption, similar to
the $H$ band data. The noise is indicated by the pixel to pixel variations as in the $H$
bandpass data.

	The v=2-0 vibration-rotation bands of CO at
$\lambda > 2.29~\mu$m are detected in 4/11 sources (see Table 3) via the edge at 
2.29~$\mu$m and marginally detected in one other source. Only one source,
053-503, shows the multiple bandheads clearly at 2.29, 2.32, 2.35 and 2.38~$\mu$m but 
they are also visible in 014-413. The weakness or absence of CO absorption is 
surprising for the low temperature photospheres indicated by the strong water 
absorption. However, it appears to be a feature of the youth and low gravity of 
cluster members, as demonstrated by the AMES-Dusty-1999 models in Section 4. Veiling by 
hot circumstellar dust cannot explain it, since the spectra decline steeply toward 
2.5~$\mu$m, showing that water absorption is not significantly veiled at these 
wavelengths. 

	At $K$ band we might expect that hot dust would redden the spectra but if this 
is occurring the effect is too small to be unambiguously identified, except in the 
cases of 014-413 and 047-436, the only sources in Figure 8(a-b) showing much more flux in
the $K$ bandpass than at $H$. 014-413 appears to be associated with a weak wind collison 
shock front in the data of Bally, O'Dell \& McCaughrean (2000) but shows no evidence of 
emission lines in the spectrum shown in Figure 8(a).
The possible effect of hot dust is shown in Figure 11, see Section 4.
We do not have $K$ band photometric data but Hillenbrand \& Carpenter (hereafter HC)
published $HK$ photometry of the central region of the Trapezium Cluster. In 
combination with extinction values derived from our $IJH$ data, these data
should identify any sources with large $K$ band excesses. 
The HC data contains photometry for only 2 of the 11 sources in our $K$ band
spectroscopic sample (which lies to the west of the cluster core). 
Their $H$ band fluxes of HC agree with our data to within 0.18 magnitudes for the 2 
sources, which is broadly consistent with the photometric uncertainties but may
indicate T Tauri-like variability. Neither of these has any $K$ band 
photometric excess, which is consistent with the spectra. Some sources
in the much larger photometric dataset do appear to have K bandpass excesses (see
Section 6 and Table 4). Wilking, Greene \& Meyer calculated the 
contribution of hot dust using $JHK$ photometry for low mass stars and brown dwarfs 
in the $\rho$ Ophiuchus star forming region. They found that hot dust contributed 
0\% to 20\% of the total flux in the $K$ bandpass in the majority of cases,
which would have only a small effect on spectral profile. Star formation
is nearly complete in the Trapezium Cluster and the average age of this
sample of low-extinction sources is probably greater than that of sources in 
$\rho$ Ophiuchus so it is not too surprising that we do not see many large infrared 
excesses here.

	The NaI line at 2.21~$\mu$m is the strongest atomic aborption line
in the $H$ and $K$ band spectra of field dwarfs (see Jones et al. 1994). It is 
clearly visible in the $K$ band spectra (see Figure 16) but is not seen in the 
Trapezium sources. We would expect it to be seen in the high quality spectrum of 
053-503 (see Figure 16) but there is only a small local minimum near that wavelength, 
which cannot be distinguished from noise. This weakness of this feature is also 
predicted by the AMES-Dusty-1999 models as a signature of low gravity, as described in 
Section 4.1.2. 

	The $K$ band spectrum of 047-436 (Figure 7(b)) has a very red continuum in 
which water absorption is weak or absent, following the trend seen at $H$ band. 
However, there is a rich sequence of H$_{2}$ emission lines, with strong lines at 
2.12, 2.41 and 1.96~$\mu$m and several weaker lines apparent also (see Table 2).
As mentioned above, this is an anomalously blue source in the $I$ band and these
are usually also resolved as proplyds by the Hubble Space Telescope (O'Dell \& 
Wong 1996), as discussed in Paper I. This particular source is listed as being stellar in 
the data of O'Dell and Wong however, perhaps because it is located a large distance 
(3 arcmin, or 0.4~pc) from the cluster centre, leading to less excitation of the 
optical emission lines that mark a proplyd, and insufficient contrast with the faint 
nebulous background for a silhouette disk to be observed. 047-436 lies 
$<1$ arcmin north of the complex of bow shocks associated with 
LL Ori., observed in H$\alpha$ by Bally et al.(2000).

	The H$_{2}$ emission is the signature of a very young system with plentiful 
circumstellar gas. Such a signature can be caused by shocked gas in a molecular outflow 
or by UV excitation of gas in a circumstellar accretion disk or envelope i.e. a proplyd 
similar to those discovered by O'Dell \& Wen (1994). In Orion the UV radiation is supplied 
mostly by $\theta_{1}$~Ori~C. Chen et al.(1998) resolved the 2.12~$\mu$m H$_{2}$ from 
2 proplyds with NICMOS and found that it was confined to the subarcsecond disk 
silhouettes seen in continuum imaging. However, the emission from the 3 strongest H$_{2}$ 
lines in 047-436 is extended along the slit of the spectrograph by at least 3 arcsec
(1300 au) in a southerly direction, the peak of the emission being coincident with
the continuum source. There is no extension in the continuum and seeing conditions
were fairly good ($\sim 0.6$ arcsec) at the time the spectrum was taken. The slit was 
oriented 18$^{\circ}$ west of north. The large extension of the H$_{2}$ emission in 047-436,
relative to the Chen et al. proplyds, may be due to the high extinction and much larger 
projected distance from the Trapezium stars, which would allow molecular material 
to survive to survive for a greater period of time.
The reduced group image is not very sensitive to larger scale 
emission because a nod length of only 10 arcsec was used in taking the spectra and
subtracting the background. However, the extended emission appears to be weaker in the 
northerly direction, extending for $<1.5$ arcsec. The strength of the emission in the 
southern direction declines monotonically with distance from the peak, dropping by a factor 
of $\sim$2 at a distance of 2 arcsec from the central source.
There do not appear to be any H$_{2}$ emission line spectra of 
proplyds in the literature, although there are spectra of shocked gas in the vicinity
of the BN object. Theoretical models of optical proplyd spectra, eg. Storzer \& 
Hollenback (1999) predict that some of the lines, eg. the (1-0) S(1) line at 
2.12~$\mu$m, may be collisionally excited in the dense environment of the circumstellar
disk, not excited by UV fluorescence. This makes it more difficult to understand the
relative strengths of the emission lines and a full analysis is beyond the scope of this
paper. However, the ratio of the (2-1) S(1) and (1-0) S(1) line strengths in Table 2 does 
suggest UV fluorescence.

\begin{figure*}
\psfig{file=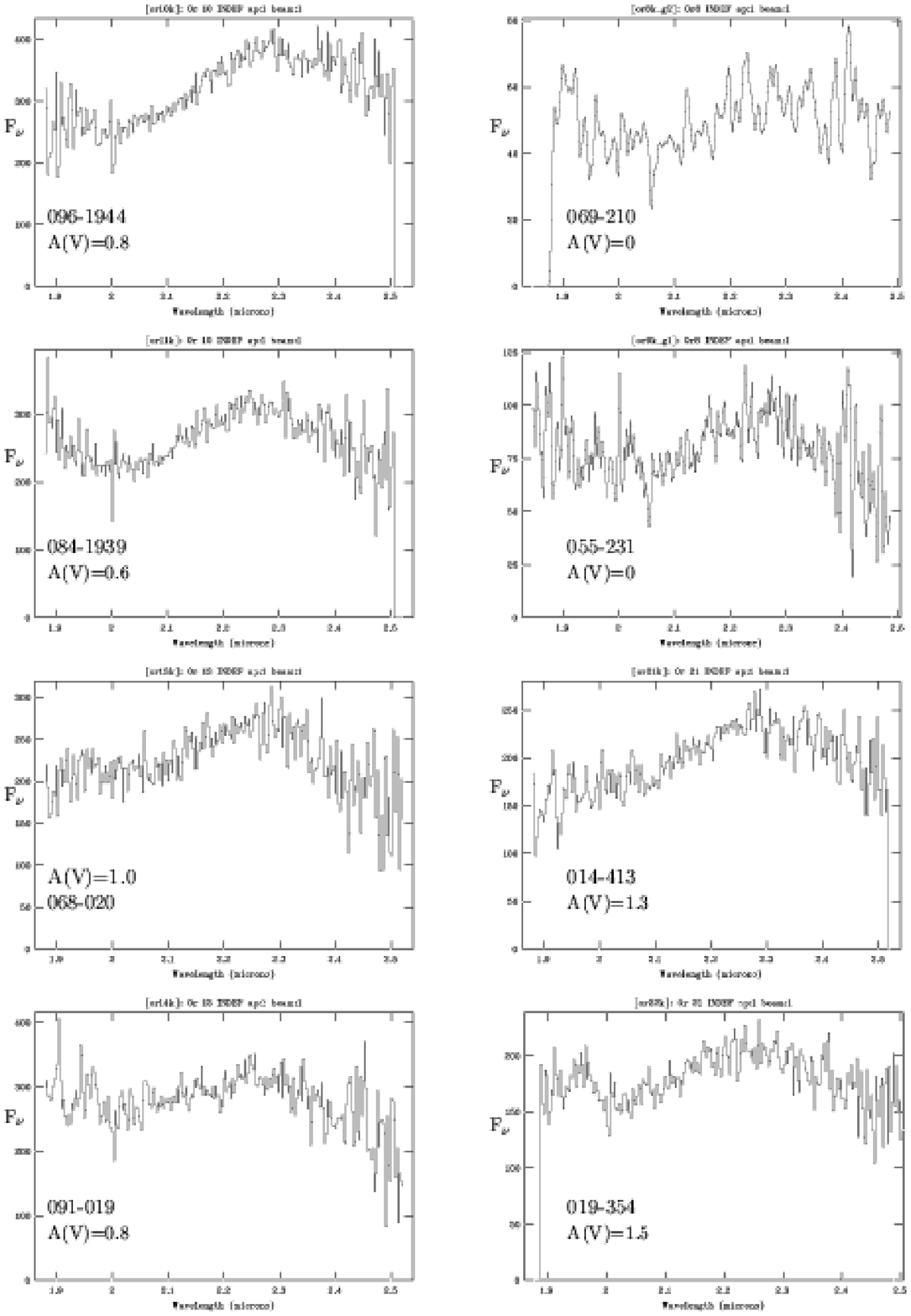,clip=,height=23cm,bbllx=0cm,bblly=1cm,bburx=20cm,bbury=28cm}
Figure 7(a): Observed $K$ bandpass spectra of Trapezium sources. The noise is indicated
by pixel to pixel variations, as in Figure 4(a-c). 069-210 and 055-231 are gaussian 
smoothed, because they were observed in poor conditions.
\end{figure*}

\begin{figure}
\begin{center}
\begin{picture}(200,530) 
\put(0,0){\includegraphics{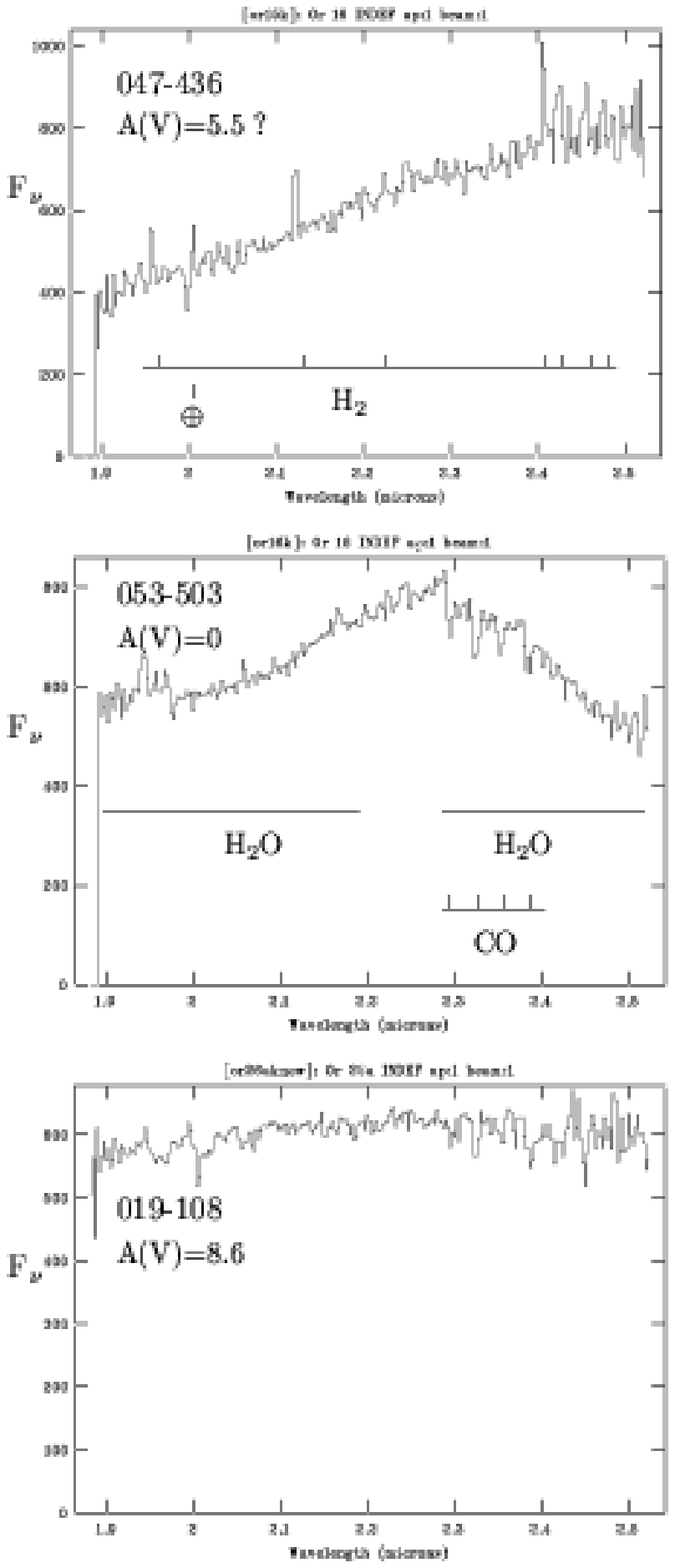}}
\vspace{-2cm}
\put(-20,0){Figure 7(b): Observed $K$ bandpass spectra of Trapezium sources.} 
\put(-20,-10){The stronger $H_{2}$ emission lines in 047-436 are marked, together} 
\put(-20,-20){with a telluric feature at 2.0~$\mu$m.}
\end{picture}
\end{center}
\end{figure}


We conclude that 047-436, the only source entirely lacking $H$ bandpass water absorption, is 
definitely a member of the Trapezium cluster. The slope of the 'dereddened' spectrum 
in Figure 8(b) is not very reliable because it is an anomalously blue source  
\it{(I-J)} \rm source.
It is quite possible that hot dust contributes to the $K$ band spectrum in this
very young system but the bulk of the very strong reddening in the $H$ bandpass must be due 
to extinction, perhaps by circumstellar dust. A very young system (age of
order 0.1~Myr) would have a hotter photosphere than the rest of the sample,
so any water absorption would be weak and in this case veiled by extinction. The 
mass of 047-436 can be estimated as 0.06~M${_\odot}$ from $IJH$ 
photometry in Paper I, assuming an extinction of A(V)=5.5. A 0.1~Myr source
would be less massive than a 1~Myr source for a given luminosity but an extinction 
of A(V)=10 is required to produce a flat F$_{\nu}$ spectrum, which might be more
appropriate. Hence, the mass is highly uncertain and it is possible that it is a 
very low mass star rather than a brown dwarf. 

	Object 019-108 also has high extinction and shows only weak water absorption 
at $H$ but its $K$ bandpass spectrum shows no sign of emission by H$_{2}$ or by hot 
dust. This suggests that it is a background star rather than a cluster member, as 
discussed in Section 3.1. However the evidence is not conclusive: circumstellar
matter could be entirely swept away by the photoionising UV radiation field
in $\ll 1$~Myr (eg. Storzer \& Hollenback), so it might be a very young
cluster member.
  
\section{Analysis}

	We analyse the spectra by comparison with both the AMES-Dusty-1999 model spectra 
calculated by Allard et al.(2000,2001) and with local field dwarfs. Since the spectra are
dominated by water absorption bands, a water index of some sort is the natural
tool to use to derive spectroscopic effective temperatures (T$_{eff}$). We do not fit the
full 1.4-2.5~$\mu$m spectra to those of field dwarfs or giants, after the fashion of
Luhman et al.(1998a) at optical wavelengths since the $H$ bandpass 
profiles are clearly different (see Section 1). The synthetic
spectra are believed to be most accurate at $\lambda \le 1.7~\mu$m (see Section 4.1). In 
addition, the most strongly 
temperature sensitive broad band feature at the observed wavelengths is the wing of
the 1.35~$\mu$m water band in the 1.4-1.7~$\mu$m region, the other bands being nearly 
saturated for all late M and L spectral types. Hence we define a water index 
$W=$F$_{\nu}$(1.50~$\mu$m)/F$_{\nu}$(1.682~$\mu$m). The model spectra are rebinned
to the same dispersion as the data and convolved with the instrumental profile
as measured in the Arc spectra. For the model spectra the 1.5 and 1.682~$\mu$m fluxes 
are averaged over 0.02~$\mu$m (or 9 data points) at both wavelengths, while 
the observed spectra have their 'continuum' fitted with high order cubic splines to 
reduce the measurement errors. The 1.50-1.682~$\mu$m range covers most of the depth of 
this water absorption band, the data being less precise at shorter wavelengths. The 
model spectra peak at 1.682~$\mu$m in the 2500-3300~K range and the dereddened data 
also peak at or very close to this wavelength.

	The choice of water index is always somewhat arbitrary, so we also
compute a reddening independent index of the type used by Wilking, Greene \& Meyer
at $K$ band, but applied here to the $H$ band spectra. This index is 
defined by $Q=(F1/F2)(F3/F2)^{1.219}$, where F1 is the F$_{\nu}$ flux at 1.57~$\mu$m, 
F2 is the F$_{\nu}$ flux at 1.682~$\mu$m and F3 is the F$_{\nu}$ flux at 1.79$~\mu$m. 
The 1.57 and 1.79~$\mu$m wavelengths are chosen because they must be roughly 
equidistant from the flux peak for good precision and because the data is often of
poor quality at $\lambda \gtsimeq 1.8\mu$m, and therefore not used in the continuum
fitting process. We also tested an index similar to that of Delfosse et 
al.(1999), based on the slope in the 1.51-1.57~$\mu$m region. However, the 
measurement errors for the slope in this narrow region are generally too large for that
index to be very useful here and $K$ bandpass data is required to normalise the
index as defined by Delfosse et al. The measured $W$ and $Q$ indices are listed in Table 3,
along with spectroscopic temperatures derived by fitting the $W$ indices to those
of the models. Temperatures are rounded to the nearest 50~K, since the model
grid has an interval of 100~K.

	We assign approximate spectral types to each source (see Table 3), by 
the comparison of their $W$ and $Q$ indices with those of the local field dwarfs 
observed by Leggett et al.(2001), whose spectra types are defined using the 2MASS
Kirkpatrick et al.(1999) system. The reddening independent $Q$ index should be the 
more reliable for measuring water depth and for spectral typing but the $W$ index 
generally produces very similar spectral types, as shown in the Table.

\begin{figure}
\psfig{file=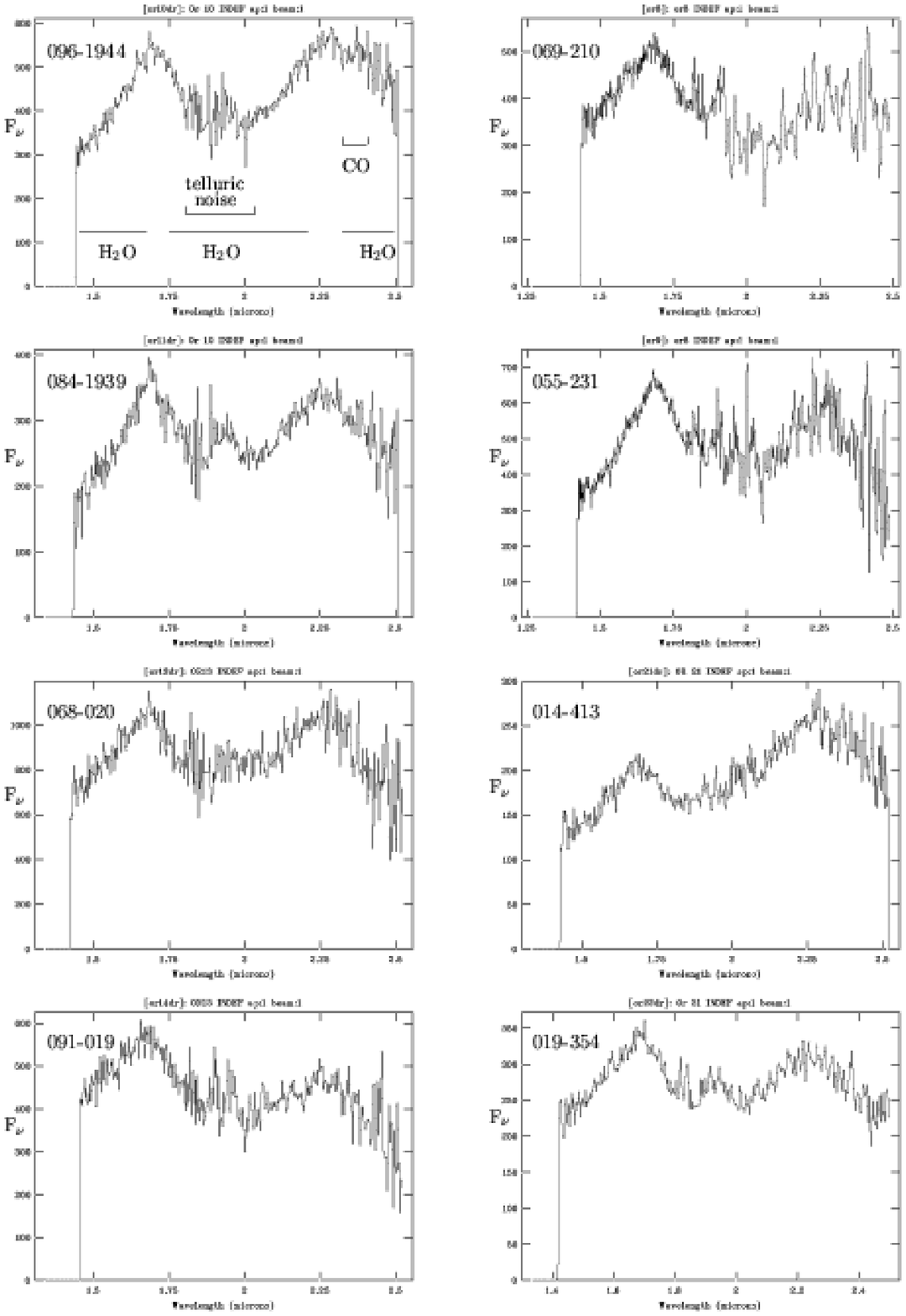,clip=,height=23cm,bbllx=0cm,bblly=1cm,bburx=20cm,bbury=28cm}
Figure 8(a): Dereddened spectra for Trapezium sources observed at both $H$ and $K$.
Noise in the 1.8-2~$\mu$m region is due to time variable telluric water absorption.
\end{figure}

\onecolumn
\twocolumn
\begin{figure}
\psfig{file=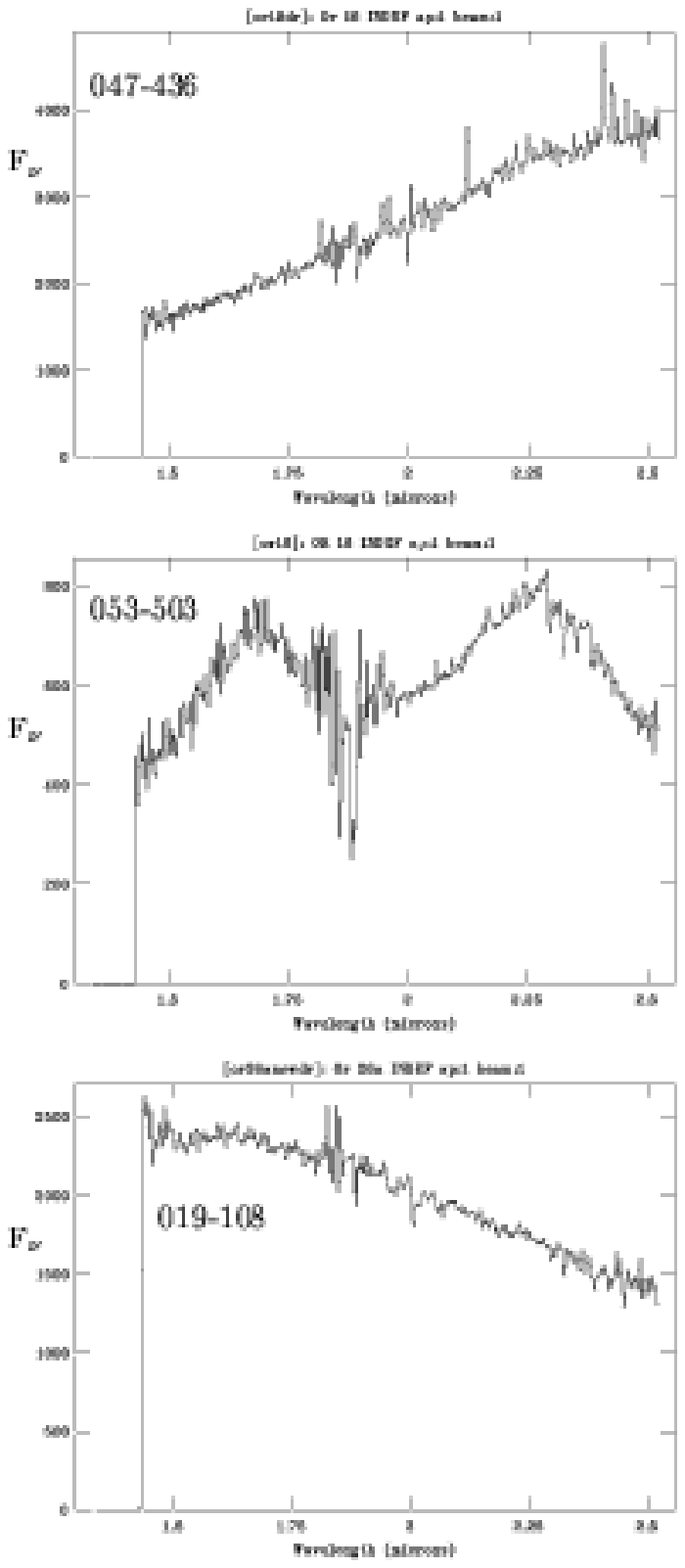,clip=,height=18cm,bbllx=6.5cm,bblly=5cm,bburx=15.5cm,bbury=23cm}
\vspace{-1cm}
Figure 8(b): Dereddened spectra for Trapezium sources observed at both $H$ and $K$.
\end{figure}

	It is important to note that these are not
real spectral types as might be defined at optical wavelengths - they are merely
the spectral types that the $H$ bandpass water depth would correspond to if the
the Trapezium sources were high gravity field dwarfs. Low gravity sources have
stronger water absorption for a given T$_{eff}$ than field dwarfs. Hence the
spectral types appear rather late for the corresponding spectroscopic temperatures in 
Table 3 and we would expect many of the L-type sources to have late M-type spectra at 
optical wavelengths. The behaviour of the $W$ and $Q$ indices as a function
of spectral type is shown in Figure 9 for the field dwarfs. Most of the
15 field sources follow a linear relation quite well but there is some 
interesting scatter (which is definitely real). Gd165b is an L4 dwarf
(in fact the prototypical object) but has very weak water absortion for its spectral
type. The 2 M9 dwarfs in the Leggett al.(2001) sample show large scatter on
either side of the best fit lines, particularly in the $W$ index. Some of these
differences might be attributed to dust effects: dust in cool metal rich
atmospheres is expected to heat the photosphere; this may be influenced by
metallicity, rotation and age, as noted by Leggett et al. However, there is no clear 
difference between the water depths of Young Disk and Old Disk sources in the sample, 
categorised kinematically by Leggett et al.(2001). Another study by Leggett et 
al.(2000) showed little difference due to metallicity in the $H$ bandpass when 
comparing Disk dwarfs and Halo subdwarfs but much weaker water absorption in 
Halo extreme subdwarfs, in which the $H$ bandpass profile is dominated by H$_{2}$ 
opacity in early M-type dwarfs.

\begin{figure}
\begin{center}
\begin{picture}(100,480)

\put(0,0){\includegraphics{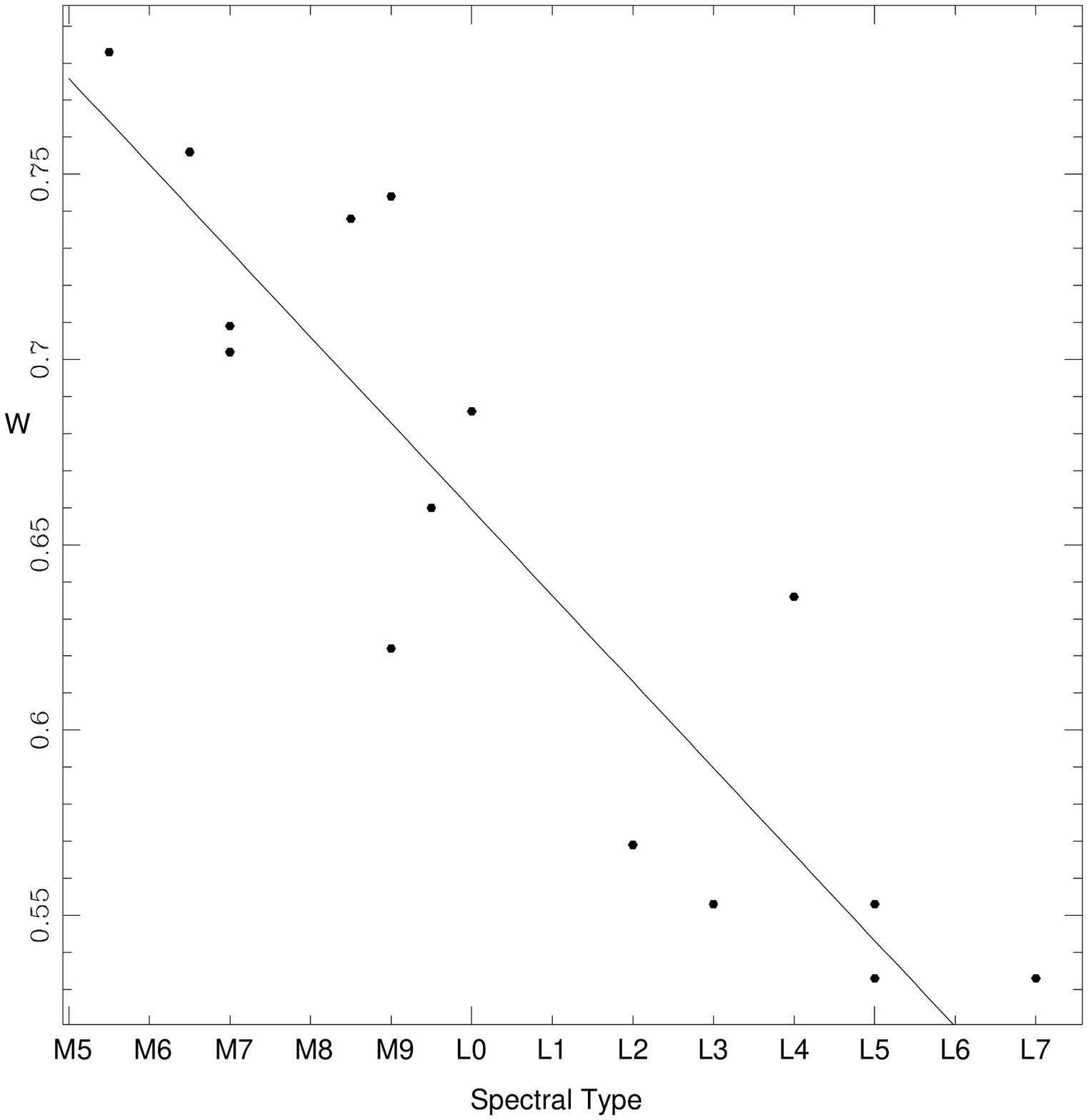}}

\put(0,0){\includegraphics{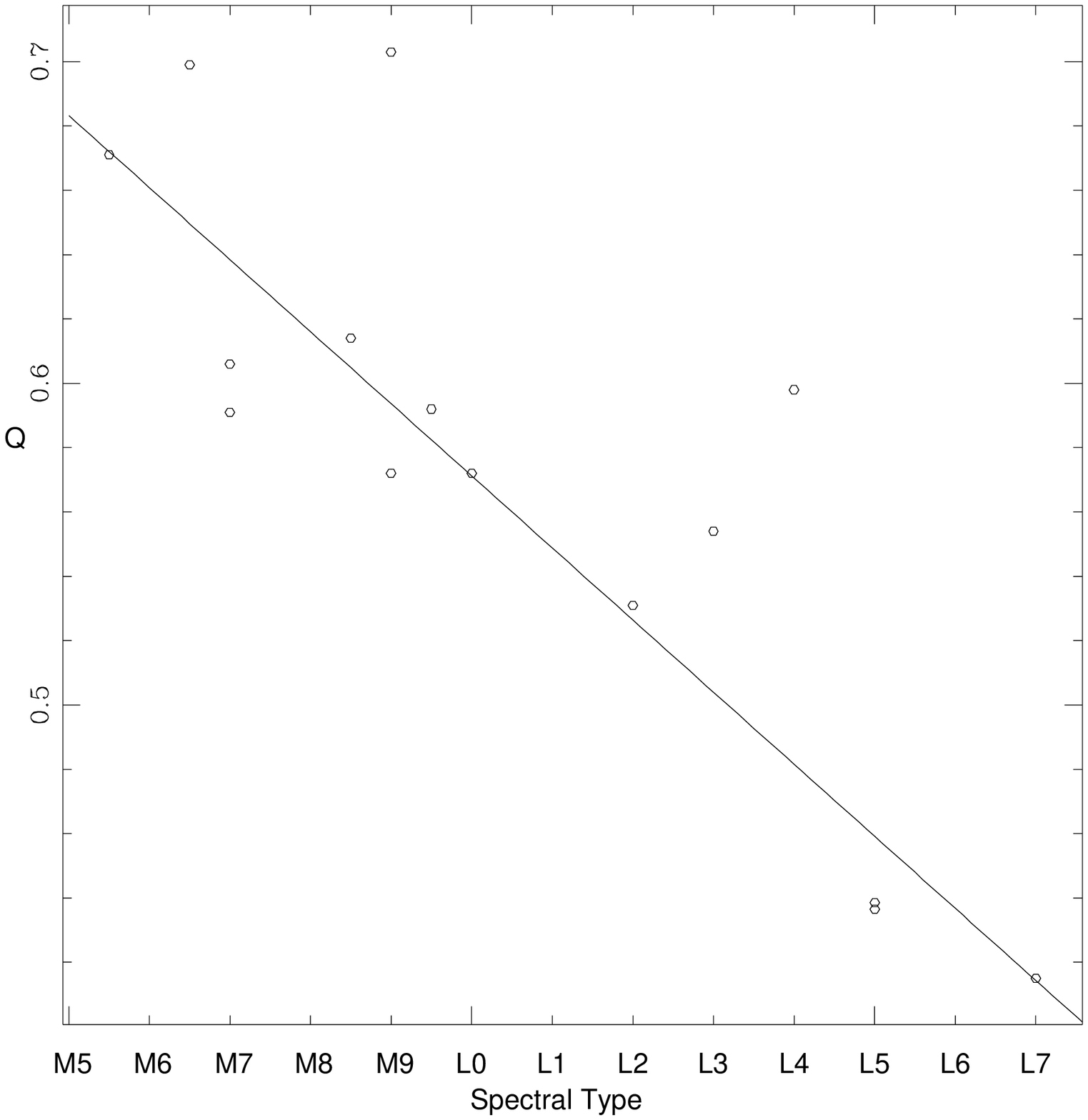}}

\end{picture} 
\end{center}
\vspace{-4mm}
Figure 9: $W$ and $Q$ spectral indices as a function of spectral type for 
Field Dwarfs. Best fit lines are plotted, excluding the divergent L4 source and
the 2 M9-types. Including these would lead to later types for the coolest
sources. Measurement errors are very small for these nearby sources.
\end{figure}

	The spectral types of the Trapezium sources lie in the range M7-L5 
for 16 of the 18 low mass sources ($<0.1~$M$_{\odot}$) which show strong water 
absorption. The 2 objects with later spectral types are faint and might have 
imperfect background subtraction (see Section 2). The fitted spectroscopic 
temperatures (2100, 2350~K) are consistent with the new
predictions for planetary mass sources by Baraffe et al. and by BM97
but are slightly low for these 2 sources with M$\sim 0.016$~M$_{\odot}$
at 1~Myr, unless they are younger, less massive objects of equivalent 
luminosity but greater 
size. It might be useful to define a new luminosity class or spectral type to 
describe these young, low gravity objects but since the gravities increase 
continuously over a period of several hundred Myr this is probably not a valuable 
concept. Instead we regard these low mass sources as a low luminosity extension to 
the region occupied by T Tauri stars on the HR diagram.

\subsection{Synthetic Spectra}

	The AMES-Dusty-1999 spectra appear to represent a significant improvement over 
previous models (such as NextGen) of cool dwarfs at infrared wavelengths, as
discussed by Allard, Hauschildt \& Schwenke (2000, hereafter AHS). Models
using the Miller-Tennyson list do not accurately reproduce
the infrared profiles of either these data or field dwarfs (see Fig.9 of 
Leggett et al.2001). By contrast the AMES-Dusty-1999 spectra, which use the newer NASA Ames
water line list, provide quite a good fit to the $H$ bandpass profiles of the Orion data
(see Figure 10(a-d)) and go some way toward reproducing to the plateau seen in
spectra of field dwarfs.
The synthetic spectra decline a little more steeply on the long side of the $H$ bandpass 
than the data, unless T$_{eff} \gtsimeq 3300K$. Such a high temperature would not be 
consistent either with the photometric temperatures obtained from the dereddened 
\it{(I-J)} \rm colours in Paper I, or the evolutionary predictions of BM97 and Baraffe et al.
As noted in Section 3.1, there appears to be
incompleteness in the water line list for the absorption band centred near 1.95~$\mu$m.
This compromises the predicted spectral profiles in the 1.7 to 2.2~$\mu$m region
and leads to small errors in the predicted $H$ and $K$ bandpass fluxes, as discussed by 
AHS. By contrast the water absorption band centred at 1.35~$\mu$m is well fitted by the 
AMES-Dusty-1999 spectra (see AHS) and the fluxes and spectral profiles
are believed to be more accurate at $\lambda \le 1.7~\mu$m. It is generally not possible to 
fit both sides of the $H$ bandpass spectra simultaneously. This is one of the reasons why
the short wavelength side is used to derive spectroscopic temperatures, the other
being the greater temperature dependence of that part of the synthetic spectra. 

\begin{figure*}
\begin{center}
\vspace{3cm}
\begin{picture}(200,480)
\small
\put(0,0){\includegraphics{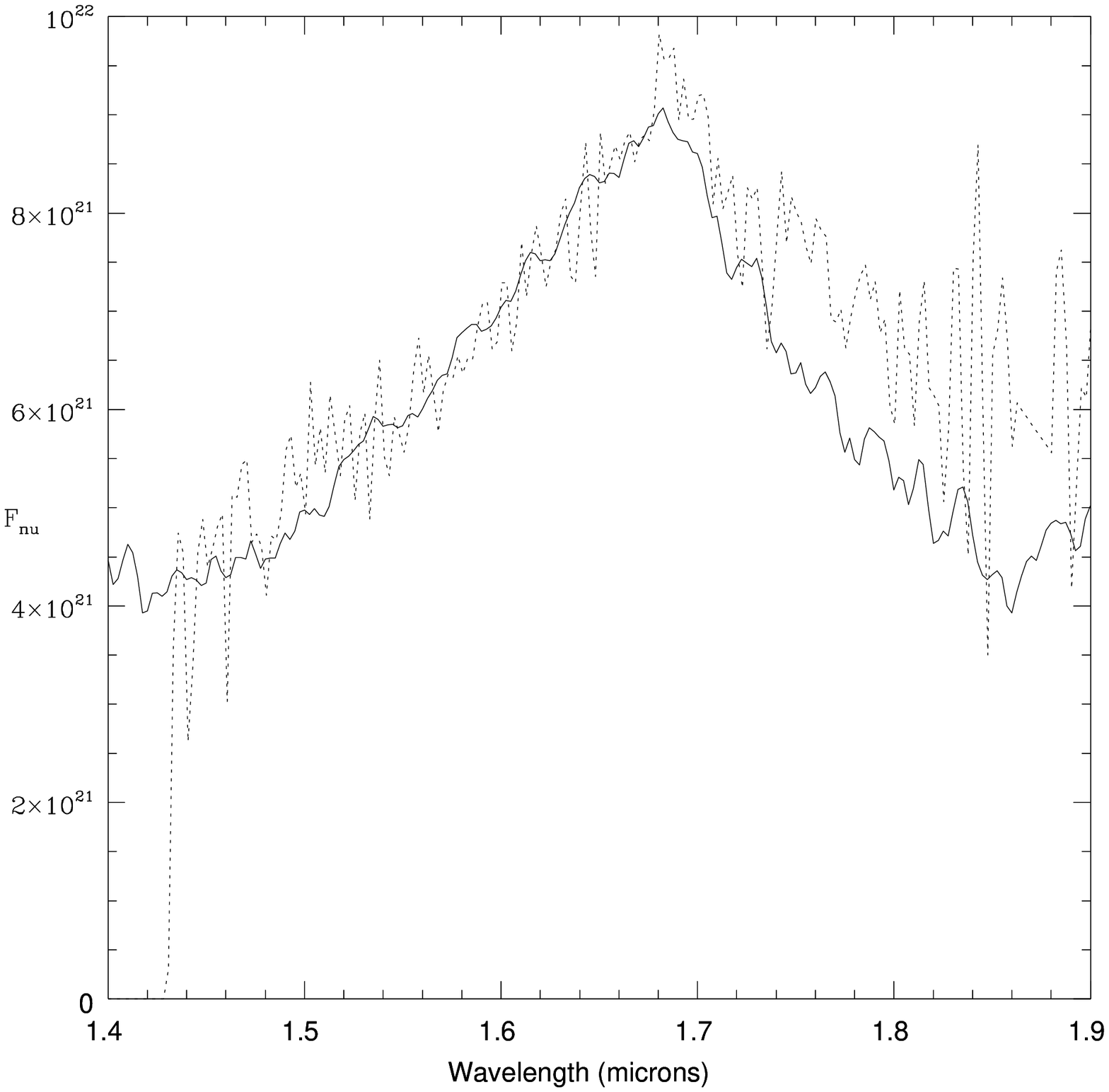}}
\put(-100,535){084-1939}
\put(-100,522){2500 K} 
\put(-100,509){[g]=3.5}

\put(0,0){\includegraphics{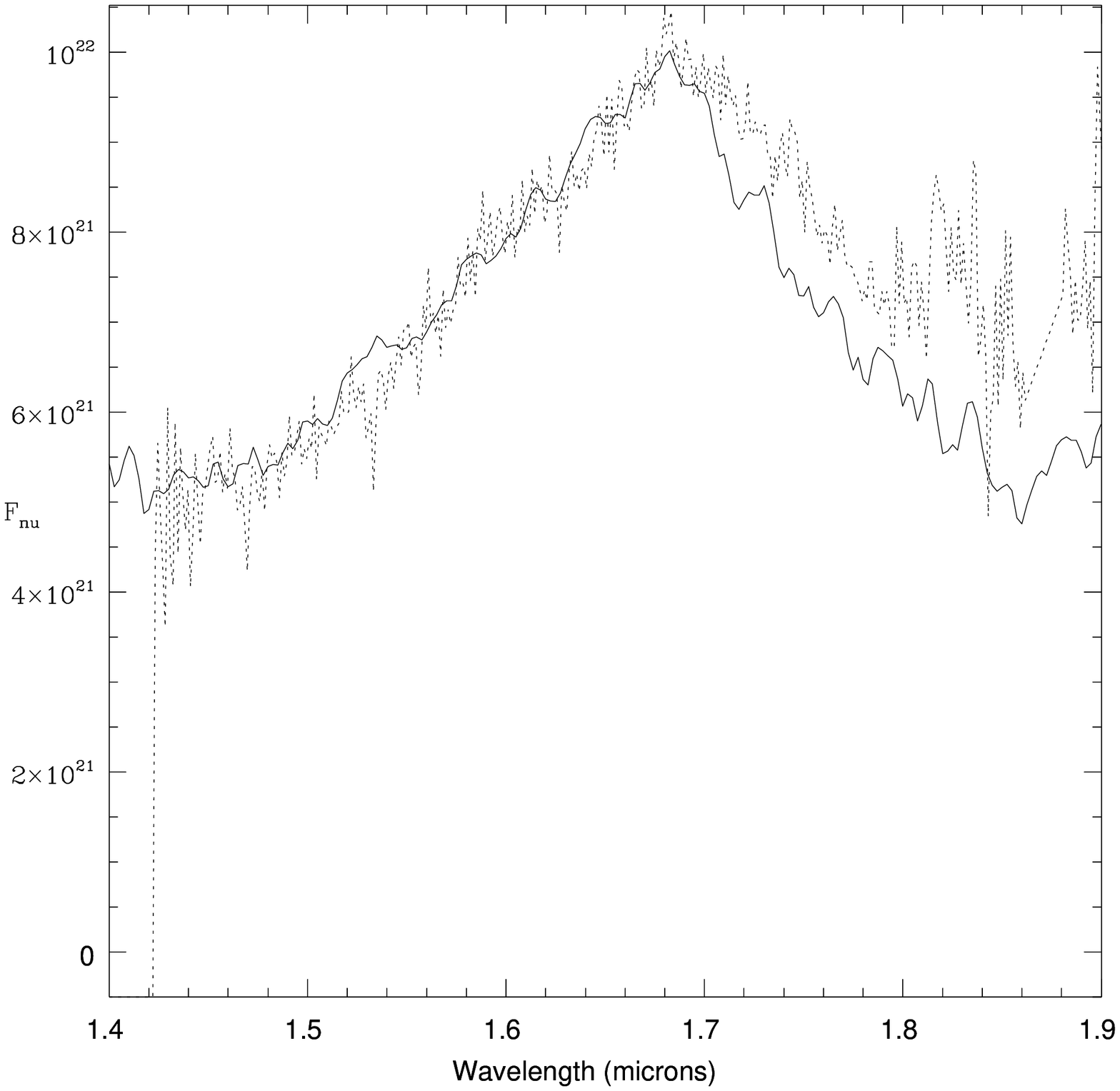}}
\put(-100,295){055-231}
\put(-100,282){2600 K}
\put(-100,269){[g]=3.5}

\put(0,0){\includegraphics{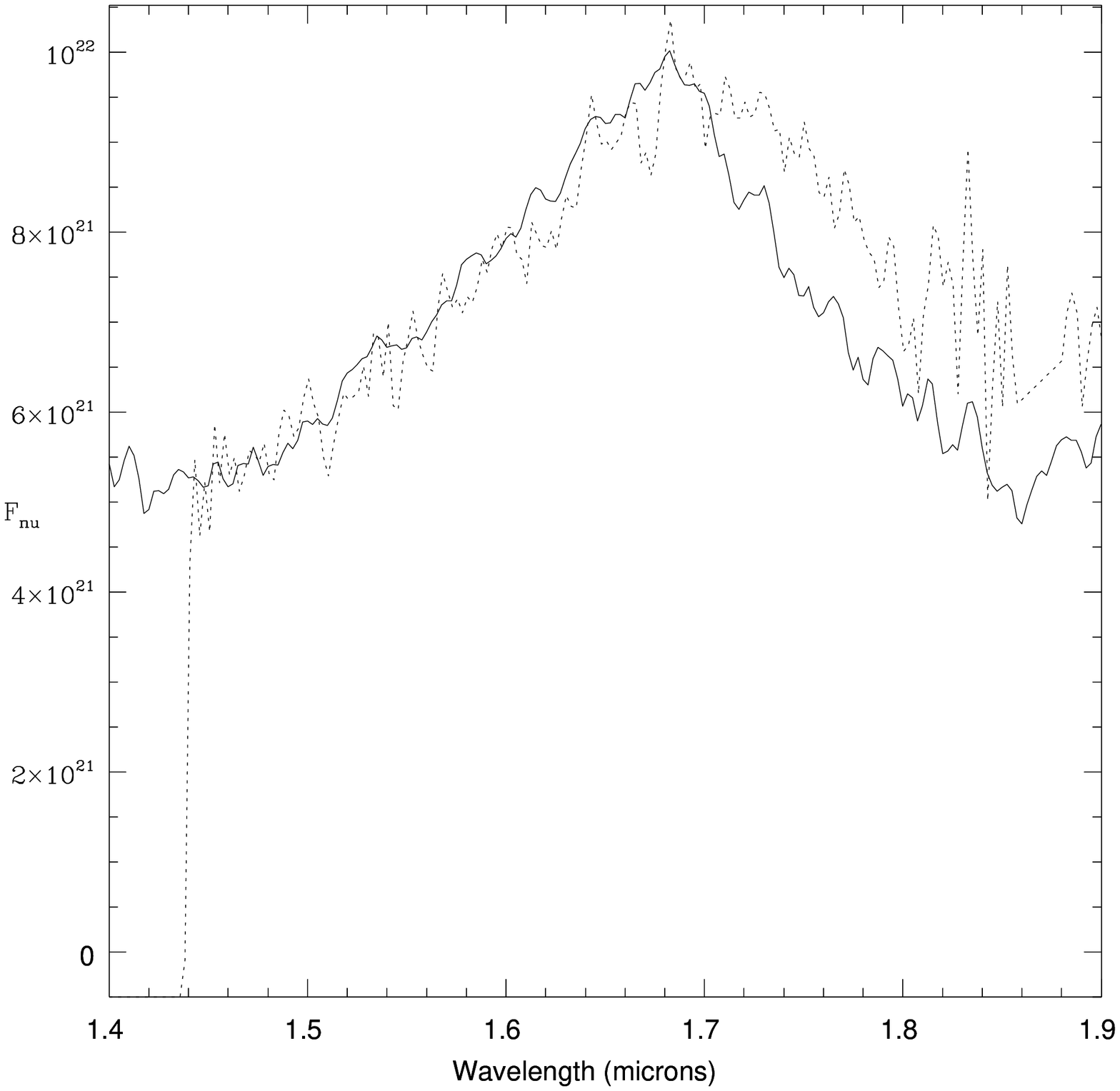}}
\put(155,535){096-1944}
\put(155,522){2600 K} 
\put(155,509){[g]=3.5}


\put(0,0){\includegraphics{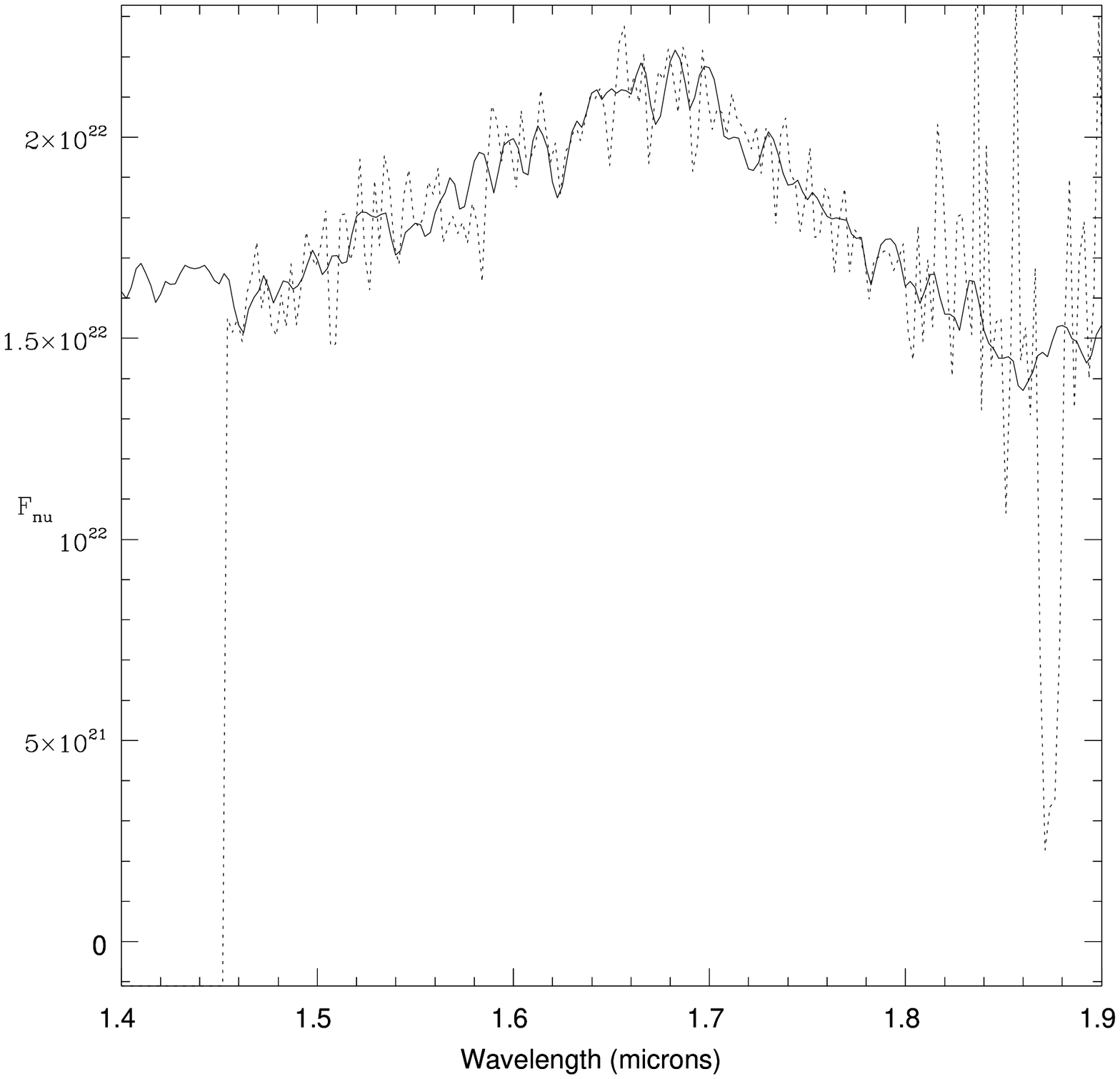}}
\put(155,295){091-019}
\put(155,282){3300 K}
\put(155,269){[g]=4.0}

\end{picture} 
\end{center}
\vspace{-26mm}
Figure 10(a-d): AMES-Dusty-1999 temperature fits (solid lines) to 4 sources
(dotted lines), fitted via the $W$ index. The 1.4-1.7~$\mu$m region and the peak 
are well fitted. The 1.7-1.9~$\mu$m region cannot generally be fitted
simultaneously, except in the warmest sources (see text). The discrepancy 
increases for higher gravities.
\end{figure*}

	The discrepancy at 1.7-1.9~$\mu$m seen in Figure 10(a-d) could in principle be 
caused by hot circumstellar dust emitting detectable flux on the long wavelength
side of the $H$ bandpass. However, as shown in Figure 11, this would 
lead to strong and obvious excess emission in the $K$ bandpass spectra, and some 
veiling of the water absorption near 2.5~$\mu$m, which is not seen in most of the 
sources in Figure 8(a-b).

The evolutionary calculations of BM97, D'Antona \& Mazzitelli (1998) and Baraffe et
al. predict surface gravities between log(g)=3.5 and log(g)=4.0 for sources of brown
dwarf and planetary mass in the 0.3-2~Myr age range expected for most Trapezium 
sources. AMES-Dusty-1999 spectra in this gravity range reproduce the constant slope 
of the data at $\lambda < 1.68~\mu$m and the wavelength of the flux peak. Higher 
gravity
synthetic spectra show a flatter profile near the flux peak, and a steeper slope at
shorter wavelengths. This is most easily seen in an F$_{\lambda}$ plot, shown in 
Figure 12, which may be compared with the composite Orion spectrum and field dwarf 
spectra in Figure 5(b). This is qualitatively more similar to the profile of 
field dwarfs (with log(g)$\approx 5.5$), although the abrupt change in slope observed 
in the 1.55-1.60~$\mu$m interval is not reproduced. The higher gravity models also 
decline more steeply on the long wavelength side of the $H$ bandpass than 
the low gravity ones, which leads to a greater discrepancy with the data at those 
wavelengths than is seen in the low gravity fits in Figure 10(a-d). 

	In the $K$ bandpass the behaviour of synthetic spectra is somewhat similar to 
that in the $H$ bandpass but less clear cut. The low gravity models have a fairly 
constant slope on the wing of the 
water absorption band in the 2.05-2.25~$\mu$m region, whereas high gravity models
show a curved profile similar those in the $H$ bandpass. The water vapour absorption is
less deep at $K$ bandpass and the profile of the flux peak is apparently affected by 
H$_{2}$ absorption longward of 2.2~$\mu$m in some field L dwarfs (Tokunaga \& 
Kobayashi 1999). One might argue that some of the spectra 
shown in Figure 8(a-b), such as the high quality spectrum of 053-503, lack the flat 
profile seen in $K$ bandpass F$_{\nu}$ spectra of late M and L dwarfs (eg. Jones et al.1994)
but there is more variation between sources than at $H$ band, both in our data and in 
field dwarf spectra. Owing to these factors and the relatively poor quality of 
most of our $K$ band spectra the water absorption bands are not clearly 
distinguished from those of field dwarfs.

	The spectroscopic temperatures fitted in Table 3 generally lie in the range 
2150-3300~K for the 18/21 sources in the sample which have strong water absorption.
All but 2 sources have $T_{eff} \ge 2500$~K, so there is insufficient dust in the 
photospheres to have a significant effect on the synthetic spectra. 
The fitted values of T$_{eff}$ differ by some 450~K between log(g)=3.5
models and log(g)=4.0 models at the high temperature end, with the higher
gravity models giving lower temperatures. At 2500~K and below the same temperatures
are fitted for both gravities. These temperatures are in good agreement with the 
evolutionary calculations of BM97. in the 0.008-0.080~M$_{\odot}$ mass range, for which 
they predict $2150 < T_{eff} < 3230$ at 1~Myr, $2310 < T_{eff} < 3290$ at 0.3~Myr
and $2030 < T_{eff} < 3180$ at 2~Myr. The Baraffe et al. and D'Antona \& Mazzitelli
models generally predict very similar temperatures to BM97 (within 50-150~K) which
in our view is less than the uncertainty in the calibration of the absolute 
temperature scale with observed spectral type. The log(g)=4.0 models produce better 
agreement for the higher mass sources of the sample than the log(g)=3.5 models. 

The spectroscopic temperatures fitted at log(g)=4.0 also agree well with the
photometric temperatures derived from the dereddened \it{(I-J)} \rm colours from 
Paper I, by comparison with the colours of field dwarfs (predicted for main 
sequence stars by BCAH using the NextGen model atmospheres). This is illustrated in 
Figure 13, which shows an excellent correlation between the two independently derived
temperature scales but a systematic offset toward lower temperatures of about 250~K
in the photometric temperatures. The scatter is largely due to photometric 
measurement errors magnified by dereddening rather than the spectroscopic 
measurements, and fits in well with the estimated uncertainty of 0.25 mag in
the dereddened \it{(I-J)} \rm colours given in Paper I. The NextGen atmospheres
are known to fit main sequence infrared colours quite well, (better then the AMES-Dusty
models at T$_{eff} > 2500$~K) despite the very poor fit to the spectral profiles at $H$ 
and $K$ seen by Leggett et al.(2001). Hence the 250~K offset must be due either
to the lower gravity of the sample or errors in the AMES-Dusty models.
The NextGen spectra predict that \it{(I-J)} \rm colours differ by only of order 
0.1 mag between log(g)=3.5 models and log(g)=5.5 sources, which corresponds to an 
interval in temperature of $\sim$50~K. However, the lower gravity also changes the 
predicted \it{(J-H)} \rm colours by $\sim$0.1 mag, which leads to a larger difference 
in the \it{(I-J)} \rm colours via the dereddening process. Hence, it is quite 
possible that the offset of 250~K is due solely to gravity effects, and the spectroscopic 
temperatures at log(g)=4.0 are correct. They are more consistent with the 
evolutionary calculations for age 1~Myr than the photometric temperatures
and are derived from the well-fitted 1.4-1.7~$\mu$m region. 
Broad band \it{(J-H)} \rm fluxes produced by the NextGen and AMES-Dusty model 
atmospheres at these gravities are not yet more reliable than the assumption of main 
sequence colours, owing to the problems with the water line list referred to above. 
(The AMES-Dusty models produce most accurate colours for high gravity field dwarfs below 
2500~K, while the NextGen models work best at higher stellar temperatures, above 3000~K).
Hence, at present we continue to use empirical main sequence colours in the
dereddening procedure.

\begin{figure}
\begin{center}
\begin{picture}(160,180)

\put(0,0){\includegraphics{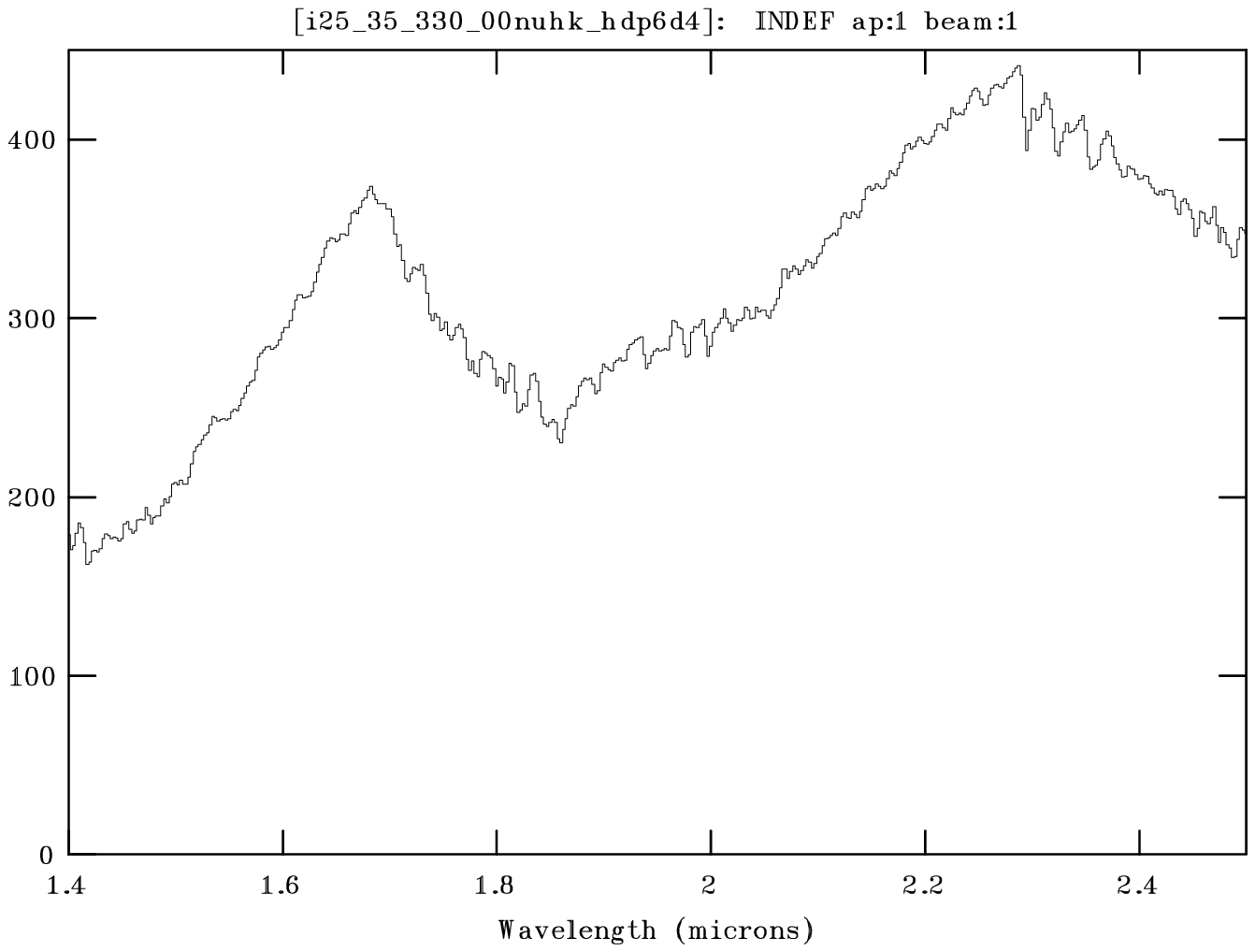}}

\put(0,0){\includegraphics{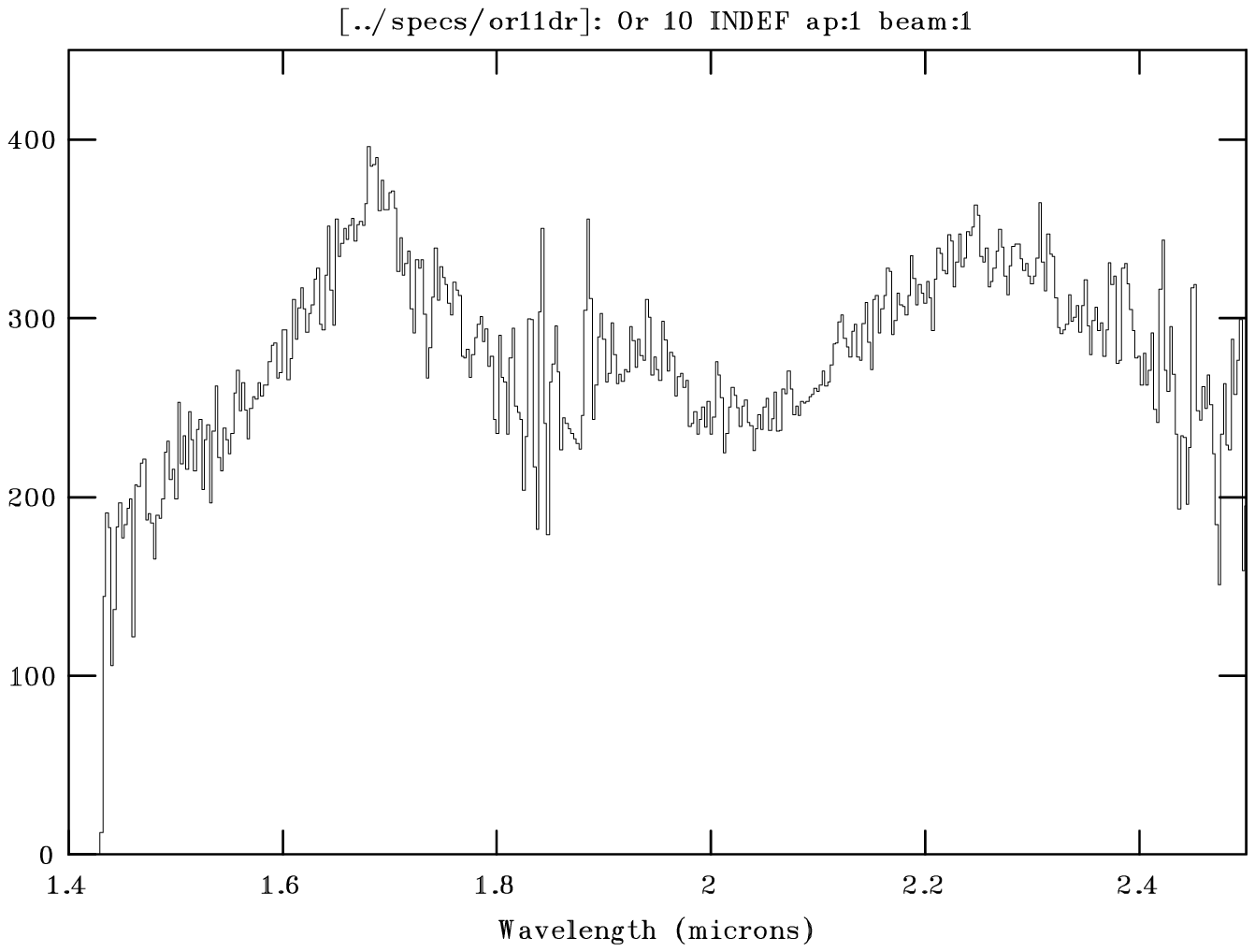}}

\put(0,0){\includegraphics{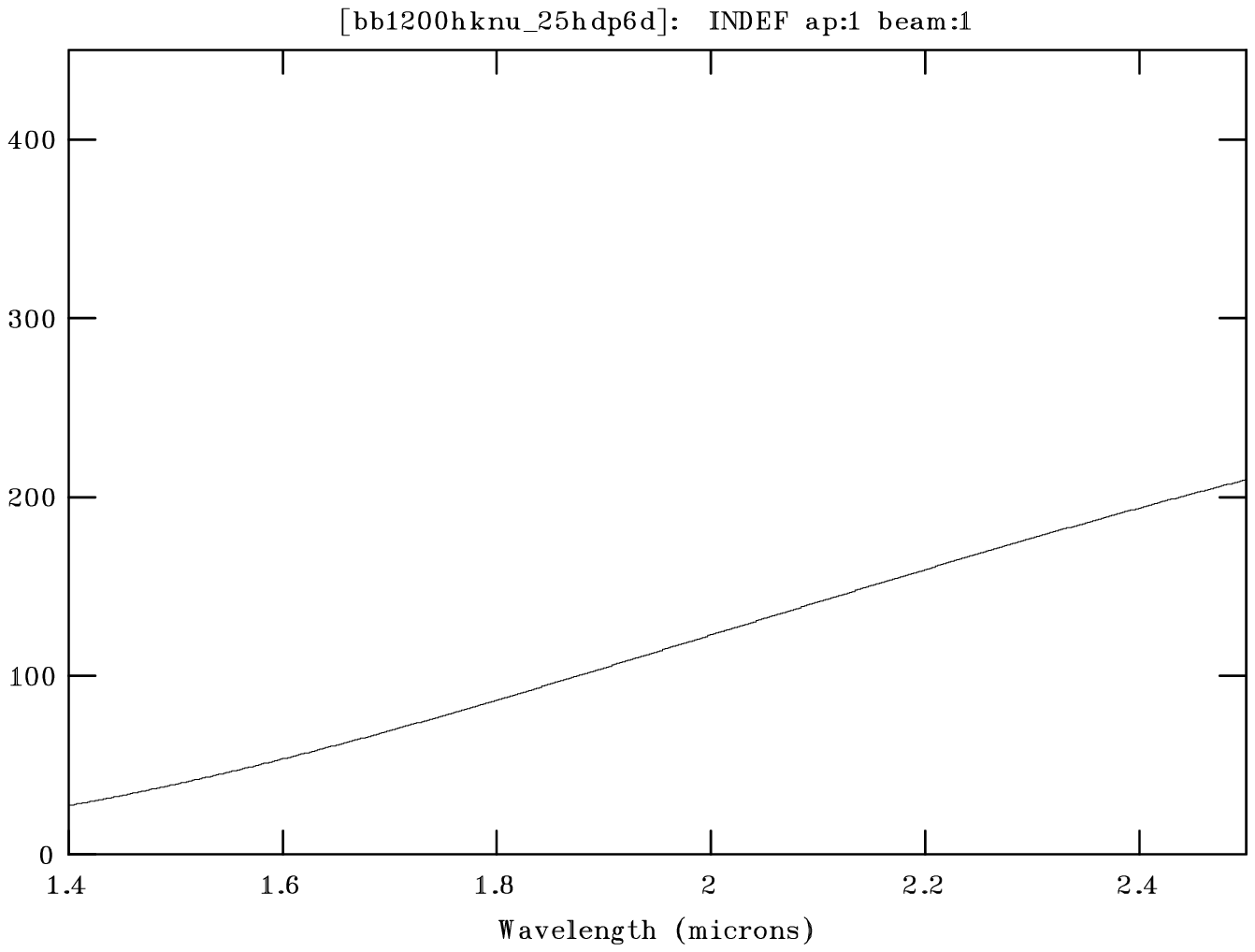}}

\put(90,55){\small{Hot Dust}}
\put(45,165){\small{Model + Dust}}
\put(120,110){\small{084-1939}}
\put(-50,100){\small{F$_{\nu}$}}

\end{picture} 
\end{center}
\vspace{-4mm}Figure 11: Hot dust. Strong emission by dust at 1200~K, corresponding to
excess E(H-K)=0.5 is added to an AMES-Dusty model. This can fill in the 
discrepancy at 1.7-1.9~$\mu$m but leads to a very obvious excess in the $K$
bandpass.
\end{figure}

\begin{figure}
\begin{center}
\begin{picture}(200,440)

\put(0,0){\includegraphics{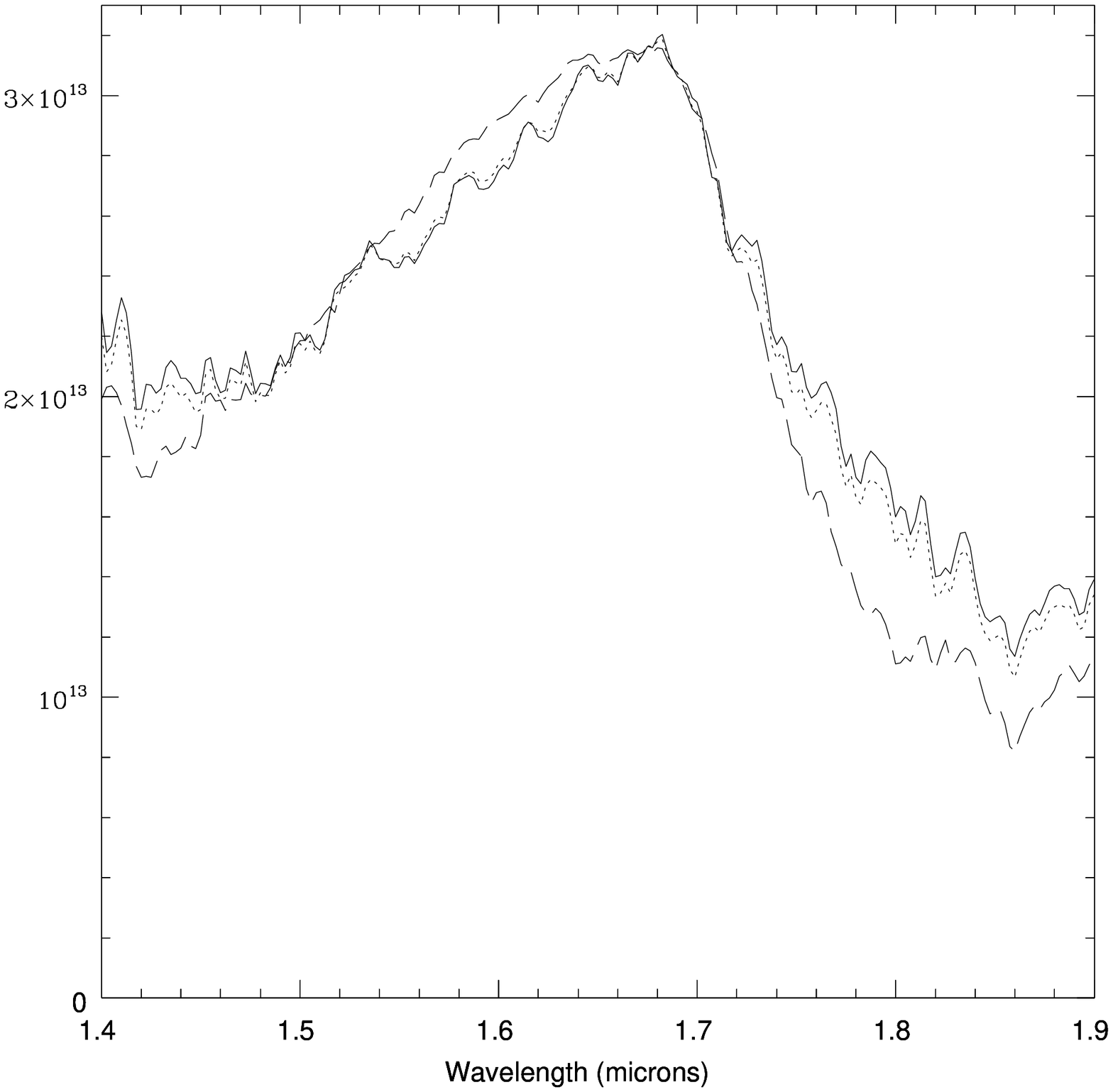}}
\put(-25,385){\small{F$_{\lambda}$}}

\put(0,0){\includegraphics{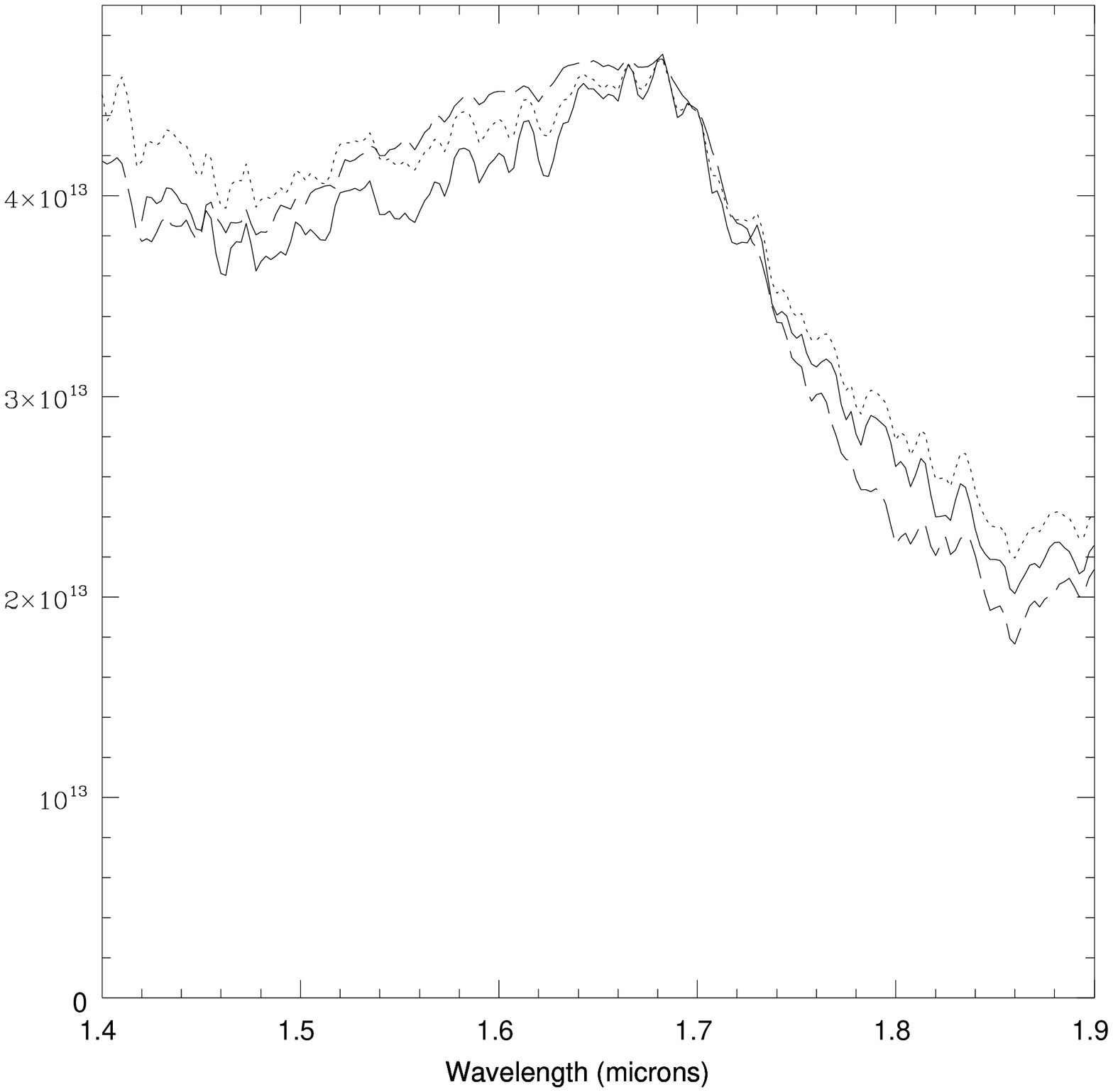}}
\put(-25,125){\small{F$_{\lambda}$}}

\end{picture} 
\end{center}
\vspace{-1cm}
Figure 12: Effect of gravity on AMES-Dusty models at $H$ (F$_{\lambda}$ spectra). 
(top) at 2500~K the [g]=3.5 spectrum (solid line) and [g]=4.0 spectrum (dotted line) 
are very similar but the [g]=5.5 model (dashed line) has a flatter peak. 
(bottom) at 2900~K the [g=3.5] and [g=4.0] spectra begin to diverge.
The [g=5.5] spectum requires a lower temperature, 2800~K, to produce
similar water absorption and crosses the lower gravity spectra.
\end{figure}

	There is a trend for less luminous, less massive sources to be cooler 
(see Figure 14), as indicated by the trend for water absorption to increase with 
decreasing mass. The BCAH models predict that water absorption 
increases monotonically with decreasing temperature in these substellar pre-main sequence 
objects. A mass-temperature relation correlation is predicted by all the 
published pre-main sequence isochrones. The derived luminosities, and hence 
masses, are nearly independent of temperature owing to the very weak temperature 
dependence of the \it{(J-H)} \rm colour so this is not a trivial result.
The correlation is imperfect, presumably because of the dispersion in ages.

\subsubsection{Carbon Monoxide and other molecules}

	The sythetic spectra also show a strong gravity dependence in the strength of the
CO absorption bands near 2.3~$\mu$m, which is largest at low temperatures. 
This is illustrated in Figure 15, with $K$ bandpass spectra of high and low gravity
model atmospheres at 2500~K and 2900~K. We define a a CO index, 
$K_{CO}=(F22-F23)/F22$ where F23 is the integrated F$_{\lambda}$ flux in the 
2.295-2.315~$\mu$m region and F22 is the integrated F$_{\lambda}$ flux in the 
2.2675-2.2875~$\mu$m region. The indices for models and data are listed in Table 3. 
All of the Trapezium sources have $K_{CO}<0.12$, i.e. an average absorption depth 
across the first CO band of $<12\%$. The observed CO indices of 0.09-0.12 observed in 
053-503, 014-413 and 055-231 are in good agreement with the model predictions for 
log(g)=3.5 to 4.0 and $2500 < T_{eff} < 3200$~K. The data are much less consistent with 
high gravity models, particularly for 
the cooler sources. For example, we observe $K_{CO}=0.119\pm0.055$ for 055-231, 
consistent with $K_{CO}=0.12$ at T$_{eff} = 2575$~K (with log(g)=3.5) but not 
$K_{CO}=0.21$ (with log(g)=5.5). 
Similarly, we observe $K_{CO}=0.059\pm0.035$ for 096-1944, roughly consistent 
$K_{CO}=0.106$ at T$_{eff} = 2700$~K (log(g)=3.5) but not $K_{CO}$=0.167 
(log(g)=5.5). The model predictions for log(g)=3.5 and log(g)=4.0 are almost
identical.

	In 6/11 sources CO is not unambiguously detected in these low resolution 
spectra, since the drop in flux seen in all sources at $\lambda > 2.3~\mu$m may be 
due to H$_{2}$O as well CO, and the bandheads are not sufficiently resolved. We place 
upper limits on the $K_{CO}$ index for these sources in Table 3. Again these are 
consistent with low gravity models, but not with high gravity models in the case of 
the cooler 
sources, since CO depths of order 20\% would be detected. This strong absorption
is observed in local dwarfs, as evidenced by our spectra of 3 red standards in 
Figure 16. 

	In field dwarfs a fairly strong feature is often observed near 1.62~$\mu$m,
with uncertain identification at present. The AMES-Dusty-1999 spectra resolve this 
into a complex feature with possible contributions from several lines of OH, CaI and FeI, 
with an OH doublet at 1.624~$\mu$m predicted to dominate in low gravity atmospheres. 
Giant spectra show the (v=6-3) CO bandhead at 1.619~$\mu$m in low resolution spectra,
which increases in strength in very low gravity supergiants (Origlia, Moorwood
\& Oliva 1993), but this is not predicted to be important at $log(g)=3.5$ to 4.0.
An absorption feature is marginally detected at the 1.624~$\mu$m wavelength in the 
composite Orion spectrum in Figure 5(a-b) but it is not unambiguously distinguished from 
noise and is much weaker than the $log(g)=3.5$ models indicate for the OH dominated 
features at T$_{eff} \ge 2800~K$. At $log(g)=4.0$ the 1.624~$\mu$m feature is no 
stronger than at $log(g)=5.5$, for the same temperature, so at that gravity the absence 
of a clear detection would be consistent with the data.
(The AMES-Dusty-1999 grid has a sampling interval of 0.5 dex in gravity). 

	Considering that the $log(g)=4.0$ models
also produce fitted spectroscopic temperatures more consistent with the evolutionary 
tracks than $log(g)=3.5$ models, this might be taken as evidence that the gravities are
closer to the higher value, though a finer grid in gravity would be needed to demonstrate
this. On the BM97 1~Myr isochrone the $3.5<log(g)<4.0$ interval covers an age range from
$<0.1$~Myr to several Myr for 0.008~M$_{\odot}$ planetary mass objects, so a very 
precise gravity measurement would be needed to determine the age with useful precision.
A slightly greater than average age might be expected for this 
spectroscopic sample of sources with low extinction.

\begin{figure}
\begin{center}
\begin{picture}(200,220)

\put(0,0){\includegraphics{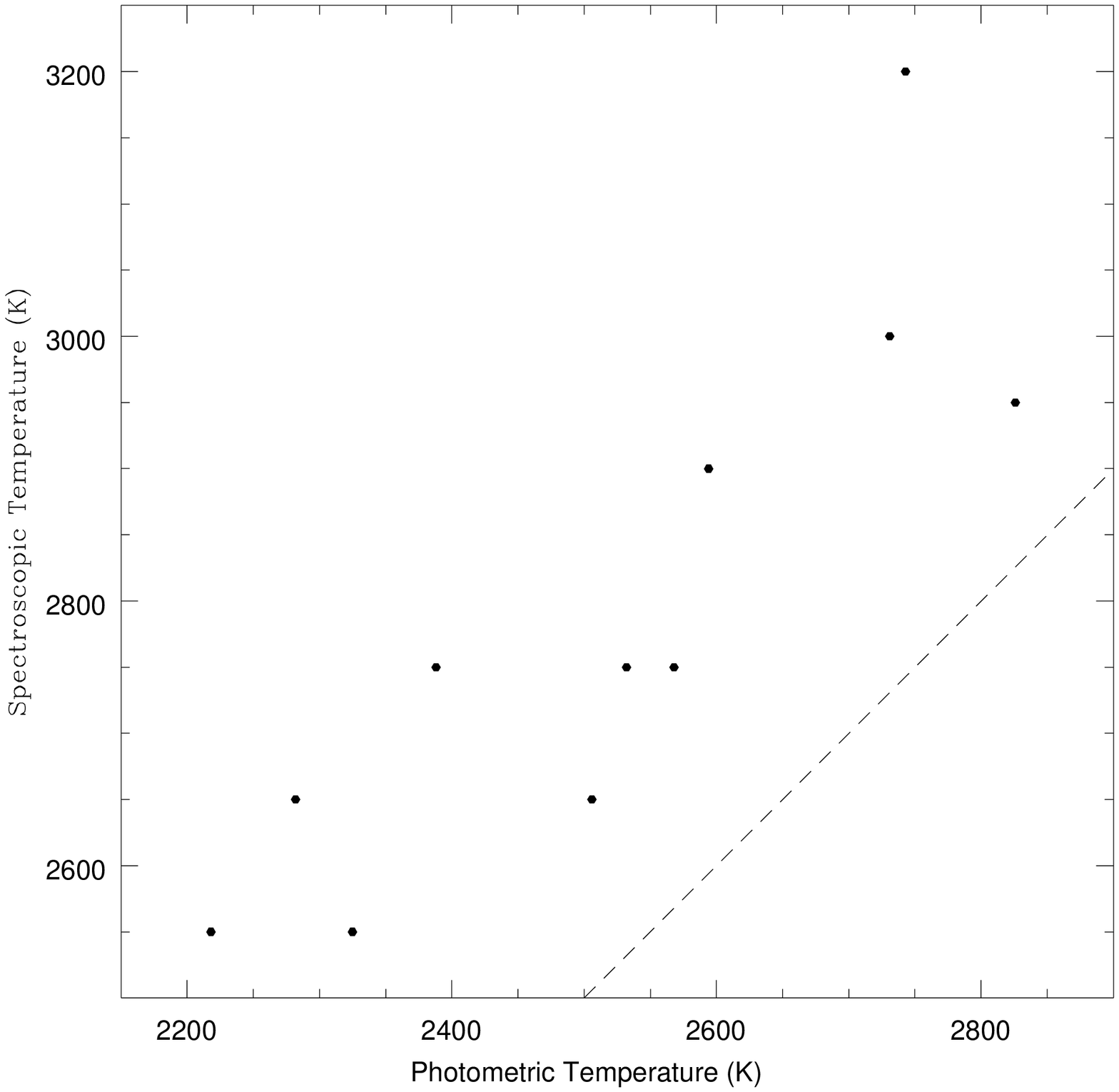}}

\end{picture} 
\end{center}
Figure 13: Spectroscopic Temperatures [g=4.0], correlated with independent
Photometric Temperatures for sources with measured (dereddened) \it{(I-J)} 
\rm colours. The correlation is excellent but the spectroscopic temperatures 
are higher by 250~K, as indicated by the dashed line for perfect agreement.
\end{figure}

\begin{figure}
\begin{center}
\begin{picture}(200,220)

\put(0,0){\includegraphics{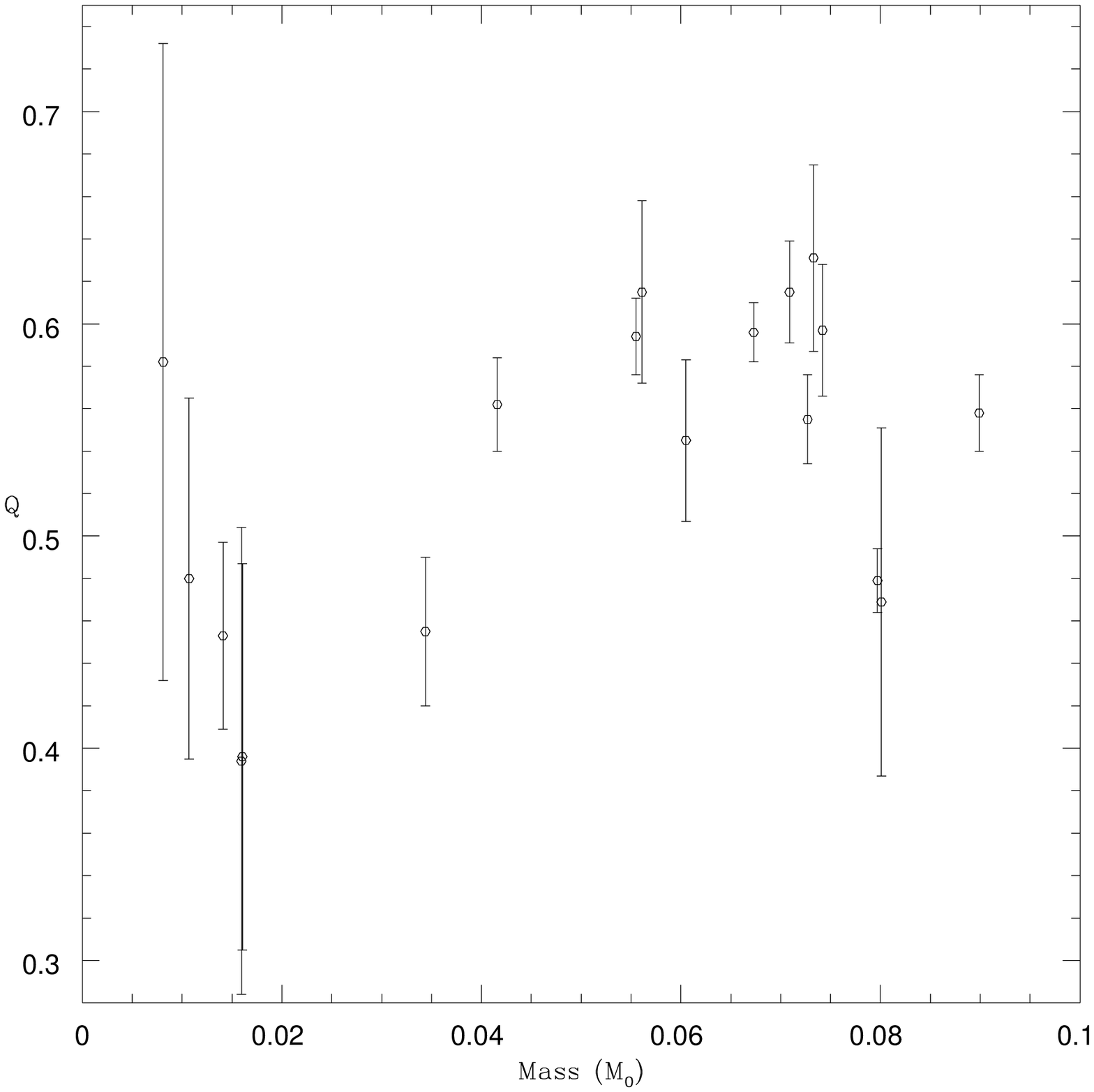}}

\end{picture} 
\end{center}
Figure 14: Spectroscopic Temperature [g=4.0] vs. Mass. Masses are from the 
BM97 1~Myr isochrone. A weak correlation is observed and the temperatures are
consistent with the essentially independent BM97 prediction.
\end{figure}

\subsubsection{Atomic Lines}

	The absence of the NaI line at 2.21~$\mu$m in 053-503 (our best $K$ 
band spectrum) is also indicative of a low gravity atmosphere, as
shown by the AMES-Dusty-1999 spectra in Figure 15. 053-503 shows only a weak local
minimum at the 1-$\sigma$ level: this cannot be distinguished from
noise but is consistent with the model. The weakness of NaI at 2.21$~\mu$m
has been observed in a low gravity YSO by Luhman et al.(1998b). In addition, 
the weakness of optical NaI and KI features in very young brown dwarfs
has been observed by Bejar et al.(1999) in the $\sigma$-Orionis
cluster; by Luhman, Liebert \& Rieke (1997) in a brown dwarf in the 
$\rho$ Ophiuchus region; and by Briceno et al.(1998) in pre-main sequence
stars and brown dwarfs in Taurus.

\begin{figure*}
\psfig{file=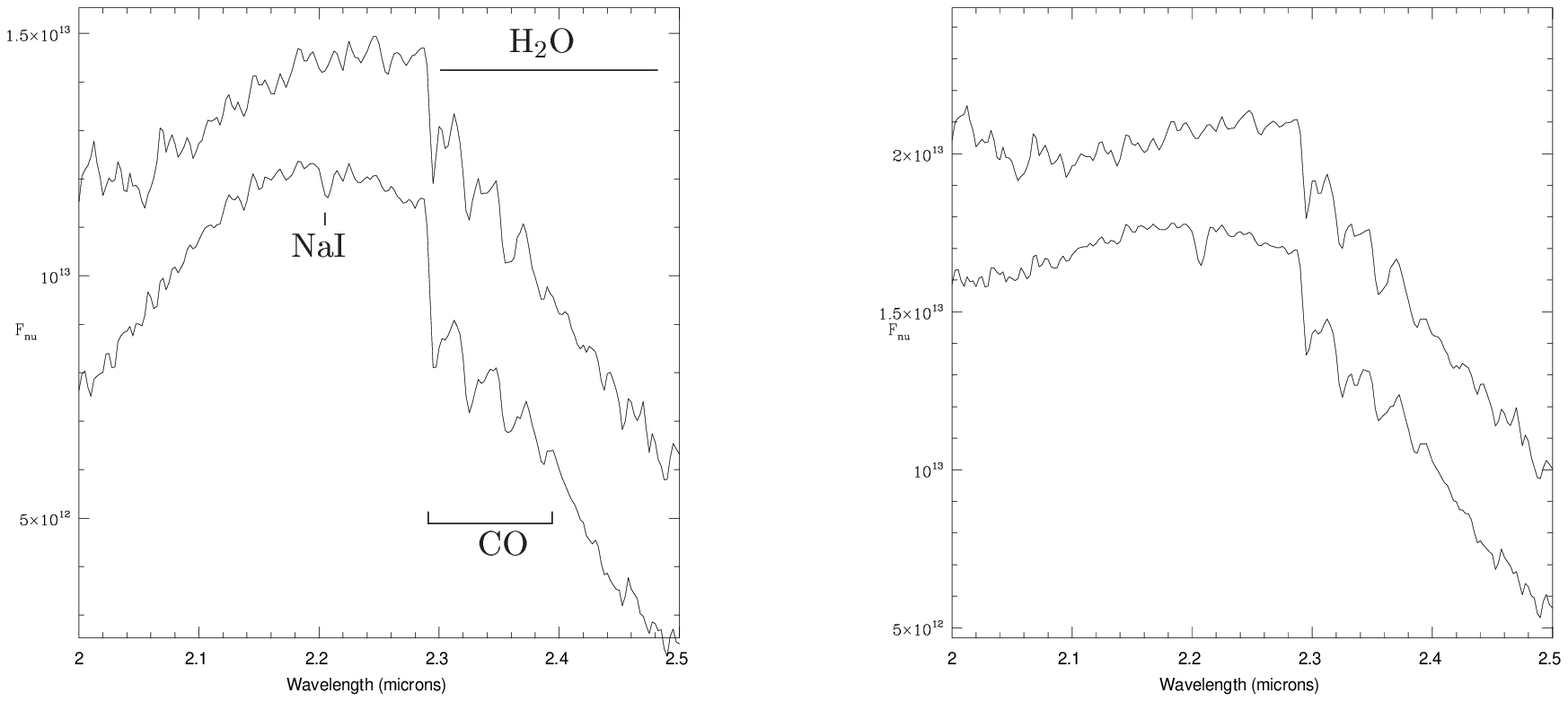,clip=,height=8cm,bbllx=0cm,bblly=17cm,bburx=20cm,bbury=27cm}
\vspace{-4mm}
Figure 15: Gravity effects on CO and NaI (F$_{\nu}$ spectra). (left) 2500~K; 
(right) 2900~K. Both CO and NaI are much weaker in the low gravity [g=3.5] 
AMES-Dusty spectra (offset upper plots) than the [g=5.5] atmospheres. 
[g=3.5] and [g=4.0] are nearly identical at these wavelengths.
\end{figure*}

\begin{figure*}
\psfig{file=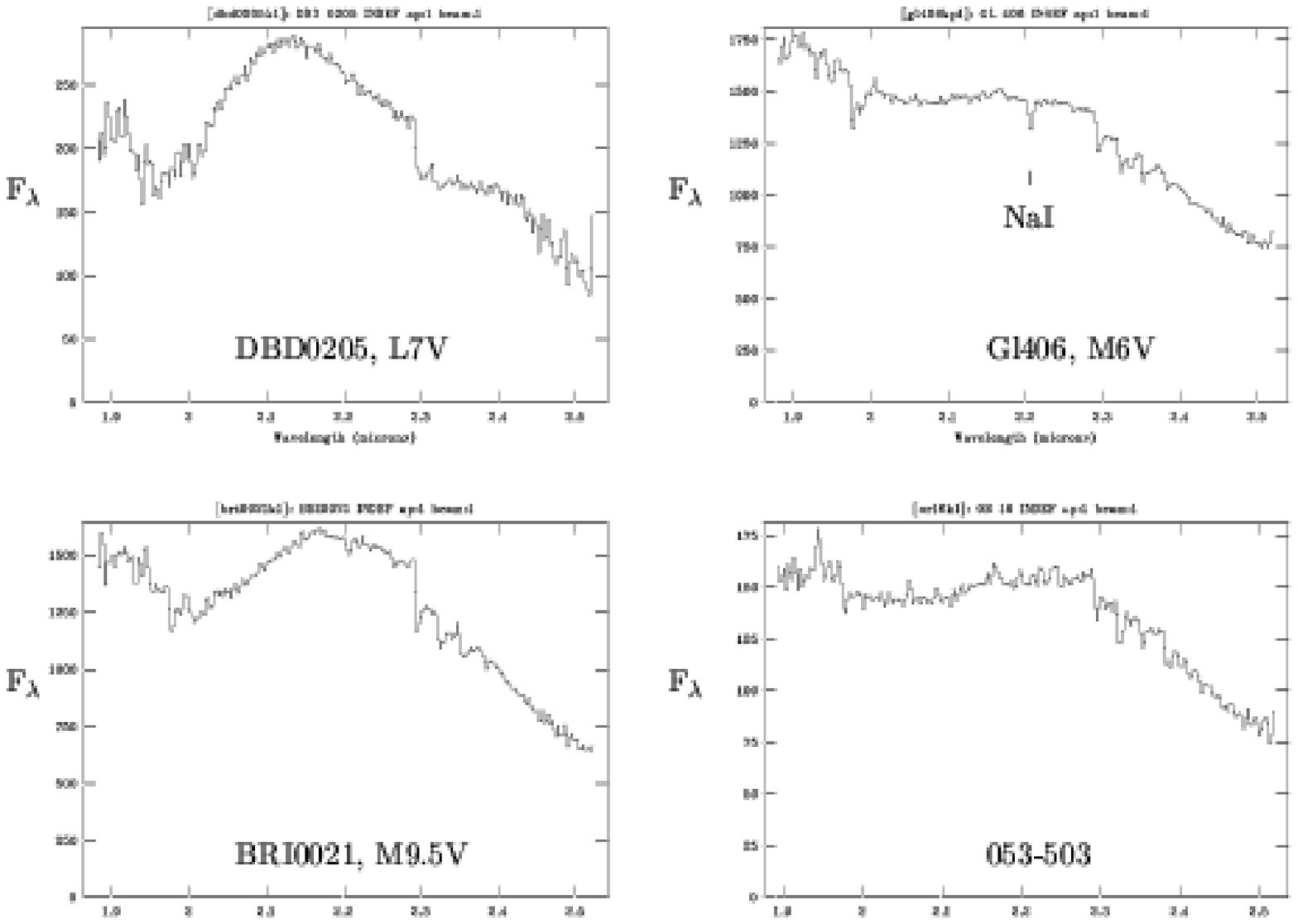,clip=,height=11cm,bbllx=2cm,bblly=9.3cm,bburx=20cm,bbury=20.3cm}
\vspace{1.5cm}
Figure 16: Field Dwarf spectra and 053-503. Late M and L type spectra
of high gravity field dwarfs show very strong CO. 053-503 has similar CO to
the M6 field dwarf but shows no NaI and stronger water absorption.
\end{figure*}

\subsection{Comparison with $\sigma$-Orionis Spectra and IC348 data}

	We note that the $H$ and $K$ bandpass spectra of 2 very low mass candidates in the 
$\sigma$-Orionis cluster (estimated age $\sim$5~Myr), plotted as low signal to noise 
smoothed spectra by Zapatero-Osorio et al.(2000), do not resemble those presented 
in Figure 4(a-c). At $H$ they show the plateau seen in field dwarfs and they 
also appear to be redder in the ratio of $K:H$ bandpass fluxes than all of the 
sources plotted in Figure 8(a-b) with strong water absorption, except 014-413,
which appeared to have an infrared excess due to hot circumstellar dust. After converting 
the spectra to F$_{\nu}$ the apparent ratios of flux near the bandpass peaks are 
$K:H \approx$ 1.6 and 1.2 for S Ori 47 and S Ori 60 respectively.
The presence of the plateau feature might be taken as evidence that these 2 
$\sigma$-Orionis objects are field L dwarfs, given that Zapatero-Osorio et al. expect 
$\sim$33\% contamination by these in their much larger survey area. However, there is 
separate evidence for the relative youth of 1 of the 2, S Ori 47, via low gravity 
features in its optical spectrum (Zapatero-Osorio et al. 1999) similar to those seen in 
Pleiades dwarfs. It would have to be an unusually young ($\sim$100 Myr) field L dwarf to 
have those optical features. In our view it is more likely that the $H$ bandpass 
spectra and larger $K:H$ ratios reflect the greater age of the $\sigma$-Orionis cluster 
than the Trapezium cluster. At 5~Myr, sources at or below the deuterium burning 
threshold are predicted to be 300~K or 500~K cooler than at 1~Myr (Baraffe et al;BM97
calculations respectively), corresponding to an 
effective temperature (T$_{eff}$) of 1600-1800~K for an object with mass 8~M$_{Jup}$ and 
$\sim 2200$~K at 15~M$_{Jup}$ according to both calculations. This is substantially 
cooler than the Trapezium sources and similar 
to the temperature of L type field dwarfs which exhibit the plateau feature. This is  
well into the T$_{eff} < 2500$~K region in which dust in the photosphere starts to have a 
large influence upon the emergent spectrum, according to the AMES-Dusty-1999 models. The 
larger ratio of $K:H$ band flux is also predicted for a lower temperature, being 
caused by the underlying Planckian shift of flux to longer wavelengths rather than
circumstellar dust. There is a some indication of a plateau feature in the AMES-Dusty
models at 2200~K in the 1.64-1.68~$\mu$m region (not shown) but it is much less broad 
than that observed by (Zapatero-Osorio et al.(2000) and weakens at lower temperatures. 
This might be due either to difficulties with the treatment of water or dust in 
the model or differences in the stratification of dust at this very early evolutionary 
stage. We emphasise that our $H$ and $K$ bandpass spectra of 3 field dwarf standards in 
Figure 5 and Figure 16 are in excellent agreement with the prior observations of Leggett 
et al.(2001), Jones et al.(1994), Tokunaga \& Kobayashi and Delfosse et al. so we have no 
reason to doubt the reliability of the Trapezium spectra. Scattering in the circumstellar 
environment might affect the spectral profiles in subtle ways but would not erase features 
at specific wavelengths such as the $H$ bandpass plateau.
	
	Najita, Tiede \& Carr (2000) have discovered 6 sources at or below the deuterium
threshold in the very young cluster IC348 (typical age 3~Myr), extending the work of Luhman 
et al. (1998a). Although they did not have spectra, they were able to obtain approximate 
spectral types via narrow band imaging across the 1.9~$\mu$m water band with NICMOS on the 
Hubble Space Telescope. They came to very similar conclusions to those presented here: the 
water absorption strength implied late M or L spectral types and they also argue that their 
sources lie above the main sequence and are therefore not background stars. This foreshadows
our argument in the next section. An interesting difference is that according to the 
AMES-Dusty models their 1.9~$\mu$m water index is stronger in low gravity pre-main sequence 
objects than in high gravity field stars of the same temperature. This apparent contrast
with our earlier statement is due to the different broad band profiles of low and high 
gravity objects. The behaviour of water absorption strength as a function of gravity
depends on the wavelength at which the water absorption is measured. We will have to wait
for the NICMOS Cooling System to be installed before measurements such as those of Najita 
et al. can be taken elsewhere, since telluric water absorption generally precludes good
quality ground based observation at these wavelengths.

\section{Cluster Membership}

	The reality of the population below 0.02~M$_{\odot}$ has been
called into question by Hillenbrand \& Carpenter (2000), who suggest
that lightly reddened background field dwarfs are present in the 
data, and are not distinguished from the proposed planetary mass 
population. They estimated that the total contamination
by non-cluster members is 5\% to K=18, using a modified Wainscoat
model (Wainscoat et al. 1992) of the entire galactic stellar population. 
This type of calculation
is difficult, owing to substantial uncetainties in: (a) 
the infrared extinction on lines of sight through the cluster; (b) the scale
height of the galactic M and L dwarf populations. This is relevant
because the Trapezium lies $\approx 122$ pc above the plane, relative
to the sun, which is more than 1 scale height above the Young Disk 
population but probably not the Old Disk.

	To address this issue we first look at the spectroscopic data
to see whether they are consistent with a field dwarf population.
Five dereddened $H$ bandpass spectra of sources with 1 Myr masses
M$\le 0.016$~M$_{\odot}$ and $0 < A(V) < 2.3$ are shown in Figure 4(c). Although 
the signal to noise ratio (SNR) in these spectra is low, it is clear that all 
possess deep water absorption on either side of the peak near 1.68~$\mu$m,
which would correspond to L-type spectra if they were main sequence
dwarfs (see Table 3). The spectral types are uncertain by a few subtypes due to the 
low SNR and the possibility of imperfect background subtraction but they are derived
by a reddening independent method, so extinction does not affect the typing.
Alternative water-based derivations of spectral type could lead to earlier 
classifications, e.g. the Delfosse slope-based water index (see Section 4) which samples 
the steepest part of the water absorption in field dwarfs and not the plateau. 
However, we are confident that the $H$ bandpass spectra of the very low mass 
candidates are of type M7 or later by any logical definition.
For example, spectra of M0-M6 dwarfs do not decline significantly in flux/pixel 
(i.e. F$_{\lambda}$) on the short wavelength side of the $H$ bandpass, going from 
1.68-1.5~$\mu$m (see eg. Leggett et al.2000, 1996), whereas the very low mass candidates 
all do, as illustrated in Figure 17 for the candidates with weakest water absorption. 
Early M-type spectra rise from 1.7 to 1.25~$\mu$m, since water absorption is weak at 
these types in Disk population dwarfs and even weaker in Halo extreme subdwarfs. 

On the CIT photometric system an M7V dwarf has an absolute 
magnitude M$_{H}=10.22$,  according to a fit to sources with known parallaxes
by Leggett (1992); or M$_{H}=10.26$, averaging 2 sources from Leggett et al.(2001) with
measured parallaxes in the Yale Catalogue (van Altena et al. 1995).
Adopting the brighter value, this leads to m$_{H}=18.44$ for a source just behind
OMC-1 at 440~pc, or m$_{H}=18.57$ on the UFTI magnitude system.
This is more than a magnitude less luminous than the 3 very low mass brown dwarfs in 
Figure 4(c) and 0.6 to 0.9 magnitudes less luminous than the 2 planetary mass candidates
(with H$_{dr}$=17.71, 17.95 on the UFTI system). Uncertainties in the dereddened
$H$ band fluxes for these 5 sources are expected to be $\approx$0.2 magnitudes via the 
random photometric error and 0.15-0.2 magnitudes possible systematic error due to the
uncertainty in the \it{(J-H)} \rm colour of the isochrone. 
Later type dwarfs decline rapidly in brightness,
eg. M$_{H}=10.90$ at M9, and 10.98 at L0 on the CIT system, as fitted by Kirkpatrick
et al.(2000) using sources with measured parallaxes. Hence, our very low mass sources 
are almost certainly too luminous to be background main sequence dwarfs, even assuming 
zero extinction by the OMC-1 cloud. In fact, the backdrop of the OMC-1 cloud provides 
substantial extinction in the region of the spectroscopic sample, as discussed below, which
further reinforces this result. The luminosities are higher because they are physically 
much larger and have higher effective temperatures than field dwarfs with equivalent 
water absorption depths.

	The profiles shown in Figure 4(c) also resemble those of
higher mass Trapezium objects rather than that of a late M or L type field dwarf, 
illustrated in Figure 17 with noise added to aid the comparison. The 
field dwarf has a plateau from 1.57-1.70~$\mu$m, whereas the very low mass candidates
appear to peak near 1.68~$\mu$m like their brighter counterparts.

\begin{figure}
\begin{center}
\begin{picture}(200,430)

\small
\put(-10,385){F$_{\lambda}$}
\put(-10,230){F$_{\lambda}$}
\put(-10,100){F$_{\lambda}$}
\put(65,335){061-401}

\put(65,190){038-627}

\put(65,45){Field M9.5}

\put(0,0){\includegraphics{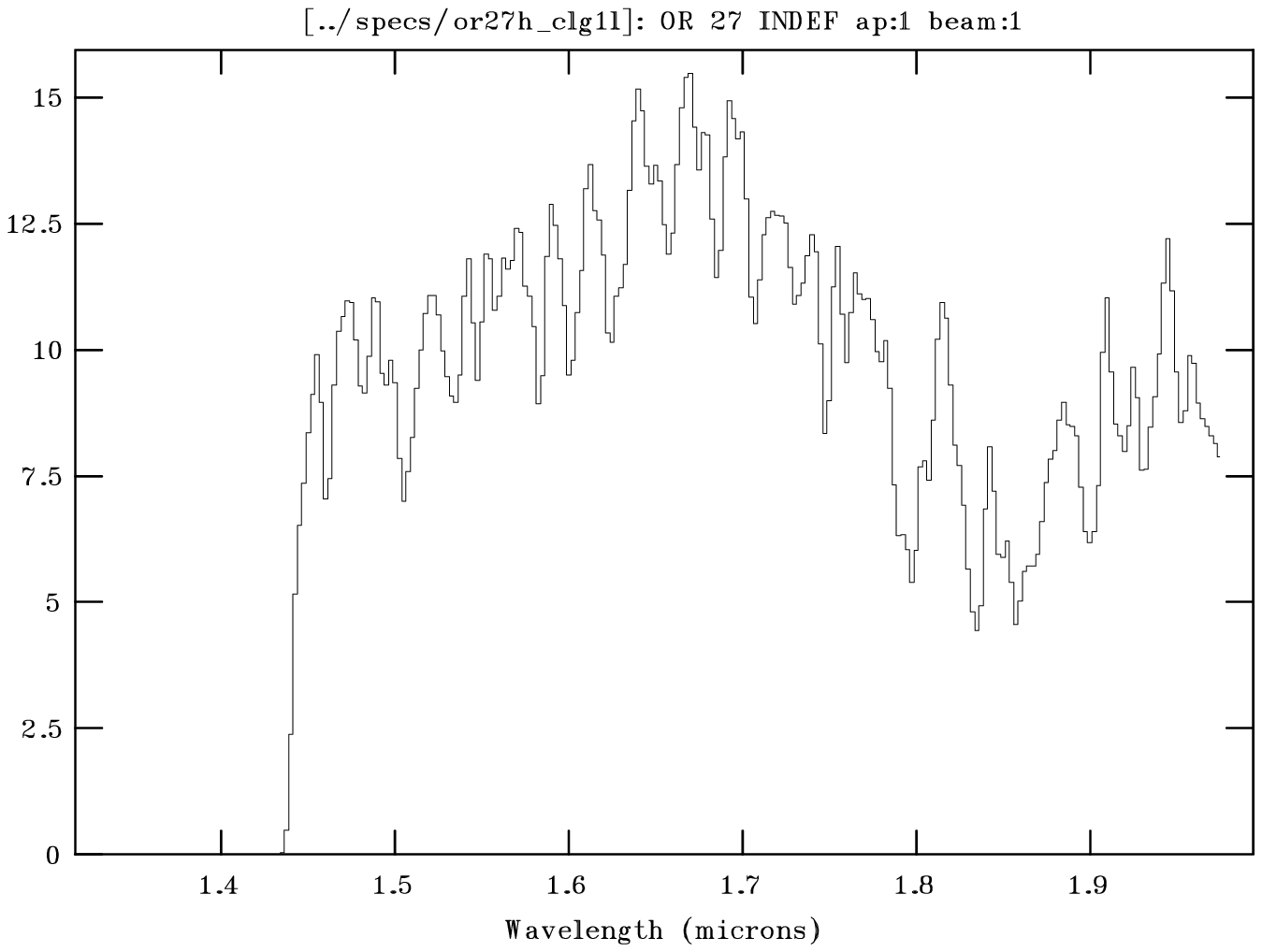}}

\put(0,0){\includegraphics{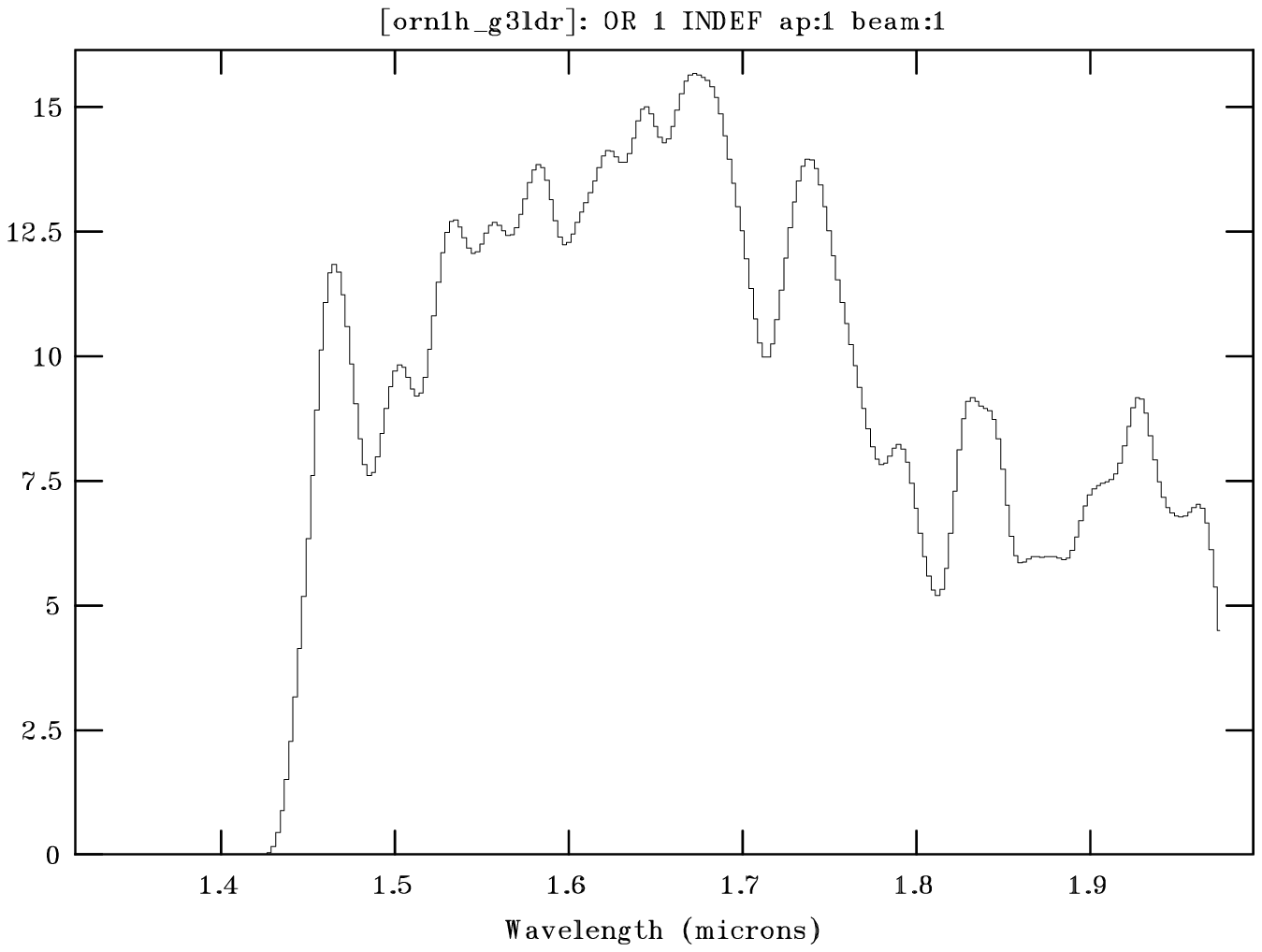}}

\put(0,0){\includegraphics{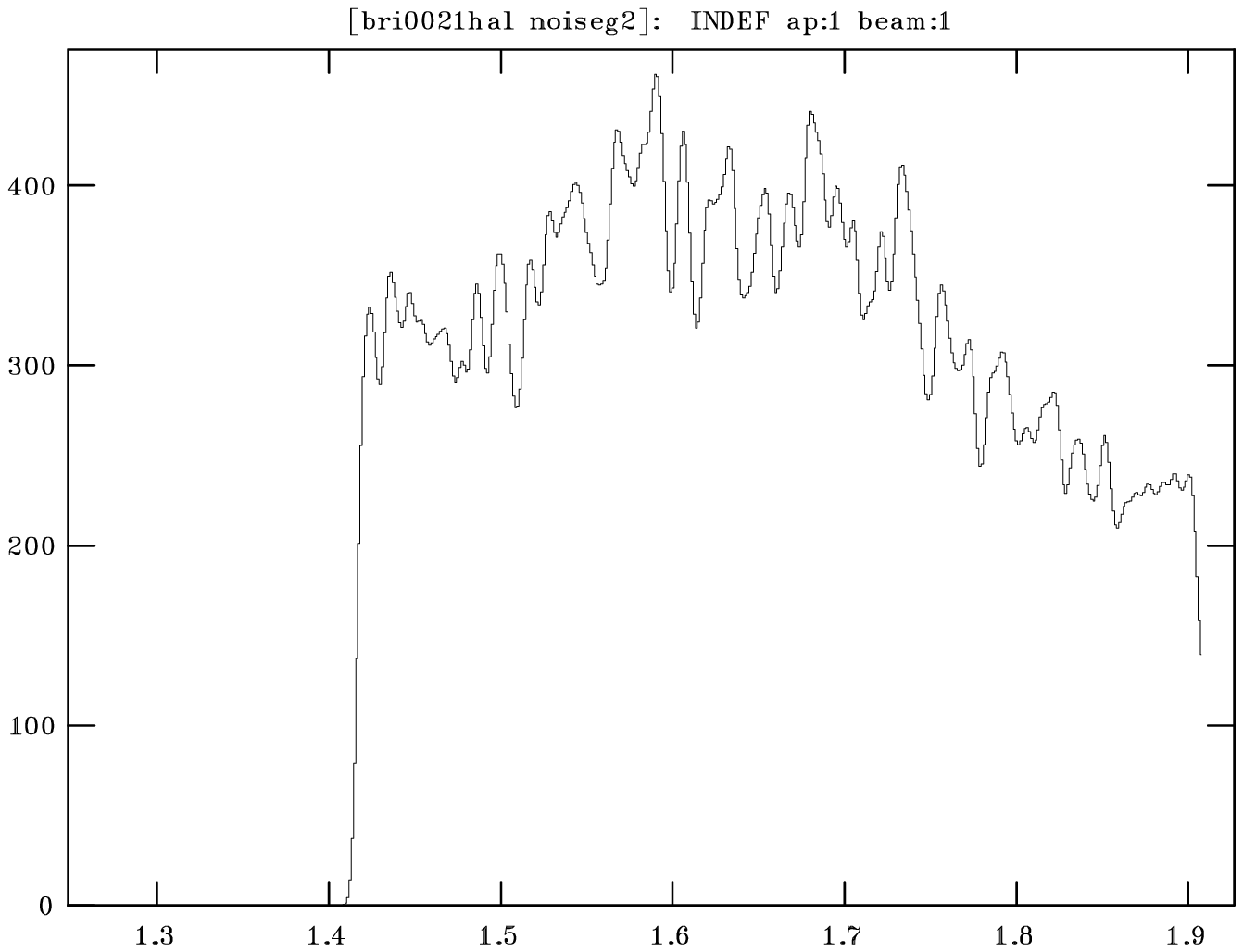}}

\end{picture} 
\end{center}
\vspace{-4mm}
Figure 17: F$_{\lambda}$ spectra of 061-401 (0.011~M$_{\odot}$) and 
038-627 (0.014~M$_{\odot}$) compared with field dwarf BRI0021-0214.
The field dwarf spectrum is given added noise and smoothed for comparison.
\end{figure}

	If background contamination were significant we would expect a 
range of lightly reddened M0-M6 dwarfs to appear in our sample, since these are the 
commonest stars which are luminous enough to be detected at d$\ge 400$pc, assuming very 
low extinction by OMC-1 in the outer regions of our survey. The earlier
M-types are less common than the later types but would be detectable
to a greater distance, so the distribution would be fairly even across spectral
types M0-M6. Since they are not seen among our planetary mass candidates
we suspect that HC have underestimated the extinction
for 2 reasons.  The first reason is that the extinctions derived from our \it{(J-H)} \rm 
colours show that the A(V) column extends beyond the spatial pattern indicated by C$^{18}$O 
maps, which are only sensitive to high density gas. High extinction (up to A(V)=14)
is measured toward many sources at the edges of our survey (see Figure 18), 
which we attribute mainly to lower density molecular gas and also to the neutral 
cluster medium (O'Dell \& Yusef-Zadeh 2000). The near infrared extinction measurements 
are direct, whereas values derived from radio measurements are always uncertain by 
a factor of a few. Nevertheless, there is radio waveband confirmation of the extinction. 
The $^{13}$CO and CS(1-0) maps of Tatematsu et al.(1993) sample the less dense molecular gas 
and show that the OMC-1 cloud has a low density wing extending far to the west of the 
main ridge in the cluster centre, beyond the location of our sources. The
column density of the wing corresponds to A(V)$\approx 4$ in the outer parts of the 
wing (using the transformations of Lada et al. 1994) for $^{13}$CO and may be higher in 
the inner part where our sources lie: this region is not well resolved spatially from the 
very high density ridge in the map of Tatematsu et al. In addition, the cold dust is seen 
directly in the 850~$\mu$m SCUBA data of Johnstone \& Bally (1999), who measure continuum 
emission of 0.5-2.0 Jy/beam in the spectroscopic survey region. Applying their adopted 
emissivity index ($\beta$=2), temperature (20~K) and 100:1 ratio of gas:dust, we calculate 
2-8$\times10^{29}$~Kg per beam (14 arcsec diameter). Adopting the relation 
A(V)/N(H)=5.3$\times 10^{-26}$~m$^{2}$ leads to $4.8 < A(V) < 19$ in the survey region, where
the very low mass sources are detected. The uncertainties in grain emissivity, temperature and
A(V)/N(H) mean that this is only an order of magnitude estimate (eg. Ward-Thompson et al.1994)
but it is consistent with the near infrared and $^{13}$CO data. The second reason to
expect significant extinction is that HC used an optical:infrared extinction conversion 
based on the interstellar medium, which significantly underestimates the infrared 
extinction in Orion (Cardelli et al.) because of the presence of slightly larger dust 
grains in the environment of the molecular cloud. The substantial contribution of these 
slightly larger grains to extinction can be seen from calculations using Mie theory (eg. 
Figure 15 of Lucas \& Roche 1998). In some very young sources the measured 
infrared extinction may be local to the young star. However, comparison of Figure 18 and
Figure 3 does not show any less extinction in the outer region of our survey than
in the full survey sample, which indicates that there is strong extinction 
by the low density molecular gas throughout the survey region. Moreover, the
$L$ bandpass data of Lada et al.(2000) show that very young, embedded protostars 
are strongly concentrated towards the high density ridge, so the case for 
significant extinction by lower density gas in the outer cluster is clear.
It therefore appears that any background stars of type M7 or later would appear more 
than a magnitude fainter than all the low mass candidates in Figure 4(c) and even background
binaries would not be bright enough to cause confusion.

\begin{figure}
\begin{center}
\begin{picture}(200,290)

\put(0,0){\includegraphics{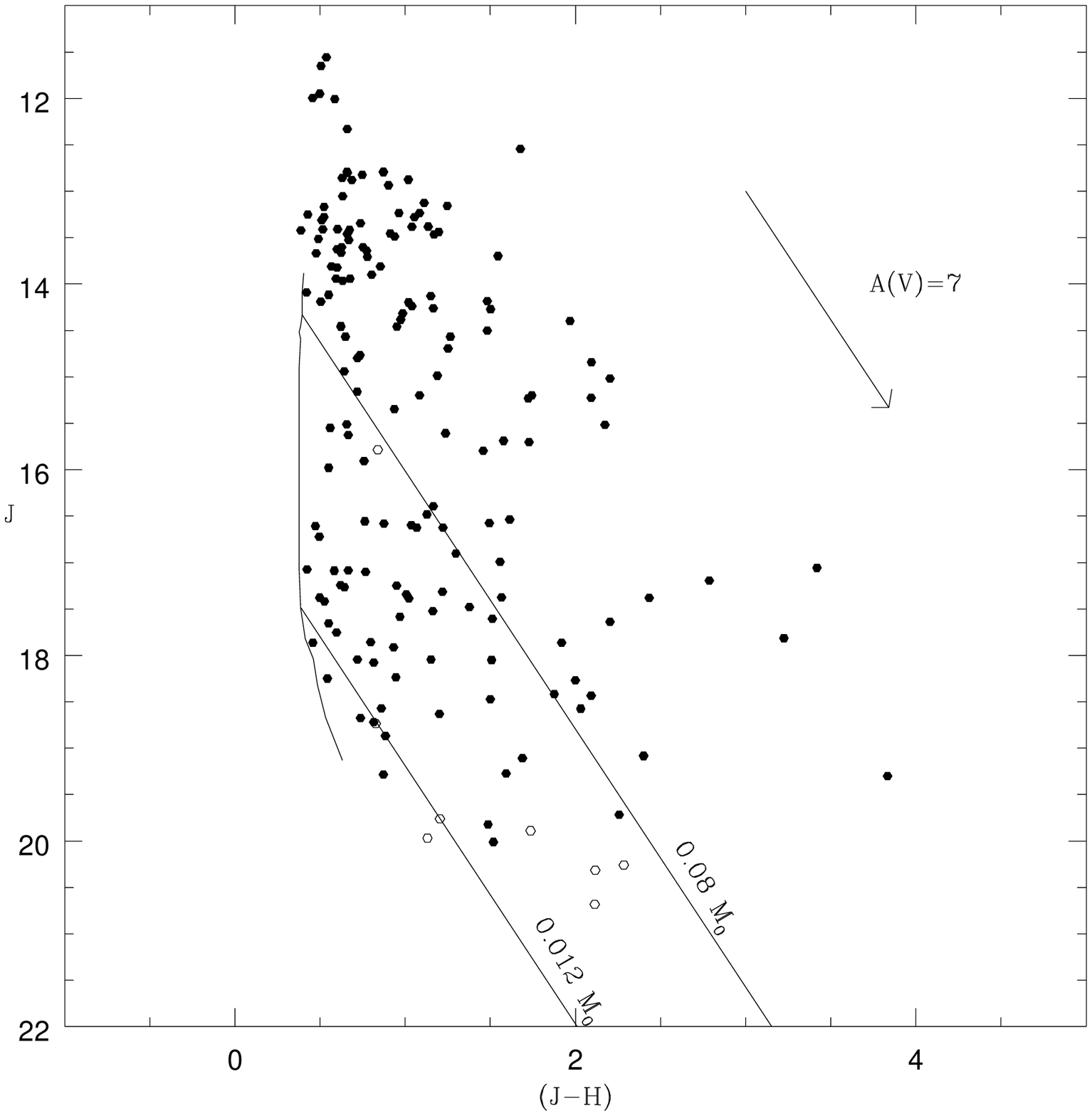}}

\end{picture} 
\end{center}
\vspace{-2cm}
Figure 18: Colour magnitude plot for the region of the spectroscopic sample.
High reddening is observed toward many many sources, similar to Figure 3. 
The new Baraffe et al. isochrone is plotted at left.
\end{figure}

Our assumption in Paper I that the OMC-1 cloud prevented any background contamination
in the survey region at $J$ band may prove to be an overstatement however. 
In Section 3.2.1, we discussed the 2 candidates near the stellar/substellar boundary
(019-108, 066-433) in our spectroscopic sample which suffer 
high extinction (A(V)=8.6, 12.6 respectively) and have weak water absorption, consistent 
with late K-type or very early M-type spectra. If they are embedded YSOs then we
can say that there is no background contamination in our sample, which is
highly encouraging for future work in this cluster. It is very possible that 
they are background field stars, since their fluxes are consistent with that 
interpretation and the extinction is broadly consistent with what we
might expect from the radio data. This would also be encouraging, since 
the high extinction towards these 2 sources would indicate that background 
contamination can be avoided by studying sources which have low extinction. 

	In Paper I we had assumed that lightly reddened background
sources are not present because of the large near infrared extinction of cluster
members and because the cluster profile at infrared
wavelengths is apparently identical to that at visible wavelengths. A large 
population of field sources would be expected to produce a broader
profile at infrared wavelengths, since these sources are uniformly
distributed and more easily seen at the edges of the cluster, where
extinction is lower. This argument was first used by Hillenbrand \&
Hartmann (1998) but not at such faint sensitivity limits. We apply it here 
using our $I$ and $H$ bandpass data, in Figure 19, plotting the number of 
detections in each filter as a function of distance from $\theta_{1}$~Ori~C.  
Inspection of the profile gives no hint of a background population
since the ratio of $H$:$I$ bandpass detections is fairly uniform and may even 
decrease slightly between radii of 200 and 300 arcsec. 
This demonstrates that any population of faint sources ($H \gtsimeq 17$ to be 
undetected at $I$) must be small in order to have no observable 
impact on the cluster profile. This method is not precise enough to 
exclude the presence of a small background population and there are 
selection effects in Figure 19 because the detectabilty of sources 
increases more rapidly at large radii in the $I$ bandpass than at
$H$, owing to the greater effect of the nebular background at $I$. 
Nevertheless, the absence of any obvious background population in the profile
at $H$ suggests that contamination is likely to be small at $J$.

\begin{figure}
\begin{center}
\begin{picture}(200,290)

\put(0,0){\includegraphics{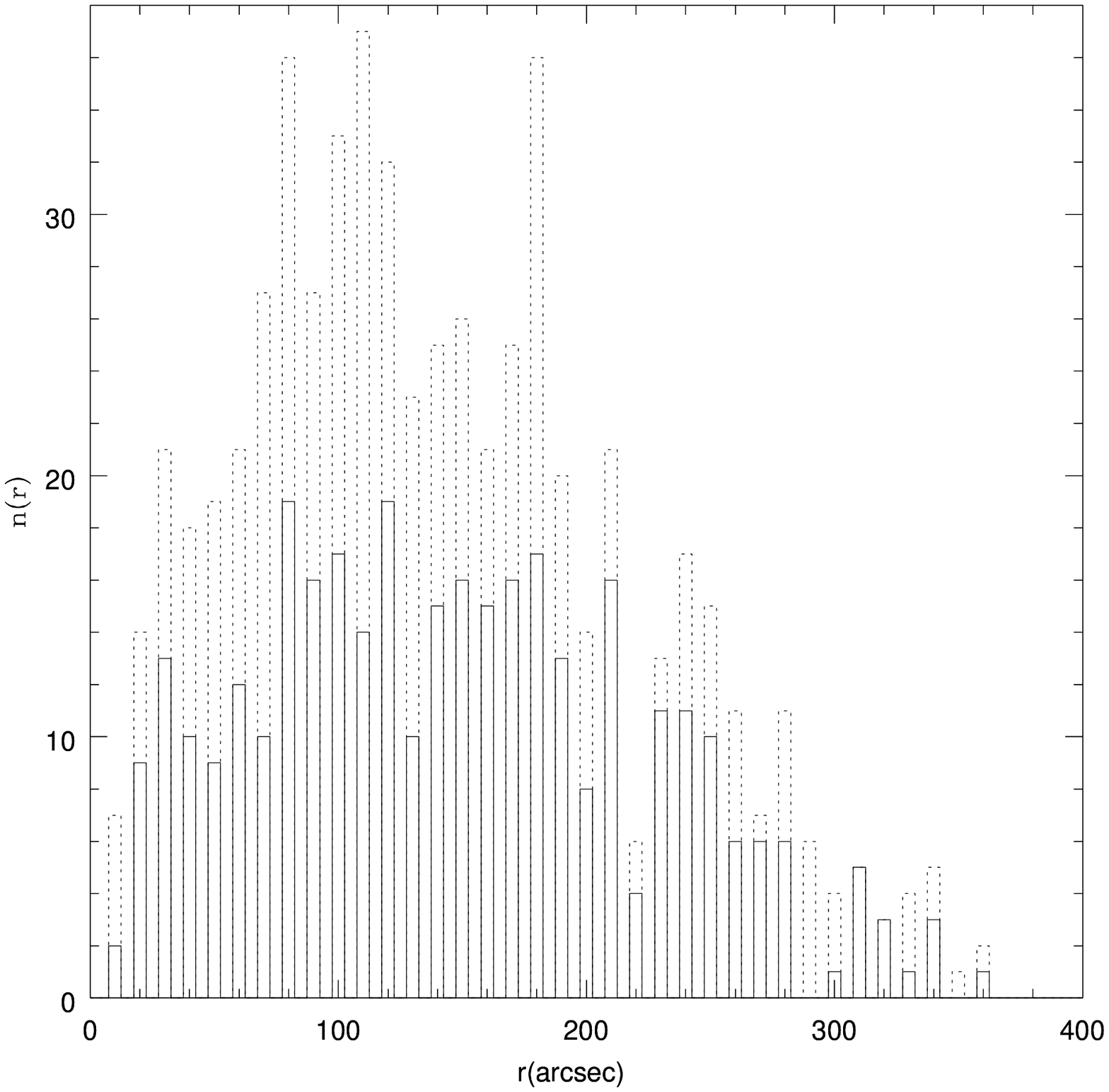}}

\end{picture} 
\end{center}
\vspace{-2cm}
Figure 19: Trapezium Cluster profile n(r)dr as a function of distance
from $\theta_{1}$ Ori C. Solid lines show $I$ band detections, dashed lines
show $H$ band detections. The profile is incomplete beyond 160 arcsec
because the survey region is not circular, but extends to the west.
\end{figure}

	Could the planetary mass objects be foreground dwarfs? About
half of the 15 candidates have sufficient reddening, \it{(J-H)} \rm $>$0.85
(UFTI system) or \it{(J-H)} \rm $>$0.94 on the CIT system, for us to be confident
that they are significantly reddened. Only field dwarfs of type L2 and later can 
reach these colours according to Kirkpatrick et al.(2000) (or L6 and later according 
to the smaller sample of Leggett et al. 2001). These provide very small contamination, 
since mid and late L-types are only detected at small distances, sampling a much smaller 
volume than late M-types m$_{J}=20.0$). Kirkpatrick et al.(2000) 
The expected number of foreground objects is only 5 or 6 in the full survey region 
($\sim 1\%$ contamination) according to both the Tinney et al.(1993) luminosity function and 
the 2MASS data (Reid et al.1999), mainly M dwarfs in magnitude range $15<J<19$. We expect to 
detect $\le 1$ foreground L dwarf in the full survey region of 40~arcmin$^{2}$, using the 
space density estimate of Zapatero-Osorio et al.(2000). Although the volume density of field 
L dwarfs is uncertain, the number cannot rise by more than a factor of 3, since 
Zapatero-Osorio et al. calculated 33\% contamination in their much larger survey area at 
very similar galactic coordinates (contamination cannot exceed 100\%). Planetary mass 
candidates ($\le 0.012$~M$_{\odot}$) are detected in no more than 60\% or our survey area, 
where the nebular surface brightness is sufficiently low. Hence, we derive an upper limit of 
1 or 2 late M and L type sources which could appear to be planetary mass objects.

\section{Age and Mass of Sources}

	The interpretation of very faint cluster members as planetary mass
objects might be incorrect if they are far older ($\ge 20~$Myr) than the
bulk of the higher mass cluster members. The luminosity of a 0.020~M$_{\odot}$
object does not decline to that of the 6 candidates with 
M$\ltsimeq 0.008$~M$_{\odot}$ until an age of 27~Myr, according to the 
BM97 calculations. The age distribution of the population has been 
studied in detail by Hillenbrand 
(1997) over the wider Orion Nebula Cluster, within a radius of 20 arcmin,
using 2 evolutionary tracks in the HR diagram. This analysis concluded that
most star formation in Orion has occurred within the past 2~Myr and that the 
youngest sources are highly concentrated towards the cluster centre, which is the 
region covered by our relatively small scale survey. The dataset did not include 
planetary mass sources but indicated that the lowest mass and highest mass sources
are most strongly concentrated toward the cluster centre, with less concentration by 
intermediate (solar mass) sources. Specifically, Hillenbrand calculated that 
$<2\%$ of the population has an age $>$20~Myr, using the D'Antona \& 
Mazzitelli (1994) tracks, which agree well with more recent calculations 
(BM97; D'Antona \& Mazzitelli 1998, Baraffe et al.). The Swenson et al.(1994)
tracks, which indicated lower temperatures for brown dwarfs than more recent 
calculations, put the proportion at $\sim 5\%$, which gives some indication
of the uncertainty in this type of calculation. Conservatively, we assume
that the old ($>20$~Myr) population is only a factor of 3 less concentrated within
our survey region (radius $\approx $4 arcmin) than the young population 
(see Figure 22 of Hillenbrand 1997). This yields a contamination of 
0.6\%-1.7\% by such old objects in our sample, or 4 to 11 objects spread over
all masses in the sample of $\sim 650$ sources detected at both $J$ and $H$
(including bright saturated sources). The more recent assessment of the age
distribution in the core of the cluster by HC gave a similar result. 
Expressed as N$(log(M))$, the IMF in Orion has been demonstrated to have
a broad maximum somewhere between 0.1 and 0.6~M$_{\odot}$ (Paper I, HC, Luhman et 
al.2000) and decline slowly at substellar masses. In the 0.03-0.08~M$_{\odot}$
interval a relation N$(log(M)) \propto M^{-1}$ is consistent with our data in 
Paper I. If this trend continues below our completeness limit to 0.013~M$_{\odot}$,
then we would expect there to be $\sim 0.16$ to 0.44 old brown dwarfs with sufficiently 
low mass ($\sim 0.013$ to $0.03$~M~${_\odot}$) to masquerade as a planetary mass object.
Even if the substellar IMF were flat we would not expect there to be more than 
1 such source, whereas there are $\sim 15$ planetary mass candidates. In the event of a 
flat substellar IMF it would be difficult to argue for an abrupt cut off above the 
deuterium threshold.

	We note that the status of 1 brown dwarf candidate, 014-413 as a youthful
Trapezium cluster member has been confirmed by direct imaging of a wind bow
shock interaction with the cluster medium in H$\alpha$ in a Hubble
Space Telescope survey (Bally, O'Dell \& McCaughrean 2000) which unfortunately
overlaps only a small part of the spectroscopic survey region. Outflow activity
is correlated with accretion rate in YSOs, which decreases with age. Hence, this is 
probably a younger than average cluster member, which is also indicated by the
presence of a large $K$ bandpass excess in Figure 8(a).

	$K$ bandpass excesses are a good indicator of youth, since they are most common
at ages $\le 1$Myr. Is there evidence for $K$ bandpass excesses in our planetary mass 
candidates ? 6/15 of our planetary mass candidates were detected in the
$H$ and $K$ bandpass data of HC, which only partially overlaps our survey
area. These can be dereddened with the addition of our $J$ bandpass data. The $H$ bandpass 
fluxes measured by HC agree well with ours in 3/6 cases,
the measurements differing by $\le 0.13$ magnitudes, see Table 5.
We note that the 3/6 sources with poor agreement all lie near the edge of 
the survey of HC, and are believed to suffer from substantially shorter
exposure times (25~s or 50~s total), whereas the other 3 are believed to have the full 
100~s exposure time. The poor agreement is therefore probably due to low SNRs in the HC 
data, given that all these faint sources are near the sensitivity limits of both surveys. 
Alternatively, large amplitude variability may occur in a small minority of sources, which 
has been demonstrated at stellar mass by Carpenter, Hillenbrand \& Skrutskie (2001). If we 
accept that the 3 sources with good 
agreement have well measured $H$ bandpass fluxes, then there appear to be large $K$ band 
excesses in 2/3 cases. It is risky to compare photometry from different surveys in this 
way but these colour excesses, E(H-K)$\approx 0.6$ magnitudes, are large enough to be fairly 
convincing despite the uncertainty of order 0.1 mag in the \it{(J-H)} \rm colour of the 
photospheres. If they are genuine then this indicates that these are very young sources, 
probably with ages of $<$1~Myr, and therefore are very good planetary mass candidates. 

	Oasa et al.(1999) were cautious in interpreting very faint sources 
in the Chamaeleon I cloud as planetary mass objects, despite the detection of 
large $K$ bandpass infrared excesses requiring the presence of a substantial mass of hot 
dust in the accretion disk. However they found that even at 10~Myr, their presumed 
upper age limit for sources with substantial quantities of hot dust, a planetary mass 
was still indicated for their faintest source.

	At present, there is no easy way to actually measure the ages of individual 
sources because luminosity and temperature are degenerate with respect to mass
and age, and because of the need to deredden sources. In principle the tracks of more
massive sources might be separated on an HR diagram (older, more massive sources 
should be smaller and hotter than planetary mass objects of equal luminosity). At 
present, however the uncertainty in temperature measurement and prediction at these very 
early ages is too large to make such a distinction (see Table 3). 
A small bias toward greater ages might be expected for the spectroscopic sample
of sources with low extinction (see Section 4.3).
Precise measurements of surface gravity via high resolution spectroscopy of 
absorption line profiles may permit this at some future date but this will be 
difficult owing to the weakness of atomic lines in low gravity atmospheres.
A promising alternative would be to use well modelled lines of simple molecules
such as CO, and perhaps the K$_{CO}$ index, because some molecular features 
increase in strength with decreasing gravity. Another option might be to search 
for deuterium, the defining attribute of these 'free floating planets'. Deuterium is 
predicted to be absent in sources more $10^{7}$ years old, for masses 
M$>0.03$M$_{\odot}$. This is illustrated by Bejar et al., using the calculations of 
D'Antona \& Mazzitelli (1997), and examined in more detail by Chabrier, Baraffe, 
Allard, \& Hauschildt (2000).

\section{Gemini Data}

	We have conducted a pencil beam search for sources in the Trapezium
even fainter than those deetected in our UKIRT survey, using the Gemini North
Telescope with the Adaptive Optics system Hokupa'a and the infrared camera QUIRC. 
The data will be described more fully in a future paper but preliminary indications
are that no such ultra low luminosity sources were detected in 5 QUIRC fields
containing $\sim 25$ stars and brown dwarfs. This tends to support our 
previous suggestion thet the Luminosity Function (LF) in the Trapezium turns down
somewhere near 8~M$_{Jup}$, but this is far from conclusive because of the very
small area surveyed and the existence of small scale spatial variation 
in the IMF, as seen in Figure 2). The decline of the LF should be investigated 
further to see whether it is a feature of the IMF or merely of the mass-luminosity 
relation. 
(An attempt at a much larger companion survey with the NAOMI Adaptive Optics system at 
the William Herschel Telescope was prevented by severe weather.) 
There are theoretical arguments for expecting few sources below 10~M$_{Jup}$
(Silk 1977; Adams \& Fatuzzo 1996) but the physical processes in crowded clusters and 
multiple star systems may circumvent the obstacles described in those papers.

\section{Conclusions}

	Low resolution $H$ and $K$ bandpass spectra of brown dwarfs and
planetary mass candidates show strong water absorption bands, equivalent
in depth to those seen in L-type and late M-type spectra, which is 
entirely consistent with the
predictions of evolutionary tracks (Baraffe et al., BM97, DM98) at ages near 1~Myr.
The new evolutionary tracks provide self consistent flux predictions, permitting
luminosities and masses to to be derived without the need for bolometric corrections. 
The results are very similar to our previous work.
However, the $H$ bandpass profiles shown here are very different from those
of field dwarfs with strong water absorption, which appears to be a
clear signature of youth and low mass. Strong evidence for the 
expected low surface gravities is provided by fitting the spectra
to the AMES-Dusty-1999 models. The low gravity models are quite successful
at reproducing (a) the $H$ bandpass profiles; (b) the weakness of CO 
absorption in the $K$ bandpass spectra; and (c) the absence of Na I 
absorption at 2.21~$\mu$m, though with stringent limits in only one source. 
There is an excellent correlation between independently derived spectroscopic 
temperatures and photometric temperatures, with a relatively small systematic offset
of 250~K.

	Spectra of 2 planetary mass candidates and 3 very low mass brown dwarfs
($\mathrm{M \le 0.016~M_{\odot}}$ at 1~Myr) show very strong water
absorption consistent with L-type spectra, and have lower spectroscopic temperatures
than the higher mass brown dwarfs, which is also predicted by the evolutionary 
tracks. The late spectral types confirm that these are cluster members, since they 
are too luminous to be background field dwarfs and the modest extinction seen in most
cases demonstrates that they are not foreground dwarfs. We would expect many of the 
L-type sources to show M-type spectra at optical wavelengths, since water absorption
is deeper in low gravity dwarfs than high gravity dwarfs, according to the AMES-Dusty-1999
spectra. The fitted temperatures are higher than those usually associated with
L-type sources. The trend of CO absorption to be significantly weaker at low gravities, 
for effective temperatures in the range 2100-2900~K, is a potentially powerful tool for 
identifying very young brown dwarfs, provided there is an independent means of
measuring the effective temperature. This trend has been observed before in low gravity
giant and supergiant stars (eg. Origlia, Moorwood \& Oliva) at both 
2.3~$\mu$m and 1.62~$\mu$m.

	While the ages of the
individual sources cannot be precisely measured, it is very unlikely
the 15 photometric planetary mass candidates could be much more massive sources
with ages $>20$~Myr, since statistically we would expect $<1$ such source in
our dataset. There is also some evidence for large $K$ bandpass excesses in
2 of the planetary mass candidates, obtained by combining our $JH$ data with the
$HK$ data of Hillenbrand \& Carpenter 2000; however this must be confirmed by a
single deep $JHK$ survey.

\section{Acknowledgements}

        We wish to thank the staff of UKIRT, which is operated by the Joint 
Astronomy Centre on behalf of the UK Particle Physics and Astronomy Research 
Council (PPARC). We are very grateful to Sandy Leggett for providing her
spectroscopic data of field dwarfs electonically, permitting a detailed 
comparison between these sources and the our substellar candidates without
the need to repeat all of the same observations. We are indebted to Isabelle
Baraffe for providing self consistent theoretical isochrones, without which
the interpretation of the photometry would be considerably less secure. We 
thank Hugh Jones, Antonio Chrysostomou and Suzanne Ramsay-Howat for helpful discussions 
and thank the referee for a useful report. We thank the UKIRT, Gemini and WHT Panels 
for the Allocation of Telescope Time for supporting this project and the University of 
Hawaii's Hokupa'a team and Gemini Observatory staff for carrying out Gemini observations 
for us. PWL is supported by PPARC via a Post Doctoral Fellowship at the University of 
Hertfordshire.

\pagebreak
\onecolumn

\large{\textbf{References}}\textmd{}\normalsize
\setlength {\parskip} {2mm}
\setlength {\parindent} {0mm}

Adams F.C. \& Fatuzzo M. 1996, ApJ 464, 256\\ 
Allard F., Hauschilt P.H., Alexander D.R., Starrfield S. 1997, Ann.Rev.A\&A 35, 137\\
Allard F., Hauschildt P.H., \& Schweitzer A. 2000, ApJ 539, 366\\
Allard F., Hauschildt P.H., \& Schwenke D. 2000, ApJ 540, 1005\\
Allard F., Hauschildt P.H., \& Alexander D.R., Tamanai A., \&  Schweitzer, A. 2001, 
  ApJ (submitted)\\
Bally J., O'Dell C.R., \& McCaughrean M.J. 2000 AJ, 119, 2919\\ 
Baraffe I., Chabrier G., Allard F., \& Hauschildt P.H. 1998, A\&A 337,403 (BCAH)\\
Bejar V.J.S., Zapatero-Osorio M.R., \& Rebolo R. 1999, ApJ 521,671\\
Briceno C., Hartmann L., Stauffer J., \& Martin E. 1998, AJ 115, 2074\\ 
Burrows A., Marley M., Hubbard W.B., Lunine J.I.,
 Guillot T., Saumon D., Freedman R., Sudarsky D.,
 \& Sharp C. 1997, ApJ, 491,856 (BM97)\\
Cardelli J.A., Clayton G.C., \& Mathis J.S. 1989, ApJ, 345,245\\
Carpenter J.M., Hillenbrand L.A., \& Skrutskie M.F. 2001, AJ (in press), astro-ph 0102446\\ 
Chabrier G., Baraffe I., Allard F. \& Hauschildt P. 2000, ApJ 542, L119\\
Chen H., Bally J., O'Dell C.R., McCaughrean M.J., Thompson R.L., Rieke M., 
   Schneider G., \& Young E.T. 1998 ApJ 492, L173\\
Comeron F., Rieke G.H., Burrows A., Rieke M.J. 1993, ApJ 416, 185\\
Comeron F., Rieke G.H., \& Rieke M.J. 1996, ApJ 473, 294\\
D'Antona F., \& Mazzitelli I. 1994, ApJS 90, 467\\ 
D'Antona F., \& Mazzitelli I. 1997, MmSAI 68 ,607\\ 
Delfosse X., Tinney C.G., Forveille T., Epchtein N., Borsenberger J., Fouque P.,
   Kimeswenger S., \& Tiphene D. 1999, A\&A Supp. Series 135,41\\
Hauschildt P.H., Allard F., Baron E., Schweitzer A. 1999, ApJ 312. 377\\
Hillenbrand L.A. 1997, AJ 113,1733\\ 
Hillenbrand L.A., \& Hartmann L.W. 1998, ApJ 492, 540\\
Hillenbrand L.A. \& Carpenter J.M. 2000, ApJ 540, 236 (HC)\\ 
Johnstone D. \& Bally J. 1999, ApJ 510, L49\\ 
Jones H.R.A., Longmore A.J., Jameson R.F., \& Mountain C.M. 1994, MNRAS 267, 413\\
Kirkpatrick J.D., Reid I.N., Liebert J., Cutri R.M., Nelson B.,
   Beichman C.A., Dahn C.C., Monet D.G., Gizis J.E., \& Skrutskie M.F. 
   1999, ApJ 519, 802\\
Kirkpatrick J.D., Reid I.N., Liebert J., Gizis J.E., Burgasser A.J., Monet D.G.,
   Dahn C.C., Nelson B., \& Williams R.J. 2000, AJ 120, 447\\
Lada C.J., Muench A.A., Haisch K.E., Lada E.A., Alves J.F., 
   Tollestrup E.V. \& Willner S.P. 2000, AJ 120, 3162\\  
Lada C.J., Lada E.A., Clemens D.P., \& Bally B. 1994, ApJ 429, 694\\
Lancon A. \& Rocca-Volmerange B. 1992, A\&A Supp. Series 96, 593\\
Leggett S.K. 1992, ApJS 82, 351\\
Leggett S.K., Allard F., Berriman G., Dahn C.C., \& Hauschildt P.H. 1996, ApJS 104, 117\\
Leggett S.K., Allard F., Dahn C., Hauschildt P.H., Kerr T.H., \& Rayner J. 2000,
ApJ 535, 965\\ 
Leggett S.K., Allard F., Geballe T.R., Hauschildt P.H., \& Schweitzer A. 2001, 
ApJ 548, 908\\
Lucas P.W. \& Roche P.F. 1998, MNRAS 299, 699\\
Lucas P.W. \& Roche P.F. 2000, MNRAS 314, 858 (Paper I)\\
Luhman K.L., Liebert J. \&Rieke G.H. 1997, ApJ 489, L165\\ 
Luhman K.L., \& Rieke G.H. 1998, ApJ 497,354\\
Luhman K.L., Rieke G.H., Lada C.J., \& Lada E.J. 1998a, ApJ 508, 347\\
Luhman K.L., Briceno C., Rieke G.H. \& Hartmann, Lee 1998b, ApJ 493, 909\\
Luhman K.L., Rieke G.H., Young E.T., Cotera A.S., Chen H., Rieke M.J.,
   Schneider G. \& Thompson R.I. 2000 ApJ 540, 1016\\
Martin E.L., Rebolo R., \& Zapatero-Osorio 1996, ApJ 469, 706\\
McCaughrean M., Zinnecker H., Rayner J., \& Stauffer J. 1995, in ``The Bottom of 
the Main Sequence and Beyond'', ed. C. Tinney, p209, (Berlin: Springer)\\
Najita J.R, Tiede G.P., \& Carr J.S. 2000, ApJ 541, 977\\ 
Oasa Y., Tamura M., \& Sugitani K. 1999, ApJ 526, 336\\
O'Dell C.R., \& Wen Z. 1994, ApJ 436, 194\\
O'Dell C.R., \& Wong K. 1996, AJ 111, 846\\
O'Dell C.R., \& Yusef-Zadeh F 2000, AJ, 120, 382\\
Origlia L., Moorwood A.F.M., \& Oliva A\&A 1993, 280, 536\\
Partridge H.,\& Schwenke D.W. 1997, J. Chem. Phys. 106, 4618\\
Reid I.N, Kirkpatrick J.D. Liebert J., Burrows A., Gizis J.E., Burgasser A.,
  Dahn C.C., Monet D., Cutri R., Beichman C.A., Skrutskie M. 1999, ApJ 521, 613\\
Saumon D., Hubbard W.B., Burrows A., Guillot T., Lunine J.I., \& Chabrier G. 1996,
   ApJ 460, 993\\ 
Silk J. 1977, ApJ 214, 152\\
Skrutskie M.F., Meyer M.R., Whalen D., \& Hamilton C. 1996, AJ 112, 2168\\
Storzer H. \& Hollenbach D. 1999, ApJ 515, 669\\
Swenson F.J., Faulkner J., Rogers F.J., \& Iglesias C.A. 1994, ApJ 425, 286\\
Tamura M., Itoh Y., Oasa Y., \& Nakajima T. 1998, Science 282, 1095\\
Tatematsu K., Umemoto T., Kameya O., Hirano N., Hasegawa T.,
   Hayashi M., Iwata T., Kaifu N., Mikamic H., Murata Y., Nakano M., Nakano T.,
   Ohashi N., Sunada K., Takaba H., \& Yamamoto, S. 1993, ApJ 404, 643\\
Tinney C.G. 1993, ApJ 414, 279\\
Tokunaga A.T. \& Kobayashi N. 1999, AJ 117, 1010\\
Tokunaga A.T. 2000, in Allen's Astrophysical Quantities; ed. A.Cox, AIP Press, New York\\ 
van Altena W.F., Lee J.T., \& Hoffleit E.D., 1995, The General Catalogue of Trigonometric
Parallaxes (New Haven: Yale O Univ. Obs.) \\
Wainscoat R.J., Cohen M., Volk K., Walker H., \& Schwartz D.E. 1992, ApJS 83,111\\
Ward-Thompson D., Scott P.F., Hills R.E., \& Andre P. 1994, MNRAS 268, 276\\
Wilking B.A., Greene T.P., \& Meyer M.R. 1999, AJ 117, 469\\
Zapatero Osorio M.R., Bejar V.J.S., Rebolo R., Martin E.L., \& Basri G. 1999, ApJ 524, 
L115\\
Zapatero Osorio M.R., Bejar V.J.S., Martin E.L., Rebolo R., Barrado y Navascues D., 
   Bailer-Jones C.A.L., \& Mundt R. 2000, Science 290, 103

\pagebreak

\onecolumn

\begin{table}
\begin{center}
\hspace{2cm}\textbf{Table 1 - Log of Spectroscopic Observations}
\vspace{2mm}

\begin{tabular}{lccccccl}

Source & $H^{a}$ & A(V) & Mass$^{b}$ (M$_{\odot}$) & \multicolumn{2}{c}{Exposure (s)} & 
Paired with & Source Notes\\  

 & & & &$H$ band &$K$ band & & \\

095-100 & 13.25 & 0.0 & 0.090 & 2700 & - & 084-104 &\\
019-108 & 15.08 & 8.6 & 0.090 & 5376 & 5376 & 023-115&\\
055-231 & 13.27 & 0.0 & 0.080 & 960 & 640  & 069-210 & 25 nm resolution, $K$ signal reduced by cloud \\
069-210 & 13.67 & 0.0 & 0.074 & 960 & 640  & 055-231 & 25 nm resolution, $K$ signal reduced by cloud \\
068-020 & 13.83 & 1.0 & 0.073 & 720 & 560  &091-019 &\\ 
053-503 & 13.57 & 0.0 & 0.073 & 1080 & 1080 &047-436&\\
091-019 & 13.68 & 0.0 & 0.071 & 720 & 560  &068-020&\\
019-354 & 14.08 & 1.5 & 0.067 & 900 & 960  & 014-413&\\ 
066-433 & 16.54 & 12.6 & 0.066 & 5376 & -  & 061-401&\\
043-014 & 13.92 & 0.0 & 0.061 & 896 & - &  -&\\
047-436 & 14.39 & 5.5 ? & 0.06 ? & 1080 & 1080 & 053-503 & anomalously blue (I-J) colour\\
014-413 & 14.44 & 1.3	& 0.056 & 900 & 960 &  019-354 &\\
013-306 & 15.61 & 6.4 & 0.056 & 3584 & - & 016-309&\\
096-1944 & 14.96 & 0.8 & 0.042 & 2240 & 1792 &084-1939 & \\
084-1939 & 15.15 & 0.6 & 0.034 & 2240 & 1792 &096-1944&\\
084-104 & 16.89 & 0.0	& 0.016 & 2700 & - & 095-100 & \\
016-319 & 17.26 & 1.9 & 0.016 & 3584 & - & 013-306&\\
038-627 & 17.71 & 2.2 & 0.014 & 3584 & - &\\
061-401 & 17.71 & 0.0	& 0.011 & 5376 &  - & 066-433&\\
023-115 & 18.41 & 2.1 & 0.008 & 5376 & 5376 & 019-108 & $K$ spectrum corrupted by nebulosity \\ \hline
017-410 & 12.23 & 1.2 & 0.30 & 900 & 960 & 014-413& serendipitous in slit\\
073-205 & 15.94 & 11.8 & 0.080 & 960 & - & 055-231& serendipitous in slit\\
178-232 & 12.17 & 14.5 ? & 0.4 ? & 240 & - & - & proplyd, higher mass\\ \hline
Gl406 & 6.57 & 0.0 & - & - & 24 & - & M6V dwarf\\
BRI0021-0214 & 11.85 & 0.0 & - & 960 & 480 & - &M9.5V dwarf, 25 nm resolution at $H$.\\
DBD0205-1159 & 13.59 & 0.0 & - & 720 & 720 & - &L7V dwarf\\

\end{tabular}
\end{center}

Notes:\\
(a) $H$ magnitudes are in the UKIRT - UFTI system.\\
(b) The quoted masses are derived from the BM97 tracks, assuming an age of 1~Myr.

\end{table}

\begin{table}
\begin{center}
\hspace{2cm}\textbf{Table 2 - H$_{2}$ emission in 047-436}
\vspace{2mm}

\begin{tabular}{lcc}

Measured $\lambda$ ($\mu$m) &   Line  &  Equivalent Width ($\AA$)\\

1.945 &   (2-1) S(5)  &  5 $\pm$ 2 \\
1.956 &   (1-0) S(3)  & 18 $\pm$ 2 \\
2.032 &   (1-0) S(2)  &  4 $\pm$ 2 \\
2.122 &   (1-0) S(1)  & 15 $\pm$ 2 \\
2.202 & (3-2) S(3), (4-3) S(5) & 4 $\pm$ 2 \\
2.223 &   (1-0) S(0) &   5 $\pm$ 2 \\
2.249 &   (2-1) S(1) &   4 $\pm$ 2 \\
2.345 &   (4-3) S(3)? &  5 $\pm$ 2 \\
2.407 &   (1-0) Q(1) &  15 $\pm$ 3 \\
2.423 &   (1-0) Q(3) &   9 $\pm$ 3 \\
2.453 &   (1-0) Q(5) &   9 $\pm$ 3 \\
2.474 &   (1-0) Q(6) &   7 $\pm$ 3 \\   

\end{tabular}
\end{center}

\end{table}

\pagebreak

\begin{table}
\begin{center}
\hspace{2cm}\textbf{Table 3 - Spectral Types and Temperatures}
\vspace{2mm}

\begin{tabular}{lcccccccl}

Source$^{a}$ & \multicolumn{2}{c}{Water Indices$^{b}$} & \multicolumn{2}{c}{Teff(K)$^{c}$} 
& \multicolumn{2}{c}{Sp.Type$^{d}$}  & Mass (M$_{\odot}$) & Source Notes\\ 

      & W  &   Q  &   [g]=3.5 & [g]=4.0 & W & Q &  & \\ \hline       

095-100 & 0.667 $\pm$.011& 0.558 $\pm$.018&    2950  &  2750  &  L0 & L1 & 0.090 &\\
019-108 & 0.998 $\pm$.008 & 0.936 $\pm$.015&   4850  &  4900  & $<$M1 & $<$M1 & 0.090&high A(V), v. young source or background star \\
055-231  &  0.585 $\pm$.013& 0.479 $\pm$.015&  2600  &    2550  &  L3 & L4 & 0.080&\\
069-210  &  0.724 $\pm$.017& 0.597 $\pm$.031&  3400  &    2950  &  M7 & M9 & 0.074&\\ 
068-020  & 0.734 $\pm$.027& 0.555 $\pm$.021& 3450  &    3000  &  M7 & L1 & 0.073 &\\
053-503 &  0.671 $\pm$.040& 0.631 $\pm$.044& 3050  &    2750  &  L0 & M7 & 0.073&\\
091-019 &  0.776 $\pm$.023& 0.615 $\pm$.024& 3550  &    3200  &  M5 & M8 & 0.071&\\
019-354 &  0.717 $\pm$.013& 0.596 $\pm$.014& 3400  &    2900  &  M8 & M9 & 0.067& \\
066-433 & 0.945 $\pm$.051 & 0.937 $\pm$.066& 4450  &    4500  & $<$M1 & $<$M1 & 0.066&high A(V), v. young source or background star\\
043-014  & 0.631 $\pm$.044& 0.545 $\pm$.038& 2700  &    2650  &  L1 & L1 & 0.061&\\
047-436 &  0.785 $\pm$.019& 0.921 $\pm$.047&   -    &  -  &   - & -  & 0.06 ?& H$_{2}$ emission source\\
014-413 &  0.672 $\pm$.014& 0.615 $\pm$.043&  3100  &    2750  &  M9 & M8 & 0.056&\\
013-306 & 0.618 $\pm$.031& 0.594 $\pm$.018 &  2650  &    2600  &  L2 & M9 & 0.056& \\
096-1944 &  0.632 $\pm$.031& 0.562 $\pm$.022& 2700  &    2650  &  L1 & L0 & 0.042& \\
084-1939 &  0.581 $\pm$.050& 0.455 $\pm$.035& 2550  &    2550  &  L3 & L5 & 0.034& \\
084-104 &  0.471 $\pm$.066& 0.396 $\pm$.091& 2350  &    2350  &  L8 & L8 & 0.016& \\
016-319  & 0.359 $\pm$.094& 0.394 $\pm$.110& 2150  &    2100  & $>$L8 & L8 & 0.016& poor sky subtraction ?\\
038-627  & 0.601 $\pm$.072& 0.453 $\pm$.044& 2600  &    2600  &  L3 & L5 & 0.014& \\
061-401  & 0.552 $\pm$.054& 0.480 $\pm$.085& 2500  &    2500  &  L5 & L4 & 0.011& \\
023-115  & -     & 0.582 $\pm$.150 &  &  - &  -  & L0 & 0.008 \\ 
\hline
017-410 & 0.851 $\pm$.047& 0.617 $\pm$.030&  3850 & 3900 & $\sim$M0 & L2  & 0.30 & higher mass, unusual profile\\
073-205 & 0.693 $\pm$.280 & 0.469 $\pm$.082&  3300 & 2800 &  M9 & L5  &  0.080 & uncertain 
W index due to high A(V)\\
178-232 & - & 0.834 $\pm$.017 & -  & - & - & & 0.4 ? & higher mass, proplyd, A(V) uncertain\\
\end{tabular}
\end{center}
Notes:\\
(a) The 3 sources below the line are not in the primary sample, see Section 3.1.\\
(b) The W, and Q water indices are defined in the text. The quoted errors are based on
apparent noise and do not include uncertainties in dereddening or the background subtraction.\\
(c) Effective Temperatures are derived from the best fit of the AMES-Dusty-1999 spectra 
to the 1.45-1.7~${\mu}$m data at two different gravities.\\
(d) Spectral types are derived from the W and Q indices, calibrated by field dwarfs from 
Leggett et al.(2001)

\end{table}

\begin{table}
\begin{center}
\hspace{2cm}\textbf{Table 4 - CO v=2-0 detections}
\vspace{2mm}

\begin{tabular}{lcc}

Source & K$_{CO}$ & CO Detection\\ \hline

053-503 & 0.095 $\pm$0.005 & yes\\
014-413 & 0.097 $\pm$0.030 & yes\\
096-1944 & 0.059 $\pm$0.035 & yes\\
055-231  & 0.119 $\pm$0.055 & yes\\
068-020 & 0.063 $\pm$0.030 & ?\\
084-1939 & 0.032 $\pm$0.047 & no\\
091-019 & 0.023 $\pm$0.053 & no\\
047-436 & 0.034 $\pm$0.026 & no\\
019-108 & 0.039 $\pm$0.080 & no\\
069-210 & 0.153 $\pm$0.082  & no\\
019-354 & 0.058 $\pm$0.040 & no\\ \hline
DBD0205-1159 & 0.196 $\pm$0.025  & yes\\
BRI0021-0214 & 0.154 $\pm$0.007  & yes \\
Gl406 & 0.102 $\pm$0.007 & yes \\
\end{tabular}
\end{center}

Note: Sources with no CO detection typically have K$_{CO}$=0.02-0.03 due to
the wing of the H$_{2}$O absorption band H$_{2}$O centred at 2.7~$\mu$m.

\end{table}

\pagebreak

\begin{table}
\begin{center}
\hspace{2cm}\textbf{Table 5 - $K$ bandpass excesses in Planetary Mass Candidates}
\vspace{2mm}

\begin{tabular}{lcccccccl}

Source & $J$ & $H^{a}_{LR}$ & $H^{a}_{HC}$ & $K$ & A(V)$^{b}$ & E(H-K)$^{c}$ & $\Delta H^{d}$ & $K$ band excess \\ \hline

115-106 & 19.32 & 18.41 & 18.30 & 17.11 & 2.49 & 0.72 & 0.11  & yes \\ 
076-147 & 18.67 & 17.94 & 17.93 & 16.98 & 1.11 & 0.49  & 0.01  & yes \\
084-322 & 18.74 & 17.91 & 18.04 & 17.40 & 1.90 & -0.02 & 0.13  & no  \\ \hline
066-229 & 19.97 & 18.84 & 18.15 & 17.59 & 4.31 & ? & 0.69  & ? \\
137-531 & 18.89 & 17.90 & 18.32 & 17.28 & 3.38 & ? & 0.42  & ? \\
092-531 & 18.72 & 17.91 & 18.46 & 17.29 & 1.78 & ? & 0.55  & ? 

\end{tabular}
\end{center}

Notes:\\
(a) $H_{LR}$ and $H_{HC}$ are the $H$ bandpass fluxes measured in Paper I and HC 
respectively.\\
(b) Visual extinctions are derived from \it{(J-H)} \rm colours, with uncertainty of at 
least 1 mag.\\
(c) An intrinsic colour of $(H-K)=0.38$ is assumed for the photospheres.\\
(d) $\Delta H$ is the discrepancy between the Paper I and HC $H$ bandpass fluxes.

\end{table}

\end{document}